\documentclass[preprint]{aastex}
\usepackage{graphicx}
\usepackage{longtable}
\usepackage{latexsym}
\usepackage{amssymb}
\usepackage{lscape}
\usepackage{natbib}
\shorttitle{Environmental Effects on Late-Type Galaxies in Nearby Clusters}
\shortauthors{Boselli and Gavazzi}

\begin{document}
\newcommand{\dgr}{\mbox{$^\circ$}}   

\title{{Environmental Effects on Late-Type Galaxies in Nearby Clusters}}

\author{Alessandro Boselli\altaffilmark{1}} 
\email{alessandro.boselli@oamp.fr}
\affil{Laboratoire d'Astrophysique de Marseille, BP-8, Traverse du Siphon, F-13376 Marseille, France}
\author{Giuseppe Gavazzi\altaffilmark{2}}
\email{giuseppe.gavazzi@mib.infn.it}
\affil{Universit\'a degli studi di Milano-Bicocca, Piazza delle scienze 3, 20126 Milano, Italy}


\begin{abstract}

The transformations taking place in late-type galaxies
in the environment of rich clusters of galaxies at $z=0$ are reviewed. 
From the handful of late-type galaxies that inhabit local clusters,
whether they were
formed in-situ and survived as such, avoiding transformation or even destruction or 
if they are newcomers that recently infall from outside,
we can learn an important lesson
on the latest stages of galaxy evolution. 
We start by reviewing the observational scenario, covering the broadest possible 
stretch of the electromagnetic spectrum, from the gas tracers (radio and optical), the 
star formation tracers (UV and optical), the old star tracers (Near-IR) and the dust (Far-IR). 
Strong emphasis is given to the three nearby, well studied clusters
Virgo, A1367 and Coma, representative of different evolutionary stages,
from unrelaxed, spiral rich (Virgo) to relaxed, spiral poor clusters (Coma).
We continue by providing a review of models 
of galaxy interactions relevant to clusters of galaxies. 
Prototypes of various mechanisms and processes are discussed and their typical 
time-scales are given in an Appendix.\\
Observations indicate the presence of healthy late-type galaxies falling into nearby clusters individually
or belonging to massive groups. More rare are infalling galaxies belonging to compact groups  
where significant pre-processing might take place.
Once entered the cluster, they loose their gas and quench their star formation activity, 
becoming anemics.
Observations and theory agree in indicating that the interaction 
with the intergalactic medium is responsible for the gas depletion. 
This process, however, cannot be at the origin of the cluster
lenticular galaxy population. 
Physical and statistical properties of S0 in nearby clusters and at higher redshift, 
indicate that they originate from spiral galaxies transformed by gravitational interactions.
\end{abstract}


\newpage
\section{Introduction}
\label{Introduction}

On scales $<100$ Mpc, the size of the largest known coherent matter aggregates,  the 
density of galaxies in the local universe spans from $\sim$  0.2 $\rho_0$ in voids to 
$\sim$ 5 $\rho_0$ in superclusters and filaments, $\sim$ 100 $\rho_0$ in the cores of 
rich clusters, up to $\sim$ 1000 $\rho_0$ in compact groups, where $\rho_0$ is the 
average ``field'' density \citep{GELH89}.\\
It has been known for decades \citep{HUBH31} that galaxy Hubble type and local density 
are not independent quantities. In their analysis of nearby clusters, \cite{DRE80} and 
\cite{WHIG93} agreed that the fraction of spiral galaxies decreases from 80\% 
in the ``field'', to 60\% in the outskirts to virtually zero in the cores of rich clusters, 
compensated by an opposite increase of elliptical and S0 galaxies.
Morphological segregation is perhaps the clearest signature of the environmental 
dependence of the processes that govern the formation and evolution of galaxies.
Understanding the morphology segregation,  i.e. the shaping of the Hubble sequence in the 
various environments, has challenged astronomers for years: "Nature or Nurture"?
The issue is still hotly debated, as it involves many controversial fields of modern 
astrophysics, including observations, simulations and theories.\\
With the advent of 10m class telescopes we hope that  
galaxy evolution will be shortly disclosed observationally, 
much as in a rewound movie taken along the ``fossil'' sequence,  
from today's fully evolved nearby galaxies 
to the early objects forming their first stars far away in space and time.
Similarly we can imagine to shoot the movie of the morphology segregation 
by comparing frames of clusters taken at increasingly large $z$. 
Snapshots at $z=0$ would show the fully formed and evolved clusters with their final mix of 
E/S0/S. Frames taken between $z=0.2$ and 0.5 would show an unchanged fraction of E, 
with the action focused on increasingly active spirals (the Butcher-Oemler effect \citep{BUTO78, BUTO84})
and a decreasing fraction of S0 \citep{DRES97, DRES04b}. 
By $z=0.5-1$ the scene would be on clusters under formation, when large groups of galaxies coalesce. 
Effective "pre-processing" is taking place in these low velocity-dispersion environments, where 
minor merging and tidal interactions among galaxies are shaping up the S0s \citep{MIHOS04, FUJI04}. 
None has yet observed elliptical galaxies under formation at $z=3$ and beyond. But there is 
general consent that they should have formed at these early epochs, very much from 
"nature" and little from "nurture", even before clusters existed \citep{DRES04b, TREU04, NOLA04, FRAN04}.  
Maybe in a decade from now, when the JWST will provide these last photograms,
the full movie of the shaping of the Hubble sequence will be on view, 
as many groups of researchers are focusing large efforts on it (the Morphs group \citep{DRES99}, the CNOC1
group \citep{YEE96}, the ACSCS team \citep{POST96}, the EDisCS group \citep{HALL04} and the 
team of van Dokkum and collaborators \citep{VANDOK00}).\\
This is perhaps somewhat optimistic, however.
Pretending to understand galaxy evolution really means shooting a movie whose individual photograms are
composed of SEDs, obtained combining best quality data taken over a large stretch of the electromagnetic spectrum, 
each individual band providing information on the evolution of a particular galaxy component.  
Restframe UV fluxes and Balmer emission line strengths are the best 
current, massive star-formation tracers. Optical bands provide the morphology, 
NIR bands trace the oldest stars, MIR and FIR the dust 
(and extinction); the radio (continuum) tells about the nuclear activity, the supernova rate and the magnetic fields, and 
the 21 cm line the primordial gas reservoir and its dynamics.\\
The problem is that, as $z$ increases, characteristic features become unobservable or more difficult to observe.  
By $z=1$, for example, the H$\alpha$ line ends up in the H band, a noisy place to carry out observations.
The peak of the dust emission falls in the sub-mm bands where the sensitivity of telescopes will remain poor
until ALMA.
The 21 cm line drops to lower sensitivity and angular resolution bands where line 
correlators do not yet exist (we are not aware of a single HI emission measurement taken beyond $z$=0.15).
Furthermore, the movie's linear resolution decreases with $z$, so that detailed morphological 
information quickly fades away. 
Most dramatic is the lost of sensitivity with increasing $z$. 
By $z<0.2$ dwarf galaxies run completely out of the scene.
Meaningful, high-resolution SEDs in the whole luminosity range covered 
by galaxies are obtainable only at $z=0$, 
providing the boundary conditions to models and simulations of galaxy evolution.\\
The end point of the evolution is within the reach of modern telescopes. 
Significantly less evolved galaxies can only be observed, with lower resolution, if they are more luminous.  
Dwarf galaxies, Ims, BCDs and dEs, by far the most common types of galaxies in the universe,
yet the less understood, can only be observed locally. \\
Since the rate of evolution of galaxies is found to correlate primarily with their dynamical mass (infrared luminosity) 
according to a remarkably anti-hierarchical scheme (\cite{GAV03, GAVSC96, 
GAVPB96a, BOSG01, GAVB02c, 2002A&A...384..812S}) that has been lately named ``downsizing'' ,
distant surveys might undersample low-luminosity, slowly evolving systems.  
This is why studying nearby clusters still is a worthwhile activity. \\
Another pro of local clusters is that they are ideal laboratories for studying 
interaction mechanisms at high resolution. Prototypes of all relevant mechanisms can be found
in nearby clusters. Abell 1367, for example, contains perhaps the best known example of 
currently ongoing ram-pressure stripping (CGCG 97-073), a rare example of a merging system in a cluster (UGC 6697) and
even a local prototype of infalling group where 
pre-processing occurs at the present cosmological epoch (see Section \ref{prototype}).
Infall was much more effective, but unfortunately hard to observe, in the past ($z \sim 0.5$), 
when it contributed shaping the S0 galaxies (pre-processing).\\  
So far we insisted on the importance of observing nearby clusters.
Among them we have focused the present review on three specific ones: Virgo,
Abell 1367 and Coma that are among the best studied in the universe. The variety of their 
conditions, e.g. spiral fraction, X-ray luminosity, kinematic properties, makes them 
suitable ``laboratories'' for comparing clusters in different evolutionary stages:
near formation (Virgo), still significantly accreting (A1367), near virialization (Coma).\\ 
Honestly speaking, the present review was conceived while we were assembling ``GOLDmine'',
the galaxy database that we made available on the WEB (goldmine.mib.infn.it) \citep{GAVB03a}. 
Two third of the galaxies contained in GOLDmine belong in fact to these three clusters
and a significant fraction of them are among the best sampled SEDs.
This does not mean that we haven't tried to review the work of other groups.
Certainly we haven't done it as extensively as we should, and we apologize for that.\\
But why focusing the present review on late-type galaxies, that are less numerous, they evolve
slower than the more intriguing S0s and Es? 
What can we learn from the handful late-type galaxies that still inhabit local clusters?
There are several reasons for concentrating on them. One is that, due to the 
fragility of their stellar and gaseous disks, they are sensitive probes of their environment.
Another is that, since their rate of evolution (star formation history) is mild and continuous, 
they provide a linear clock for evolution,
as opposed to E whose action occurred suddenly in one brief and intense episode that occurred
early in their life.  
Spirals are also interesting because significant infall of galaxies of this type
has occurred and is still occurring (at decreasing rate due to the $\Lambda$ dominated universe)
onto clusters of galaxies. Will they sooner or later evolve into lenticulars? Can we use 
them as probes of the cluster environment? 
\begin{figure*}[!htb]
\epsscale{0.7}
\plotone{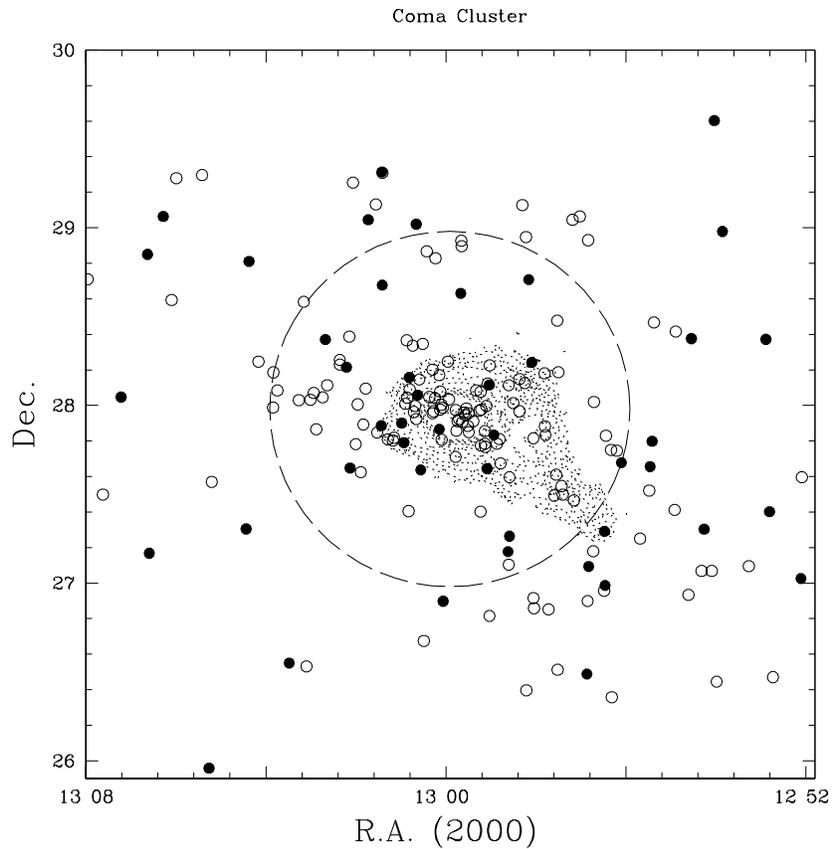}
\caption{The celestial distribution of galaxies brighter than
15.7 from the CGCG catalogue, \cite{CGCG} in a 4x4 $\rm deg^2$ box around the Coma
cluster. The 146 early-type (E+S0+S0a) galaxies (empty symbols) clearly mark the cluster density enhancement,
whereas the 49 late-type ($\geq$Sa) objects (filled symbols, including 10 unclassified spirals) hardly trace
the cluster. The circle of 1 degree radius is traced about  the X-ray center.
The X-ray contours from XMM are superposed.} 
\label{coma_map}
\end{figure*}
To help better focusing the addressed issues, let's take a look at the
Coma cluster whose bright galaxy distribution is shown in Figure \ref{coma_map}.
The early-type galaxies clearly mark the cluster density enhancement,
whereas the late-type objects hardly trace the cluster at all, making \cite{ABE65}
to doubt their very membership to the Coma cluster. 
The latter are very ``healthy'' objects, actively star forming, as witnessed by their average 
H$\alpha$ E.W. (equivalent width) of 35 \AA~ (with peaks of 78 \AA).
This perhaps explains why Abell questioned their membership to Coma,
because the majority of galaxies observed in clusters shows no signs of recent star formation activity.
He didn't know that some of these objects are not as in ``good shape'' as it looks at first glance, 
something that was realized
20 years later. The pioneering work of \cite{GIOH85}, continued by many other researchers, among them 
\cite{GAV87, GAV89}, showed in fact that galaxies at
projected distances $<$ 1 Mpc from the center of Coma suffer from significant neutral hydrogen deficiency
(see Section \ref{The atomic gas}), having lost from 30 \% to 90 \% of their original HI content. 
The first resolved map of their
HI distribution was obtained even more recently by \cite{BRAC00}. 
It shows that in most highly
or moderately deficient objects the HI distribution is asymmetric or significantly displaced from
the parent optical galaxy. This phenomenon occurs in objects projected within the
X-ray emitting gas in the Coma cluster. For the interpretation of this observation, whether is due
to ram-pressure stripping or to galaxy-galaxy or galaxy-cluster interactions,
the reader is referred to the remaining of the discussion. However, generally speaking, 
a high HI deficiency represents perhaps
the clearest signature of the interaction occurring to galaxies in the environment of rich
clusters, in other words it gives a clear clue to their cluster membership. 
Abell was wrong! These are bona fide cluster members.
What future is reserved to these galaxies? 
This is the type of issue addressed by the present review.
The reader can find many excellent review papers in the literature 
that deal with the evolution of galaxies in the cluster
environment \citep{HAYGC84,DRES84,SAR86,DRES97,POGB01a,DRID03,VANGO03}
each focused on a particular aspect or wavelength domain. 
The most recent achievements in cluster research 
are collected in the proceedings of the Carnegie Observatories Astrophysics Series, Vol. 3: 
Clusters of Galaxies: Probes of Cosmological Structure and Galaxy Evolution (2004). 
With the present review we wish, for the first time, to consider together all aspects. The 
price we pay is that we must narrow down the reviewed interval of $z$, in practice to $z$=0.  
We start by reviewing the observational scenario (Sections \ref{constituents}, \ref{obs_diff}), 
covering the broadest possible 
stretch of the electromagnetic spectrum, from the gas tracers (radio and optical), the 
star formation tracers (UV and optical), the old star tracers (Near-IR) and the dust (Far-IR). 
We continue by providing a review of models 
of galaxy interactions relevant to clusters of galaxies (Section \ref{models}). 
We conclude with a discussion on what proposed 
model(s) better interprets various aspects of the observational scenario (Section \ref{Discussion}),
once again checking the models predictions on recent data on Virgo, Coma and A1367.

\section{The constituents of late-type galaxies in nearby clusters: comparison with the ``field''}
\label{constituents}

\subsection{The atomic hydrogen}
\label{The atomic gas}
\begin{figure*}[!htb]
\epsscale{1.0}
\plottwo{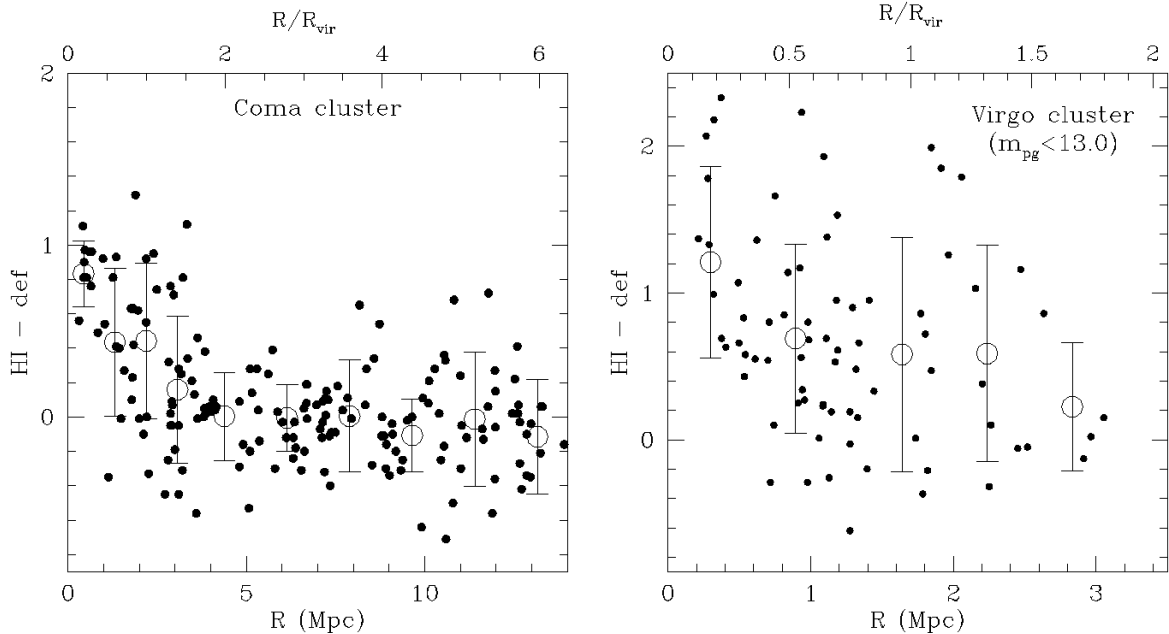}{fig02b.epsi}
\caption{The projected distribution about the X-ray center of the HI deficiency 
of similar luminosity late-type members of the Coma (left) and Virgo clusters (right). 
Large circles represent averages with one $\sigma$ uncertainties 
(adapted from \cite{GAVON05,GAVB04}).}
\label{comadef}
\end{figure*}

The atomic gas is the principal component of the ISM in late-type galaxies, 
supplying the ``fuel'' that feeds the star formation.
In normal, isolated galaxies the HI gas distribution outstands the optical disk: 
column densities of $\sim$ 10$^{20}$ atoms cm$^{-2}$ are observed at $\sim$ 
1.8 the optical diameter, with a relatively flat radial distribution, sometimes
showing a central dip \citep{CAYK94, SALH96, BROR97}. 
The HI extension in the direction perpendicular to the disk is comparable to the optical thickness \citep{LEEI97}.\\ 
Because of its distribution the HI gas is  weakly bound
by the galaxy's gravitational potential well, thus it can be easily removed.
It is therefore obvious to expect that the HI content and distribution of galaxies in clusters
differ from those of isolated galaxies (see the reviews by \cite{HAYGC84, VANGO03}).
The first study of the atomic gas properties of galaxies in the Virgo cluster
dates back to the sixties \citep{ROBK65}. Since then, several HI surveys
of this and other nearby clusters were carried out using single
dish radiotelescopes \citep{GIOH85} and interferometers \citep{WAR88, CAYG90, BRAC00}. 
Owing to the survey of \cite{HOFS96} and to recent Arecibo observations
(e.g. \cite{GAVB04}) HI data are available
for the entire late-type galaxy population in the Virgo cluster, including dwarf irregulars and BCDs.\\
All works agree on the general conclusion that cluster galaxies have on average 
a lower atomic gas content than their isolated counterparts, as first noticed by \cite{DAVL73}. 
A quantitative determination of this difference could however not be assessed before  
the systematic HI properties of isolated galaxies were known.
This fundamental step was achieved by \cite{HAYG84}
who studied at Arecibo a reference sample selected from the \cite{KAR73} 
catalogue of isolated galaxies. \cite{HAYG84} defined 
the HI-deficiency parameter as the logarithmic difference 
between the observed HI mass and the expected value in isolated objects of similar 
morphological type and linear size. Galaxies with an HI-deficiency parameter 
$\leq 0.3$ can be treated as unperturbed objects\footnote{The isolated sample by \cite{HAYG84} is
mainly composed  of giant galaxies. The HI-deficiency parameter of dwarf objects is therefore poorly calibrated}.\\
By comparing the statistical HI properties of galaxies in 9 nearby clusters
with those of ``field'' objects of similar type and luminosity, \cite{HAYGC84}
and \cite{GIOH85}
showed that clusters contain a large fraction of HI deficient galaxies, 
with the exception of unrelaxed, loose ones.
This fraction is a strong function of the angular distance 
from the X-ray center \citep{GAV89, BOS94}.\\
\cite{DRE86}, by re-analyzing the \cite{HAYG84} catalogue showed that the most deficient
galaxies are generally early-type spirals in radial orbits, 
while gas-rich galaxies have isotropic or circular orbits (confirmed 
by \cite{SOLM01}).
\cite{SOLM01}, analyzing the HI properties of $\sim$ 1900 
galaxies in 18 nearby clusters, concluded that 2/3 of the clusters are composed of
HI deficient galaxies inside the Abell radius $R_A$. The residuals of the deficiency parameter is related to
the galaxy morphology and not to their optical size: early-type spirals and probably dwarfs are 
more deficient than Sbc-Sc at any radii inside 2$R_A$. 
The deficiency increases toward the cluster center, being significant for $r<R_A$. 
They did not  confirm the relationship
between the fraction of HI-deficient galaxies and the cluster X-ray luminosity claimed by \cite{GIOH85}.
\\
\cite{GAVON05} have completed the HI observations of late-type CGCG galaxies in the Coma supercluster.
One important issue that they were able to address, given the completeness of their data-set, 
is on what scale the phenomenon of HI ablation holds.
The HI deficiency parameter of individual galaxies is given in Fig.\ref{comadef} as a function of the  projected
linear separation from the X-ray center of the Coma cluster, out to 15 Mpc, along with average values
taken in bins of 0.5\dgr~ (from 0\dgr~ to 2\dgr) and in bins of 1\dgr~ further out.  It is apparent  
that significant HI deficiency occurs out to approximately 3 Mpc radius. Little above one virial radius 
(i.e. at 2.2 Mpc, \cite{GIRG98}), the average HI content of the supercluster galaxies becomes 
indistinguishable from that of the field, in agreement with \cite{SOLM01}.
We stress that the filaments constituting the "Homunculus legs" \citep{LAPP86}, i.e. the groups/filaments 
in the immediate surrounding of the Coma cluster, perhaps infalling toward Coma (not shown in Fig.\ref{comadef}), 
are made of galaxies retaining 100 \%  of their HI content, identical to unperturbed, isolated galaxies. \\
The HI survey of the Virgo cluster has also been completed by \cite{GAVB04}, allowing to determine
the radial distribution of the HI deficiency parameter of all spirals brighter than 18.0 mag 
(only $m_p<13.0$ mag are plotted in Fig.\ref{comadef} for consistency with Coma).  
Unfortunately for Virgo the survey does not
extend out in the low density region of the local supercluster as much as in Coma, but only out to 3.2 Mpc.
We find that in Virgo there is a significant population of non-deficient galaxies at all radii, mixed 
with highly deficient ($>$1) ones. These exist within 2.2 Mpc, just above 1 virial radius (i.e. at 1.7 Mpc,
\cite{GIRG98}).
Furthermore it is found that individual groups constituting the Virgo cluster
(cloud N and S for example) contain less than 10 \% highly deficient objects, 
compared to 33\% in Cluster A (M87).
\\
High resolution HI mapping of galaxies in the Virgo cluster with the Westerbork \citep{WAR88} and VLA 
\citep{CAYG90} interferometers, and more recently in the Coma cluster 
\citep{BRAC00} revealed that cluster galaxies not only suffer from significant
HI depletion,  but that their HI distribution is less extended than
isolated objects of similar morphological type and luminosity \citep{WAR88, CAYK94, BOSL02A}.\\
While in normal isolated late-type galaxies the HI diameter (as measured at the 10$^{20}$ atoms cm$^{-2}$ isophote) 
is on average a factor of $\sim$ 1.8 larger than the optical isophotal $B$ band 
25 mag arcsec$^{-2}$ diameter (somewhat depending on the morphological type), 
in cluster objects the HI to optical diameter ratio strongly depends on the HI-deficiency
parameter and on the distance from the cluster center \citep{WAR88, CAYK94}.
The most HI-deficient objects close to the X-ray center of Virgo 
have HI diameters up to $\sim$ 5 times smaller than
the optical ones \citep{CAYK94}.  Anemic 
galaxies whose HI disks are slightly smaller than the optical ones, have significantly lower
HI column densities (by a factor of $\sim$ 10) than normal, isolated objects of similar 
morphological type \citep{CAYK94}. The HI surface density correlates tightly with the 
HI-deficiency parameter \citep{BOSL02A}.
Asymmetries in the HI distribution and displacements of the gas from the optical parent galaxies
are frequently observed in moderately HI deficient objects \citep{GAV89, CAYK94, BRAC00, VOG04a}. \\
 
The HI mass function of the Virgo cluster was found by \cite{GAVB04} to differ 
significantly from  the one in the ``field'' due to the combined effect of morphology
segregation and presence of HI-deficient objects. A similar result was recently obtained by \cite{SPRI05}
who found, on average, a decrease of $M_*$ and a flattening of the low-mass end of the HI mass function
in high density environments.\\
The recently initiated Arecibo Legacy Fast ALFA (ALFALFA) survey aims to map $\sim$7000 $\rm deg^2$ 
of the high Galactic latitude sky visible from Arecibo \citep{GIO05}. The survey
will detect on the order of 20,000 extragalactic H I line sources out to $z\sim0.06$
and enable a first wide-area blind search for local H I emitters. Environmental studies will 
greatly benefit from this survey.

\subsection{The molecular hydrogen}
\label{The molecular gas}
The molecular gas phase, even if it represents only $\sim$ 15 \% of the total gas reservoir
in normal, late-type galaxies \citep{BOSL02A}, is the component of the 
ISM that takes direct part to the process of star formation.
The HI gas has in fact to condense inside molecular clouds before collapsing to form stars.
It is thus expected that any possible external perturbation induced by the environment on the 
molecular gas can have important consequences on the star formation activity, and
thus on the evolution of late-type galaxies.\\  
Early systematic works aimed at determining the molecular gas properties of late-type galaxies 
in clusters are the CO surveys of Virgo by \cite{STAK86}, \cite{KENY88a} and \cite{BOSC95a}
and that of Coma by \cite{CASB91}. Using a $standard$ conversion factor $X$=$n(H_2)/I(CO)$ between 
the intensity of the CO line and the column density of the molecular hydrogen 
calibrated on the Milky Way, these authors estimated the molecular hydrogen mass $M(H_2)$ of 
cluster galaxies. Their immediate conclusion was that cluster galaxies have
a normal molecular gas content and the interpretation was that
molecular gas, being more centrally peaked, deep inside the potential well of the galaxy, than the atomic 
gas, cannot be easily removed by any stripping mechanism \citep{KENY89, RENI92}. 
Molecular gas removal could be effective only in low-mass galaxies with shallow potential wells,
explaining the apparent CO deficiency of low-luminosity Virgo galaxies \citep{KENY88b}.\\
This result suffers however from several biases. It firstly resides on the
hypothesis that the relation between the CO line intensity and the H$_2$ gas column density
is universal. \cite{BOSL02A} showed however that $X$ varies by more than a factor of 10 
over the sampled range of UV radiation field, metallicity, morphological type and luminosity.
The second limitation is the lack of a CO reference sample of isolated galaxies selected
according to criteria similar to the cluster \citep{BOSGL97b}. For example 
the CO survey by \cite{CASB91} is mostly FIR selected, 
thus biased toward strong CO emitters \citep{BOSG95b}.\\
To overcome these difficulties \cite{BOSGL97b} and \cite{SAU03} obtained CO observation 
of an optically selected sample of 
isolated galaxies. They applied a luminosity-dependent 
CO to H$_2$ conversion factor, calibrated on the nearby galaxies with an independent measure of $X$:
\begin{equation}
{log X=-0.38(\pm0.06)logL_H +24.23(\pm 0.24)} ~~~~   {\rm (mol~cm^{-2}(K~km~s^{-1})^{-1})}
\end{equation}
From a $H$ band luminosity-H$_2$ mass relation calibrated on the reference sample of isolated galaxies
selected following similar criteria \citep{BOSL02A}:
\begin{equation}
{log M(H_2)_e=3.28(\pm0.39)+0.51(\pm0.05)logL_H} ~~~~ {\rm (in~solar~units)}
\label{eq2}
\end{equation}
they obtained, in full analogy with the HI deficiency \citep{HAYG84} an 
H$_2$ deficiency parameter defined as:
\begin{equation}
{Def_{H_2}=log M(H_2)_e - log M(H_2)_o}
\end{equation}
where $M(H_2)_e$ is the expected molecular gas mass of
a galaxy of a given $H$ luminosity as determined from eq. \ref{eq2} and
$M(H_2)_o$ is the observed molecular gas mass \citep{BOSL02A}.\\
The lack of relationship between the HI and the H$_2$ deficiency parameters
confirms and extends to lower luminosities similar previous claims obtained in 
Coma \citep{BOSGL97b, CASB91} and in Virgo \citep{KENY89, BOS94}. 
The low luminosity, CO deficient spiral galaxies found in Virgo by \cite{KENY88b} are not necessarily
deficient in molecular hydrogen since their H$_2$  mass was probably underestimated
assuming a constant $X$ conversion factor.\\
We conclude that cluster galaxies are significantly deficient in atomic hydrogen,
but their molecular content is, on average, normal.

\subsection{The metal content}
\label{The metallicity}

Metals are produced and injected into the interstellar medium by massive stars 
through stellar winds and supernova explosions. They play a fundamental role in the process of star formation.
Since they are produced by stars during their evolution, they can be used to constrain
the star formation history of galaxies.\\
\cite{SKIK96} analyzed the effects of the environment
on the metal content of spiral galaxies by comparing the metallicity gradients observed 
in 9 objects in the Virgo cluster with similar measurements of isolated
objects of equal type. The 9 galaxies were selected to span a large range in HI-deficiency 
and angular distance from the cluster center.
HI-deficient galaxies in the center of Virgo have, on average, higher metallicity (0.3 to 0.5 dex in O/H)
than HI normal galaxies at the periphery of the cluster or isolated objects with similar morphology 
and luminosity \citep{SKIK96, PILM02}. 
Moreover \cite{SKIK96} showed that HI deficient Virgo
galaxies have shallower abundance gradients than normal objects.
Gavazzi et al. (2006, in preparation) recently analyzed the HI-deficiency metallicity 
relation using integrated long slit spectra
of a large sample of $\sim$ 300 galaxies in the Virgo cluster and in the Coma/A1367 
supercluster. 
\begin{figure}[!htb]
\epsscale{0.6}
\plotone{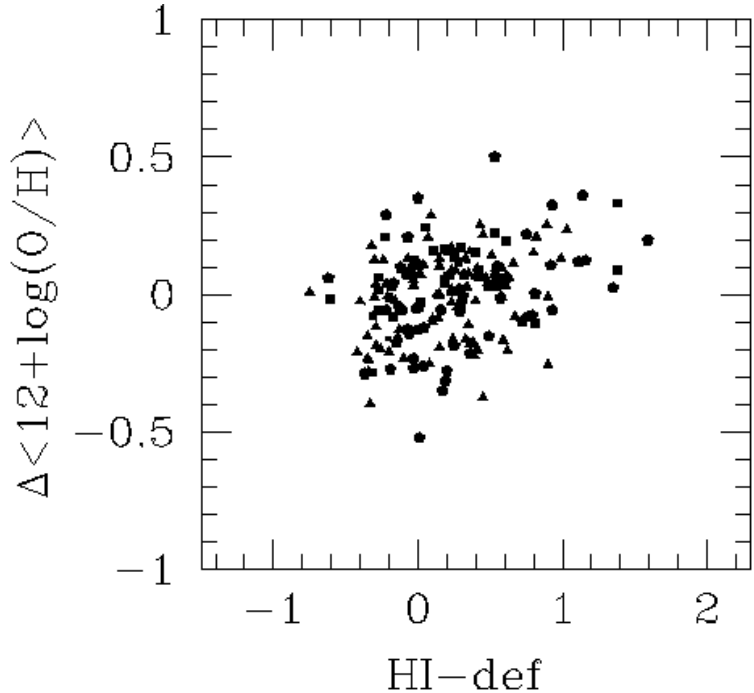}
\caption{The relationship between the residual from the metallicity (12+Log(O/H)) vs. luminosity
($L_V$) relation and the HI deficiency parameter. The
metallicity has been derived averaging the three calibrations of 
\cite{ZEES98}, \cite{DOPK00} and \cite{KOBK99}. Squares represent Sa-Sb,
triangles Sbc-Scd, pentagons Sd-Im-BCD.}
\label{metdef}
\end{figure}
Their analysis (Fig.\ref{metdef}) shows that the most HI deficient galaxies have on average 
larger metallicities than similar objects
with a normal gas content, confirming the results of \cite{SKIK96}
\footnote{Remember however that the trend between metallicity residual and HI-deficiency
is exclusively observed when metallicities are determined using the 
calibration of \cite{ZEES98} which is based on [NII]/H$\alpha$. Due to a bias,
the other two methods, requiring the detection of [OIII], H$\beta$ and [OII],
cannot be applied to highly HI deficient galaxies because these lines are not detected
in these objects.}.
A similar trend was also observed for the dwarfs, which, in dense environments, seem to have 
higher metallicities than isolated objects \citep{VIL95}. 
The unanimous interpretation of this finding is that metals in cluster galaxies are less
diluted in the unpolluted gas, due to HI ablation. 

\subsection{The dust}
\label{The dust}
Dust grains of different size and composition build up from metals
produced by stars and injected into the ISM by stellar winds and supernovae explosions. 
Dust plays an important role in the process
of star formation since atoms of hydrogen condense on dust grains   
to form H$_2$ molecules \citep{HOLW71}. 
The interstellar dust is observable in absorption at ultraviolet
and optical wavelengths and in emission in the mid- and far-IR, as
the Infrared Astronomy Satellite (IRAS) dramatically revealed \citep{SOIN87}.
While most of the flux emitted by dust in normal galaxies is in the 60-200 $\mu$m domain (e.g. \cite{BOSGS03a})
(relatively warm, $T_{dust}$=30 K big grains), the bulk of the dust mass is colder ($T_{dust}$$\leq$ 15 K), and it
radiates in the sub-mm domain at $> \rm 100 \mu$m \citep{BIAD99}. 
\begin{figure}[!htb]
\epsscale{0.6}
\plotone{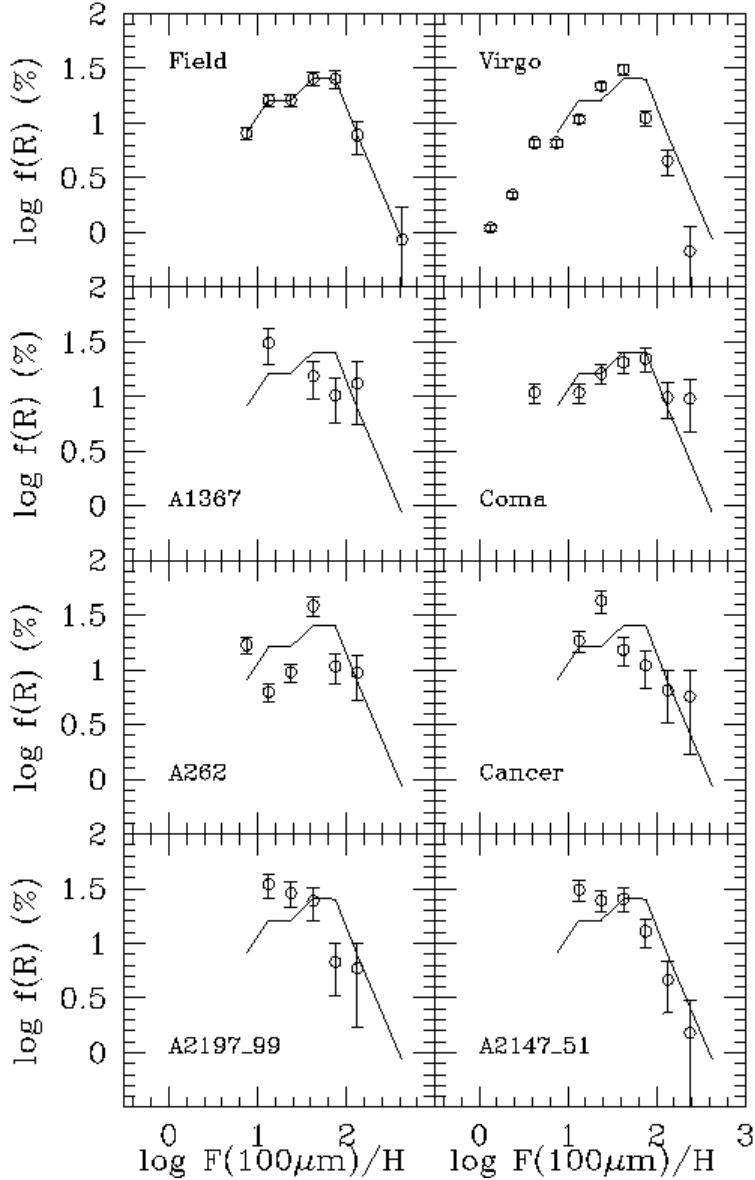}
\caption{The differential far-IR to near-IR ($F$(100 $\mu$m)/$H$) distribution in seven 
nearby clusters (dots) and in 
the reference sample of isolated galaxies (continuum line).
}
\label{firlf}
\end{figure}
At intermediate wavelengths (15-60 $\mu$m) the emission is
dominated by very small grains, while in the mid-IR (5-15 $\mu$m)  it comes mostly from
unidentified infrared bands (UIB), probably associated to planar molecules, 
called PAHs \citep{DESB90}.\\
Dust is generally associated with the gaseous component: the dust to gas column density ratio 
is relatively well correlated to the metallicity of the ISM, and thus it varies with galaxy morphology and/or 
luminosity or, within a galaxy, it follows the metallicity gradient. 
It is thus expected that part of the dust associated with the atomic gas can be removed from cluster spirals
and might contribute to the IGM pollution \footnote{For a recent review on the infrared properties of clusters
of galaxies observed by ISO see \cite{METC05}}.\\
The mid-IR (5-18 $\mu$m) emission per unit galaxy mass 
of normal, late-type galaxies in the Virgo cluster does 
not depend on the distance from the cluster center \citep{BOSL98}, thus 
it can be tentatively concluded that the carriers of the IR bands, dominating
the emission at these wavelengths, are not affected  by the cluster environment.\\
Using far-IR IRAS data for a sample of $\sim$ 200 galaxies in 7 nearby clusters, 
\cite{BICG87} showed that optically selected cluster galaxies have
a normal far-IR (60-100 $\mu$m) emission. There is however a lack of bright
($L_{FIR}>$10$^{11}$ L$\odot$) galaxies in clusters, and there is also a trend between the dust
temperature and the HI deficiency, with HI deficient objects having colder dust.
This result was confirmed in the Virgo cluster by \cite{DOYJ89}, who also 
found that HI-deficient galaxies have slightly lower 60 and 100 $\mu$m flux densities per unit optical area  
than HI normal galaxies.\\
Using far-IR flux densities normalized to the photographic magnitude,
\cite{GAV88} showed that the far-IR luminosity function 
of isolated and cluster galaxies are similar, except for the lack of objects with high far-IR to optical ratio, 
as claimed by \cite{BICG87}.\\
Using the data on 9 nearby clusters and on the Great Wall (reference sample) obtained from GOLDmine 
we re-determined the far-IR luminosity distribution (see Fig. \ref{firlf}),
but this time using far-IR (100 $\mu$m) normalized to $H$ band flux densities\footnote{
Normalizations are necessary to remove the first order dependences on luminosity (``bigger galaxies have more 
of everything''). As shown by \cite{GAVPB96a}, the proper normalization is the one 
obtained using the near-IR $H$ band luminosity 
because this parameter
traces the total dynamical mass of galaxies, it suffers from low internal extinction
and it is not affected by recent events of star formation.}.
\begin{figure}[!htb]
\epsscale{0.8}
\plotone{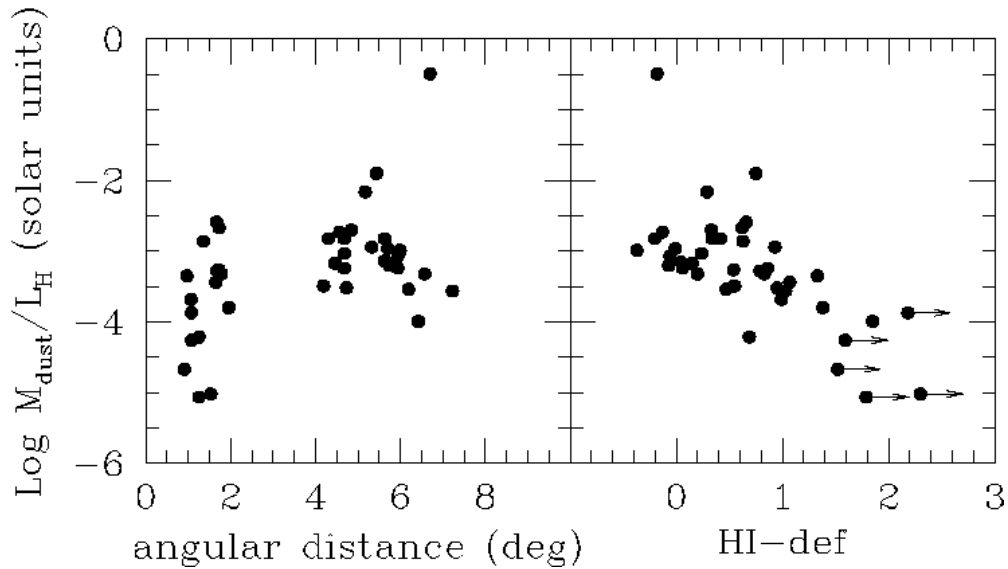}
\caption{The relationship between the total dust mass normalized to the $H$ band luminosity 
and a) the angular distance from the cluster center and b) the HI deficiency parameter 
for Virgo cluster galaxies observed at 170 $\mu$m with ISOPHOT.
Filled dots are for HI detected galaxies,  arrows for undetected objects (upper limits to the HI mass).
}
\label{dusthiang}
\end{figure}
\begin{figure}[!htb]
\epsscale{0.5}
\plotone{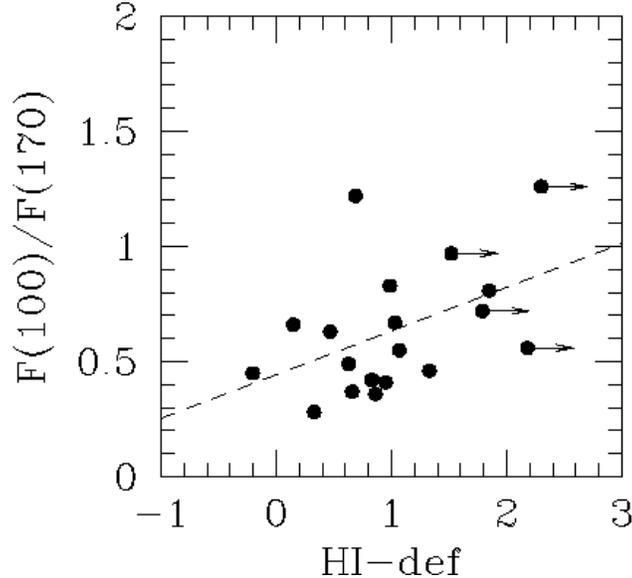}
\caption{The relationship between the 100 to 170 $\mu$m flux density ratio 
and the HI deficiency parameter for Virgo cluster galaxies.
Filled dots are for HI detected galaxies,  arrows for HI undetected objects.
The dashed line represents the best fit to the data.
}
\label{firdef}
\end{figure}
As for the far-IR to optical luminosity distribution, we do not observe any systematic 
difference between cluster and ``field'' galaxies.
The emission of normal galaxies in the 60-100 $\mu$m range is however primarily dictated by the
intensity of the UV and visible stellar radiation field, 
and only marginally by the dust mass. Therefore the above test is not sufficient to exclude  
dust removal in cluster galaxies.\\
The lack of homogeneously selected photometric data in the range 100 $\mu$m-1 mm for  
cluster and isolated galaxies  
prevents us from studying the possible effects of the environment on the cold dust component
using a similar statistical approach.
At present the only systematic observations at this wavelength are the ISOPHOT
170 $\mu$m data of 63 late-type galaxies in the Virgo cluster by \cite{TUFP02}. 
Using these data we show in Fig. \ref{dusthiang} the relationship between the total dust mass 
(defined as the sum of the cold and warm 
dust masses given by \cite{TUFP02}, normalized to the total galaxy mass, as traced 
by the $H$ band luminosity) vs. the angular distance from the cluster center
and vs. the HI-deficiency parameter.
Both figures show an interesting trend of decreasing dust content with decreasing clustercentric distance 
and with increasing HI-deficiency, that could suggest dust removal in gas stripped galaxies.
The reader should however take this result with some caution because it might be biased
by a possible dust mass vs. morphology relation, an effect that remains to be confirmed when  
more far-IR data will be available, in particular of early-type spirals.
A safer test is shown in Fig. \ref{firdef} where 
the 100 to 170 $\mu$m flux density ratio (which gives
a model independent dust temperature, thus is a tracer of the dust mass because
$M_{dust}$ $\alpha$ $T_{dust}^{-6}$, for a $\beta$ emissivity parameter equal to 2, \cite{BIAD99})
is plotted against the HI-deficiency parameter. 
The sample has been limited to the massive spirals earlier than Scd.
Figure \ref{firdef} indicates a trend of the dust temperature (and thus an inverse trend of the dust mass)
with the HI-deficiency, 
suggesting that the external cold dust might be swept together with the neutral hydrogen
in stripped galaxies 
\footnote{Notice that this trend is opposite to 
the one between the 60 to 100 $\mu$m flux density ratio and the HI-deficiency (decrease of temperature 
with the deficiency) found by \cite{BICG87}}. \\
Given the poor statistics and the
aforementioned lack of a reference sample, dust depletion in cluster galaxies requires confirmation. 
The 60-170 $\mu$m all-sky survey that will be soon carried out by the Japanese satellite ASTRO-F (spring 2006) 
will help disclosing this issue. 

\subsection{The cosmic rays}

The 1400 MHz (20 cm) radio continuum emission of normal, late-type galaxies is dominated
by the synchrotron emission of cosmic ray electrons on magnetic fields. 
The contribution of thermal emission is negligible ($\leq$ 10 \%)
in bright disk galaxies \citep{GIOG82, KLE90}, 
but it can be relevant (up to 60-70\%) in BCD galaxies \citep{KLEW91}.
The tight relationship between the radio continuum and the H$\alpha$ emission 
observed in normal galaxies
demonstrates that cosmic-ray electrons are accelerated by the
young stellar population, probably in supernova remnants \citep{LEQ71, KEN83a, GAVBK91a}.\\
\label{radio continuum}
\begin{figure}[!htb]
\epsscale{0.8}
\plotone{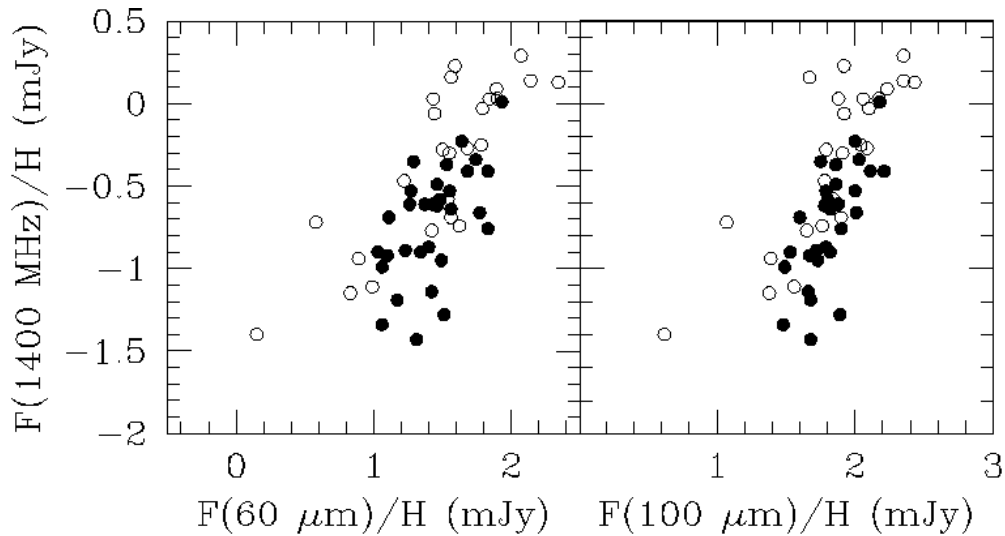}
\caption{The relationship between radio continuum (at 1400 MHz) and the far-IR at 60 (left panel)
and 100 (right panel) $\mu$m emission (normalized to the $H$ band luminosity) for galaxies
in the two clusters Coma and A1367 (open symbols) and in the ``field'' (filled symbols).
}
\label{radiofir}
\end{figure}
\cite{GAVJ86} in their first systematic comparison of the radio continuum properties of 
late-type galaxies in 
different density environments concluded that spiral galaxies in 
the Coma and A1367 clusters have their radio continuum emission enhanced by $\sim$ 5 times with 
respect to isolated objects. This result was confirmed by \cite{ANDO95}, \cite{RENK97}
and \cite{GAVB99a}. A marginal increase in the radio emissivity is observed in Virgo and A262, 
and none in loose clusters such as Cancer \citep{GAVB99a}.\\
\cite{GAVBK91a} showed that the far-IR radio correlation, shared by late-type galaxies 
in a huge luminosity interval, is different for cluster and isolated galaxies.
The cluster objects have an increased radio emissivity, at any given far-IR luminosity, 
with respect to ``field'' objects. 
This is shown in
Fig. \ref{radiofir}, where the normalized (to $H$) radio continuum emission at 20 cm   is plotted versus the 
normalized (to $H$) far-IR (60 and 100 $\mu$m)
emission for galaxies in the Coma and A1367 clusters (open symbols) 
and in the ``field'' (filled symbols).
A similar increase has been observed at other frequencies (6.3 and 2.8 cm) in 
some galaxies close to the Virgo center by \cite{NIKK95}.\\
The synchrotron emissivity $\epsilon_{\nu}$ is proportional to 
the density of relativistic electrons $n_e$ and to the magnetic field density $B$ as \citep{RYBL79}:
\begin{equation}
{\epsilon_{\nu} \propto n_e \times B^{(1+\eta)/2} \times \nu^{-\alpha_r}}~~~~~ \rm{(erg~cm^{-3} s^{-1} strd^{-1}Hz^{-1})}
\end{equation}
\noindent
with typical spectral slope $\alpha_r$=0.8 \citep{KLE90, GIOG82}.
The density of relativistic electrons $n_e$, which are accelerated in supernova remnants, 
is proportional to the number of young stars,  thus to the ongoing star formation rate.
Since both the current star formation activity and the far-IR emission of cluster galaxies 
are comparable to, or lower than those of isolated objects (see Sections \ref{The dust}, 
\ref{star formation}), the observed increased
radio emissivity and radio to far-IR ratio of cluster objects implies a factor of $\sim$ 2-3 
increase of the magnetic field density, possibly due to compression \citep{GAVB99a} or most likely due to
shock induced re-acceleration of relativistic electrons, as proposed by \cite{VOLX94}.

\subsection{The star formation activity}
\label{star formation}


The current star formation activity of galaxies can be studied using both direct
and indirect tracers\footnote{Indirect tracers of the star formation rate,  
such as the far-IR and radio continuum luminosities, should be used with caution for cluster galaxies
because, as discussed in Sections \ref{The dust} and \ref{radio continuum}, 
these quantities might not be reliable indicators of the star formation activity,
due to dust removal or magnetic field compression}. Any luminosity tracer of the young stellar population can be translated into
a star formation rate (in M$\odot$ yrs$^{-1}$) using stellar population synthesis models,
provided that the star formation activity is constant over a time scale similar to the age of the emitting population.
The H$\alpha$ emission of a galaxy is due to the hydrogen ionized in HII regions by massive ($>$ 8 M$\odot$), 
young ($<$ 4 10$^6$ yrs) OB stars \citep{KEN98}. Once corrected for dust extinction (i.e.  
using the Balmer decrement), the H$\alpha$ luminosity is the most direct star formation tracer in normal galaxies.
The UV luminosity is also a good tracer of young stars: at 2000 \AA, for instance, the emission of late-type galaxies 
is due to relatively young and massive (2-5 M$\odot$) A stars \citep{BOSG01}. The UV
luminosity is proportional to the star formation rate provided that the star formation is constant 
over time scales of the order of 3 10$^8$ yrs, which generally holds true in unperturbed disks.
UV fluxes must however be corrected for dust extinction, using for example the method proposed
by \cite{BUATB02}, \cite{BOSGS03a} or \cite{CORT05} based on the Far-IR to UV flux ratio.\\
\begin{figure}[!hb]
\epsscale{0.6}
\plotone{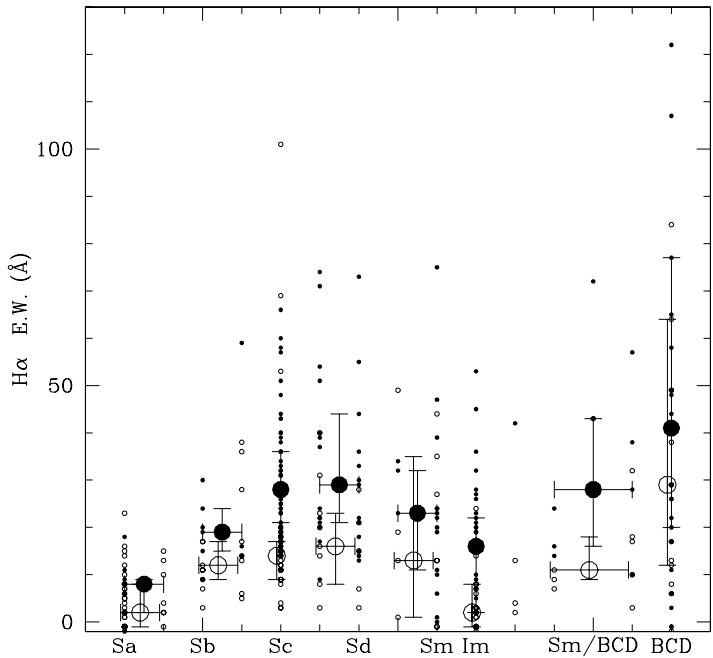}
\caption{The distribution of the individual H$\alpha $ E.W. measurements in the 
  Virgo cluster along the Hubble sequence (small dots) and 
  of the median H$\alpha $ E.W. in bins of Hubble type. Error bars are drawn at the $25^{th}$ and $75^{th}$ percentile
  of the distribution. Filled symbols represent  $HI-def\leq0.4$ (unperturbed) objects and 
  open symbols $HI-def>0.4$  (HI deficient) galaxies.}
\label{hatype}
\end{figure} 
The first attempt to study the star formation properties of galaxies in nearby clusters was carried 
out by \cite{KEN83b}. By comparing the H$\alpha$ equivalent width of 26 Virgo galaxies 
with that of isolated objects of similar Hubble 
type he showed that Virgo cluster galaxies have, on average, a lower star formation activity, 
redder colors and a lower HI gas content than their ``field'' counterparts. 
This pattern is fully confirmed by \cite{GAVB02b} and more recently by \cite{GAVPE05} who extended the 
H$\alpha$ imaging survey of galaxies in Virgo, Coma and A1367 to 545 cluster members, 
95\% complete at $m_{pg}$ $\leq$ 18.0 in Virgo and at $m_{pg}$ $\leq$ 15.7 in Coma.
Their analysis showed that 
 the current, massive star-formation rate per unit mass 
increases along the Hubble sequence from Sa to Scd, it decreases to a relative minimum for Im and it
reaches the highest value for BCDs (see Figure \ref{hatype}).  
Within each Hubble type class, galaxies with normal HI content ($HI-def<0.4$)
have H$\alpha $ E.W. \footnote{When we refer to 
H$\alpha$ E.W. we mean H$\alpha$+[NII] E.W., since measurements are affected by the contamination of
the [NII] lines bracketing H$\alpha$.} 
systematically higher than their HI deficient ($HI-def>0.4$) counterparts by a factor of two. 
Moreover the relative minimum found for Im galaxies is not due to
a particularly high HI deficiency among these objects, 
but is a characteristic of objects in this Hubble class.\\ 
Both the SFR per unit mass and the birthrate parameter $b$, 
defined as the ratio between the present day star formation rate and 
the star formation rate averaged over the entire life of the galaxy \citep{BOSG01}, 
strongly depends on the total (HI plus H$_2$) gas deficiency parameter:  
gas deficient objects have a birthrate parameter significantly lower than that expected from their
luminosity. This indicates that the present day star formation activity of
cluster galaxies is regulated by the total (HI plus H$_2$) gas content, despite the fact that
most of the total gas is located outside the optical star forming disks. The 
role of the molecular gas phase is thus probably dominant only on small scales \citep{BOSG01}.\\
\begin{figure}[!ht]
\epsscale{0.7}
\plotone{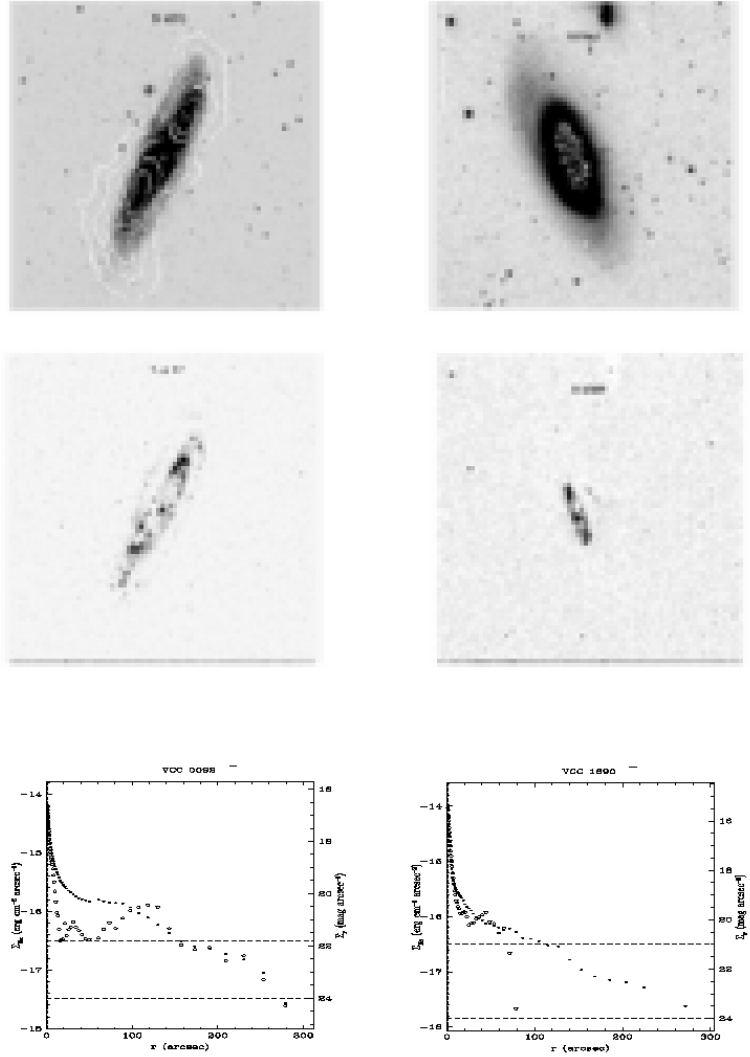}
\caption {$r$ band continuum images (upper panel) with superposed HI isophotal contours adapted from \cite{CAYG90}; 
H$\alpha$ NET images (central panel);
$r$ (filled dots) and H$\alpha$ (NET; empty dots) surface brightness profiles (bottom panel) for two 
galaxies in the Virgo cluster:
the normal VCC 92 (NGC 4192) and the deficient VCC 1690 (NGC 4569).}
\label{cayatte}
\end{figure}
Mixed with the massive, passively evolving galaxies, however, there exist a population of low-mass 
cluster objects \citep{GAVBK91a,GAVC98}  with a surprisingly high fraction of actively star
forming galaxies showing vigorous star formation activity, comparable to that of isolated galaxies,
 or even boosted by the dynamical interaction with the IGM (as in the case of CGCG 97073 
and 97079 in A1367; see sect. \ref{ram_prot})
or by tidal interactions with nearby companions (as the Blue Infalling Group (BIG) in A1367; see sect. \ref{big}).\\
Beside the global star formation activity, the morphology
of the star formation regions of galaxies in clusters has been studied in some details
\citep{MOSW00, KOOK98, KOOK04a, KOOK04b, KNAP04},
leading however to controversial results that derive from the intrinsic difficulty
of classifying the (irregular and seeing dependent) H$\alpha$ morphology.  
One aspect is to find whether cluster disks show a truncation of the H$\alpha$ emission 
in their outer profiles, as expected in the ram-pressure scenario, or if significant 
star formation excess takes place in the circumnuclear regions, as expected if nuclear sinking of gas
occurs due to tidal processes, including harassment (see Section \ref{models}) \footnote{ 
We remind however that ram pressure can induce an increase of the inner disk gas column density
up to a factor of $\sim$ 1.5 in galaxies moving edge-on close to the cluster center ($\sim$ 100 kpc; \cite{VOLC01a}), 
or the formation of an inner gas ring \citep{SCHS01}. The increase of the gas column density might induce an inner disk
star formation and feed a nuclear activity \citep{SCHS01}}.
Using an objective-prism survey of 320 galaxies in 8 nearby clusters, \cite{MOSW98} and \cite{MOSW00}) 
found an excess of galaxies with circumnuclear H$\alpha$ emission, 
showing perturbed or peculiar H$\alpha$ morphologies. 
Their frequency increases with the local (and central) galaxy density,
suggesting that they are produced by tidal (galaxy-galaxy or galaxy-cluster) interactions. 
We notice however that, because of the poor spatial resolution of their objective-prism data (4 kpc),
what \cite{MOSW98} and \cite{MOSW00} attribute to circumnuclear H$\alpha$ emission corresponds in fact  to
a truncated disk.\\   
\cite{KOOK98,KOOK04a,KOOK04b,KOOHC06}, in their H$\alpha$ imaging survey 
of 55 Virgo cluster galaxies, found that a 
severe reduction of the star formation activity takes place only in the outer disks, producing
truncated H$\alpha$ profiles, with normal or possibly enhanced activity in the inner disks. 
To illustrate this point we show in Figure \ref{cayatte} 
two galaxies: the normal VCC 92 (NGC 4192) (left) and the HI deficient, anemic, nucleated VCC 1690 (NGC 4569). 
On top of the continuum $r$ band frames, contours of the HI column density are sketched (adapted from \cite{CAYG90}).
In spite of their different gas deficiency, vigorous star formation takes place up 
to the HI radius.
Among the deficient objects the H$\alpha$ morphology is always that of a truncated disk.\\ 
\begin{figure}[!ht]
\epsscale{0.6}
\plotone{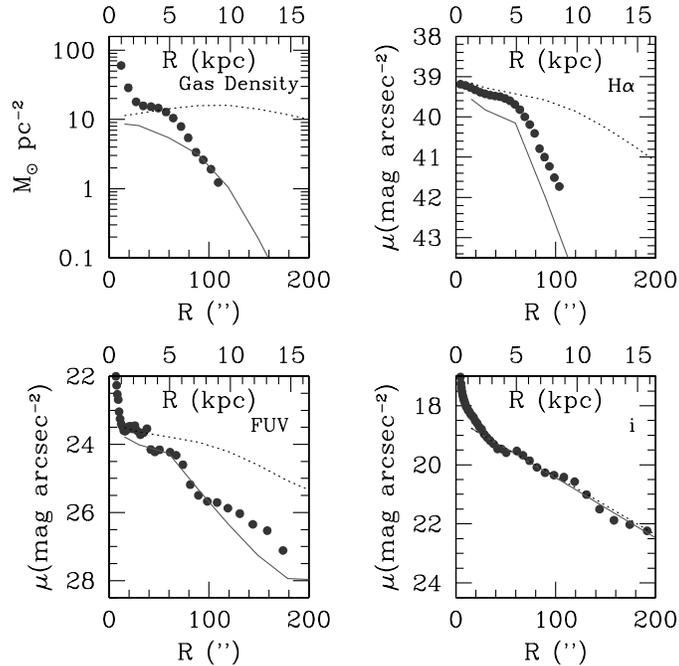}
\small{\caption{The
radial profile of the observed (filled dots) total gas, H$\alpha$, FUV
(1530 \AA), and $i$ surface brightness (corrected for dust extinction) of NGC 4569,
compared to the multizone spectrophotometric model predictions for an umperturbed
galaxy of similar total mass and velocity rotation curve (dotted line). 
The continuum line gives the radial profiles predicted by models after a 400 Myr old
gas stripping event. Strong truncation occurs to the gas and to the H$\alpha$, mild 
to the FUV and none to the $i$ band. (adapted from \cite{BOSB05b})}
\label{N4569}}
\end{figure}
Once the gas is radially removed (outside-in) by interactions of any kind, 
a radial truncation of the star formation activity, as well as a steepening of the stellar disk 
with decreasing wavelength and of the metallicity are expected to occur, 
as first claimed by \cite{LARS80}, and 
predicted by the chemo-spectrophotometric models of galaxy evolution of \cite{BOIS00}. 
The radial truncation of the star formation activity is in fact dictated by the Schmidt law, as
the decrease of the total gas column density produces an instantaneous decrease of the star formation activity. 
Multi zone chemo-spectrophotometric models show a clean dependence of the disk truncation on wavelength, given 
that, depending on their age, different stellar populations contribute more to different wavelengths. 
By comparing model predictions with observed multifrequency radial profiles it is thus possible
to date the stripping process. 
In the case of the anemic, HI-deficient Virgo cluster galaxy NGC 4569 (see Fig. \ref{N4569}) 
\cite{BOSB05b} were able to reproduce the truncation of the total gas,
of the star forming disk and of the various stellar populations assuming a ram-pressure 
stripping event that took place $\sim$ 400 Myr ago. 
Gas stripping in cluster galaxies with truncated HI and H$\alpha$ disks but unperturbed stellar disks at long wavelengths 
($\lambda$ $\geq$ 6000 \AA) is thus a relatively recent event (some 10$^8$ yr), indicating that the perturbing interaction 
took place close to the cluster center (this time scale is in fact $\sim$ 5 times smaller than the cluster 
crossing time).\\
\begin{figure}[!htb]
\epsscale{0.6}
\plotone{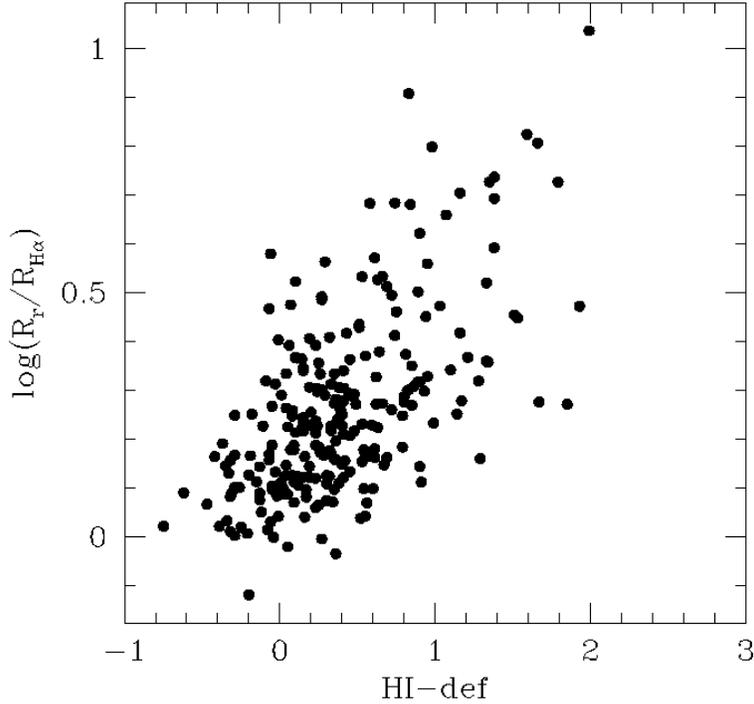}
\caption{The ratio of the optical to H$\alpha$ radius versus the HI deficiency of late-type galaxies 
in the Virgo cluster. The H$\alpha$ isophotal radius is 
computed within $10^{-16.5} \rm erg~ cm^{-2}~ s^{-1}~ \AA^{-1}~ arcsec^{-2}$
and the $r$ band isophotal radius is within $24^{th}\rm mag~ arcsec^{-2}$ (adapted from Gavazzi et al., in preparation).}
\label{trunc}
\end{figure}
Statistical evidence for a definite increasing trend of the ratio of the optical 
to H$\alpha$ radius ("truncation") and the HI deficiency parameter (see Fig. \ref{trunc}) was found by
Gavazzi et al. (2006, in preparation) in their recent analysis of the complete H$\alpha$ survey
of disk galaxies in the Virgo cluster.\\ 
\begin{figure}[!ht]
\epsscale{1}
\plottwo{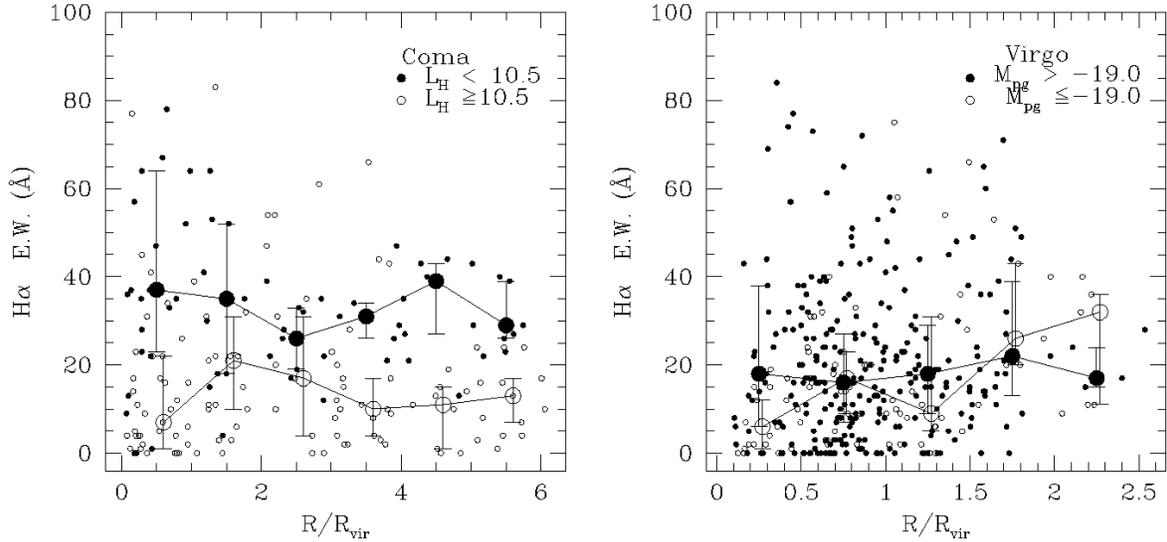}{fig12b.epsi}
\caption{The clustercentric radial distribution of the individual H$\alpha $ E.W. measurements in the Virgo cluster 
  (right) and in the Coma+A1367 clusters (left).
   High and low luminosity galaxies are given with open and filled dots respectively. 
   Medians in bins of 0.5 (1) $R/R_{vir}$ are given.
   Error bars mark the $25^{th}$ and $75^{th}$ percentile of the distribution.}
  \label{haradvirgo}
  \end{figure}
Another important issue is that the
amount and the scale over which the current star formation rate is found
quenched in nearby clusters of galaxies seems to depend significantly on luminosity.\\
For the Virgo cluster the H$\alpha$ data-set analyzed by \cite{GAVPE05} is 95\% complete at $m_{pg}$ $\leq$ 18.0
($M_{pg}$ $\leq$ -13.0) over the area covered by the VCC, i.e. up to approximately 2.5 virial radii from M87. 
Similar completeness (95\%) at $m_{pg}$ $\leq$ 15.7 ($M_{pg}$ $\leq$ -19.0) is reached in Coma,  but
over a much larger area of the Great Wall, including the bridge between Coma and A1367. \\
In bright spiral galaxies \footnote{We use two slightly different thresholds to discriminate bright from faint in Virgo and Coma.
In Virgo we use the optical limit $M_{pg}$ $\leq$ -19.0 and in Coma the Near-Infrared limit $L_H$ $=$ 10.5}
the star formation rate per unit mass decreases from $\sim 20-30 \AA$ found at approximately 2 virial 
radii projected distance from Coma and Virgo, inward (see open circles in Fig.\ref{haradvirgo}) 
to virtually zero at the cluster center. 
An even more conspicuous decrease is observed in the redshift range 0.05$<z<$0.095
by \cite{GOMN03} using a sample of $\sim$ 8500 galaxies from the SDSS, and by \cite{LEWB02} 
on $\sim$ 11000 galaxies from the 2dF. 
These surveys have shown that the star formation decreases
above a characteristic galaxy projected density of $\sim$ 1 $h_{75}^{-2}$Mpc$^{-2}$, 
corresponding to 1-2 virial radii \citep{GOMN03,LEWB02,TANA04,NICH04}, consistently with Figs.\ref{haradvirgo} and 
with the clustercentric distribution of the HI (Fig.\ref{comadef}).\\
The SDSS and 2dF surveys have the virtue of extending the study of the star formation activity  
in clusters to lower density environments and to extremely significant statistical samples.
However, they are limited by the absence of reliable morphology indicators 
for the bulk of their objects
(the morphological classification based only on light concentration and/or 
spectrophotometric parameters has been proved to be highly inaccurate \citep{SCOD02}).
They suffer from the difficulty of disentangling the radial dependence of the star formation
activity in spiral galaxies from the morphology segregation effect itself.\\
What we found most surprising in Virgo and Coma is that galaxies less luminous than some threshold 
have their star formation rate per unit mass independent on the clustercentric radius 
(see the filled circles in Fig.\ref{haradvirgo}).
This pattern could not be found neither in the SDSS nor in the 2dF because 
these surveys do not include such low luminosity objects.
The sample of nearby clusters used by \cite{GOMN03},
for instance, is limited to $M_r$ $\leq$ -20.5, while that of \cite{LEWB02} to $M_b$ $\leq$ -19.\\
The analysis of 60 clusters at $z <0.11$ in the 2dF  
by \cite{DEPR04} showed that the fraction of blue galaxies
depends on the luminosity, clustercentric distance and galaxy density, but is insensitive to the Bautz-Morgan type, 
the velocity dispersion (mass), richness, presence of substructures and cluster concentration.
Similarly, the 2dF analysis by \cite{BOWB04} shows that significant quenching of the SFR occurs above
the local galaxy density threshold of $\sim$ 1-2 $h_{75}^{-2}$Mpc$^{-2}$, but is insensitive to 
the radial clustercentric distance and to the velocity dispersion. 
These evidences indicate that significant suppression of the SFR takes place in the group environment,
where tidal mechanisms, as opposed to ram pressure, are likely to take place.
Summarizing, the present star formation rate in late-type cluster galaxies is significantly 
more quiescent than in their ``field'' counterparts and it is found to depend primarily on the residual HI
content. The hydrodynamic interaction with the hot IGM is certainly responsible of 
significant HI ablation at the present cosmological epoch, thereby producing radial truncation 
in the birth of new stars, however tidal processes might have been dominant at suppressing
the star formation during earlier preprocessing phases.

\section{Other observational characteristics of cluster galaxies}
\label{obs_diff}

\subsection{Morphology segregation}
\label{Morphology segregation}

As mentioned in the introduction,  morphology segregation is at present the strongest observational 
signature of a different nature or nurture of cluster galaxies.
In his seminal work based on photographic plates of 55 nearby clusters
including $\sim$ 6000 galaxies 
\cite{DRE80} showed that the fraction of early-type galaxies (ellipticals and lenticulars)
increases with the galaxy density and/or clustercentric radius \citep{WHIG93}. 
This relation appears universal, as it holds over 
6 orders of magnitude, from rich clusters to loose groups \citep{POST84}.
\cite{WHIG93} claim that the morphology-radius relation
is independent of the number density within the central 0.5 Mpc, of the X-ray luminosity or of
the velocity dispersion of the cluster. The fraction of ellipticals in the
outer parts of clusters is $\sim$ constant (10 to 16 \%) at a distance  $>$ 0.5
Mpc. This fraction increases to 60-70\% in the cluster center. The fraction of 
lenticulars rises moderately up to the central 0.2 Mpc, then it drops sharply. On the other hand
the fraction of spirals decreases continuously from the outskirts ($\sim$  60\%) 
to the cluster center, where it drops to virtually 0\%. Furthermore, in the Coma cluster, 
the fraction of barred galaxies increases toward the cluster core \citep{THOM81}.\\ 
\cite{BINT90}, \cite{THUA91} and \cite{SABA04} showed that segregation also affects 
dwarf galaxies,  i.e.  dwarf ellipticals are more frequent in dense environments 
while dwarf irregulars are ubiquitous.\\
\cite{VOG04a} showed that the spiral fraction depends inversely on the cluster X-ray
temperature. They also found that, while the fraction of ellipticals 
is almost constant ($\sim$ 15 \% regardless of the X-ray temperature or spiral fraction), there 
is a strong inverse correlation between the fraction of spirals and S0s constituting the remaining 85\%,
in other words the increase of the spiral fraction in the clusters outskirts compensates for the 
decrease of lenticulars. This effect holds in clusters of different richness, and extends 
beyond the virial radius. Both early- and late-type spirals are found in an envelope 
surrounding the cluster core, at a mean distance of 1.5 $h^{-1}$Mpc, while ellipticals are at $\sim$
0.85 $h^{-1}$Mpc.
Note that the morphology-density relation also affects distant 
($z$=0.5), rich, centrally-concentrated clusters, while it seems to avoid irregular ones
\citep{DRES97}. 
In distant clusters the spiral fraction is larger, the elliptical fraction is $\geq$ 
and the lenticular fraction 
is a factor of $\sim$ 2-3 smaller than in nearby clusters \citep{DRES97,FASA00}.\\
As mentioned in the introduction, morphology segregation is self-evident in the Coma cluster. 
\cite{AND96}
showed however that, while spirals are homogeneously distributed over the cluster, the early-type component
concentrates along the direction marked by the supercluster structure. Furthermore he showed evidence for
velocity segregation, with ellipticals and lenticulars having significantly smaller dispersion
($\sim$ 700 km s$^{-1}$) than spirals ($\sim$ 1300 km s$^{-1}$) (see also Section \ref{infall}).
More recently \cite{KASS98}, by analyzing CCD images covering more than 5 square degree of the 
Coma cluster, showed a strong luminosity segregation in the 
magnitude range -20 $\leq$ $M_R$ $\leq$ -16. Galaxies with high central light concentration 
have a clustering strength significantly dependent on luminosity, while objects with a low
central concentration show almost no luminosity segregation. 
Once again we remind that interpreting this evidence in terms 
of morphology is not straightforward since, as mentioned earlier, 
galaxy light profiles better correlate in shape with luminosity than with 
morphological type: both dE and low-luminosity late-type have
exponential light profiles with low concentration indices \citep{GAVF00a}.\\
The study of the galaxy morphology distribution in the Virgo cluster is made complex by projection effects
due to the elongated 3-D structure of the cluster (see Section \ref{infall}). 
\cite{SCHB99} compared the
distribution of galaxies cataloged in the VCC with that of the X-ray emitting gas from the ROSAT All Sky Survey, 
and found that the two components have a similar distributions. The cluster can be decomposed in 
three major substructures centered on M87, M49 and M86. They found 
no luminosity segregation and positive morphology segregation: spirals are more 
spread than early-types, while nucleated dwarf ellipticals are more concentrated toward the cluster center 
than their non-nucleated counterparts (see also \cite{BINT87}). \\
The SDSS and 2dF surveys made it possible to extend the study of morphology segregation 
to regions of low density contrast with respect to the ``field''. 
They confirm the increase of the fraction of the red, bulge-dominated galaxies 
with galaxy density and cluster-centric distance \citep{GOTO03a, HOGG03, BALO04, DEPR04} that was known 
in rich nearby clusters.

\subsection{The luminosity function}
\label{The luminosity function}

\begin{figure*}[!htb]
\epsscale{0.5}
\plotone{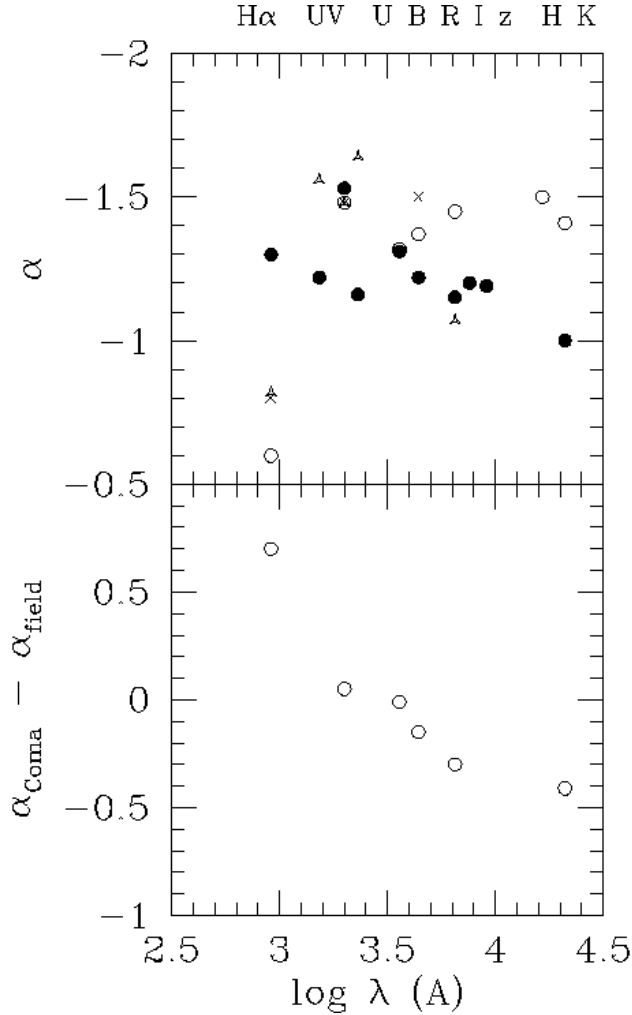}
\caption{The wavelength dependence of the slope of the luminosity function for cluster and isolated
galaxies (upper panel) and of the difference between $\alpha$ in Coma and in the ``field'' (lower panel).
Filled symbols are for ``field'' galaxies, empty circles for Coma, triangles for A1367 and crosses for Virgo.
For the clusters we took averages of the various luminosity functions available in the literature.
For the ``field'' we used \cite{BLAD01}
in the visible, \cite{GALZ95} in the H$\alpha$, Treyer (private communication) in the UV at 2000 \AA~ and \cite{WYD05}
in the UV GALEX bands.
In the $K$ band we averaged \cite{COLN01} with \cite{HUAG03} and \cite{KOCP01}).
}
\label{lumfw}
\end{figure*}

The determination of the luminosity function (LF) in distinct morphological classes (elliptical vs. spirals) 
is a difficult task both for the clusters and for the ``field'' because of the lack of high quality 
imaging material that is necessary for an accurate morphological classification.
Let us then start reviewing the most recent determinations of the global (elliptical + spirals) LF,
as derived from the extensive spectroscopic and imaging surveys carried on in the
optical in the nearby Universe (POSS, SDSS, 2dF)\footnote{
For characterizing the luminosity function we will use in the following the \cite{SCH76}
formalism:
\begin{equation}
\phi(L)dL=\phi^*(\frac{L}{L^*})^{\alpha}e^{-(\frac{L}{L^*})}d(\frac{L}{L^*})
\end{equation}
}.
\cite{PAOA01}, by combining POSS-II data of 39 nearby Abell clusters, determined that the composite 
cluster luminosity function in the high-luminosity range (-22 $\leq$ $M_R$ $\leq$ -18.5) has
a Schechter parameter $\alpha$ $\sim$ -1.1 $\pm$ 0.2, both at red and blue wavelengths.
At fainter magnitudes, however, the slope  becomes significantly 
steeper. \cite{YAGK02}, using $R$ band CCD images of 10 nearby ($z$ $\leq$
0.076) clusters, determined that the composite luminosity function in the range -23.5 $\leq$ $M_R$ $\leq$ -16 
has a slope $\alpha(R)$ $\sim$ -1.31. More recently, using the 2dF redshift survey of 60
nearby clusters ($z$ $\leq$ 0.11), \cite{DEPC03} constructed a composite luminosity function 
in the $b_J$ band. In the magnitude range -22.5 $\leq$ $M_{b_J}$ $\leq$ -15 they derive a slope  $\alpha(b_J)$ = -1.28.
These works indicate that the slope of the faint end of the optical luminosity function
is steeper in clusters ($\alpha$ $\sim$ -1.3) than in the ``field'' ($\alpha$ $\sim$ -1.2; \cite{BLAD01}). 
A steeper slope ($\alpha(R)$ $\sim$ -1.6) in the magnitude range -17 $\leq$ $M_R$ $\leq$ -14
was found by \cite{TREN04}, indicating that the steepening of the cluster luminosity function
is due to the dwarf galaxy population. \\
\cite{DEPC03} also showed that the cluster luminosity function has a characteristic 
magnitude $M_{b_J}^*$ 0.3 mag. brighter than the ``field'' one, that they tentatively attribute to an excess
of very bright galaxies in the cluster cores. Furthermore the luminosity function is similar
in clusters with high and low velocity dispersions, of different type and richness, with or without substructures. 
It was however noticed that the cluster luminosity function has a 
dip at $M_R$ $\sim$ -18.5 mostly due to the early-type component, whose presence generally correlates 
with the cluster velocity dispersion \citep{YAGK02}.\\
In their effort to disentangle the contribution of early- and late-type galaxies to the LF,
by dividing galaxies according to the shape of their radial light profiles, \cite{YAGK02}
showed that the composite cluster luminosity function of exponential disks has a 
faint end slope $\alpha(R)$ = -1.49,
significantly steeper than that of galaxies characterized by $r^{1/4}$ light profiles ($\alpha(R)$ = -1.08).
The interpretation of this finding in terms of morphology is however uncertain because it is well known
that both early- and late-type low-luminosity systems
are characterized by exponential disks, and both early- and late-type high luminosity, bulge dominated systems 
have $\sim$ $r^{1/4}$ light profiles \citep{GAVF00a}.\\ 
Using a spectral classification similar to the one used for the ``field'' by \cite{MADL02}, 
\cite{DEPC03} showed that the luminosity function of cluster galaxies with 
early-type-like spectra is brighter and steeper than in the ``field'', while
that of objects with late-type-like spectra is similar. This suggests that the observed steepening of the cluster
luminosity function is due to the early-type component. Once again, however one should use some caution since
\cite{GAVZ03c} showed that many bright cluster, late-type galaxies have spectra 
similar to ellipticals. These are systems devoid of gas with no star formation activity.\\
For galaxies in the Virgo cluster, due to its proximity (17 Mpc), both the cluster membership, based
on surface brightness criteria, and the morphological type separation are very accurate, 
even for the dwarf component. The $B$ band LF, determined with high precision by \cite{SANB85},
can be summarized as follows:
the global LF has a faint end slope $\alpha(B)$=-1.4 up to $M_B$ $\leq$ -13, 
and even steeper at lower luminosities ($\alpha(B)$=-1.6 up to $M_B$ $\leq$ -11;
\cite{TREH02}). 
Although Virgo is a spiral rich cluster, the steep rise of the luminosity function 
is primarily do to the dwarf elliptical component, which appears flatter in the ``field'' \citep{BINT90}.\\
The spirals (Sa-Sm) follow a Gaussian, peaked at $M_B$=-17.7, but when the fainter Im are added,
which follow a Schechter function of slope $\alpha(B)$=-0.25, the combined 
late-type (Spiral + Im) galaxies have a slope $\alpha(B)$=-0.80.\\
A similarly accurate type separation is not yet achieved in the ``field''
because of the poorer morphological classification. 
Using a spectroscopic type classification, \cite{MADL02}
obtained slopes $\alpha(b_J)$ = -0.99, -1.24, -1.50 for early-spirals, late-spirals and irregulars
respectively. Consistent values are obtained by \cite{HEYC97}
with similar techniques ($\alpha(b_J)$ =-0.99, -1.25, -1.37, -1.36 for
Sab, Sbc, Scd, Sdm/Starburst respectively). 
\cite{MARC98}, using a morphological
classification based on plate material for 5404 galaxies, determined a type-dependent 
luminosity function of slope $\alpha(B)$ =-1.11 for spirals and $\alpha(B)$ =-1.81 for peculiars and irregulars.
Despite the large uncertainty affecting these determination, it appears that, at least at optical 
wavelengths, the slope of the late-type galaxies luminosity 
function of Virgo is significantly flatter than that of the ``field''.\\
For nearby clusters such as Coma, A1367 and Virgo it is possible to derive separate 
determinations of the LF, each for every frequency band relevant to the stellar emission, 
from H$\alpha$ and UV to near-IR. 
The comparison with the ``field'' is however possible only for the global (all types) LFs.\\
In the spiral poor, relatively relaxed Coma cluster 
the slope $\alpha$ (in the visible) listed in the literature ranges between $\sim$ -1.7 and $\sim$ -1.2,
depending on the absolute magnitude range and on the cluster region included 
\citep{IGLB03a}. 
The slope $\alpha$ and the difference between the slope of the luminosity function in the ``field''
and in clusters are wavelength dependent, as shown in Fig. \ref{lumfw}.
\noindent
While at $\lambda$ $>$ 2000 \AA~ the slope of
the Coma luminosity function is increasingly  (with $\lambda$) steeper than that in the ``field'', 
for $\lambda$$>$ 1500 \AA ~the reverse is true. 
At long wavelengths (near-IR) the luminosity is a tracer of the dynamical mass \citep{GAVPB96a}, 
thus the luminosity function can be assumed as a mass function.
Even though significantly steeper than in the ``field'' at near-IR,  
the cluster luminosity function ($\alpha$ $\sim$ -1.5; \cite{DEPE98};
\cite{MOBT98}, \cite{ANDP00}, \cite{BALC01}) 
is still significantly flatter than the one
predicted by CDM semi-analytical hierarchical models of galaxy formation 
\citep{SOMP99}. In any cosmology, CDM models predict slopes 
$\alpha$ $\sim$ - 2 for the mass function. Assuming a mass to light ratio
in agreement with the Tully-Fisher relation, it is at present impossible 
to obtain from these models the observed slope of the near-IR LF.\\
At wavelengths shorter than $\leq$ 2500
\AA, where the emission of galaxies is dominated by the star formation activity
\footnote{In elliptical galaxies the (faint) UV emission is not due to star formation, 
but to  extreme Horizontal Branch stars and their progeny  \citep{OCON99}: their contribution 
to the faint end slope of the UV luminosity function is significant because of the large number of E
contributing at low luminosity \citep{CORG03, CORB05}} the luminosity function is representative of the late-type
galaxy population.
\cite{IGLB02} constructed the first H$\alpha$
luminosity function of cluster galaxies using a deep, wide field H$\alpha$ imaging survey 
of Coma and A1367 and compared it to that of Virgo. 
Despite their different nature (Coma is relaxed, spiral poor, A1367 relaxed spiral rich
and Virgo unrelaxed spiral rich), the three clusters have a similar H$\alpha$ luminosity 
function with slope $\alpha(H\alpha)$ = -0.70 $\pm$ 0.10.
In the bright luminosity range this slope is similar to that of the ``field'' determined by \cite{GALZ95}.
At fainter luminosities it is  
significantly flatter than the poorly constrained ($\alpha(H\alpha)$ $\sim$ -1.3) found by \cite{GALZ95}.\\  
The composite cluster UV luminosity function, obtained by combining balloon borne 
1650-2000 \AA~data of Coma, A1367 and Virgo,
has a slope $\alpha(UV)$ =-1.50 $\pm$ 0.10 \citep{CORG03}, consistent with the ``field'' 
one ($\alpha(UV)$ =-1.51, \cite{SULT00}). More recent results based on higher quality 
and deeper GALEX UV data of A1367 give a slope of the  UV
luminosity function ($\alpha (UV 2310 \AA)$=-1.64$\pm$0.21; $\alpha (UV 1530 \AA)$=-1.56$\pm$0.19; \cite{CORB05})
significantly steeper than in the ``field'' ($\alpha (UV 2310 \AA)$=-1.16$\pm$0.07; 
$\alpha (UV 1530 \AA)$=-1.22$\pm$0.07; \cite{WYD05}). If limited to star forming galaxies, however,
the cluster and ``field'' luminosity functions are similar \citep{CORB05}.\\
We conclude that the cluster luminosity function is increasingly steeper than that in the ``field''
for $\lambda$ $\geq$ 1500 \AA, and it is flatter for the ionizing radiation $<$ 912 \AA. 
The steepening of the cluster luminosity function with increasing $\lambda$ is due 
to the increasing contribution of E-dEs toward longer wavelengths. The weight of E-dEs in the
luminosity function is already significant in the UV bands while it vanishes only for $\lambda$ $<$ 912 \AA.

\subsection{ Galactic kinematic disturbances}
\label{Kinematics}

\noindent
The presence of kinematic disturbances in late-type
galaxies in clusters is controversial.
Based on the analysis of long slit spectra, \cite{RUBF88} and
\cite{WHIF88} showed that the rotation curve of cluster galaxies  strongly decline at outer
radii, in particular in those galaxies located close to the cluster center, 
in contrast with the asymptotically flat or rising rotation curve of isolated objects.
They also showed that most of the declining rotation curves are associated with HI-deficient 
galaxies \citep{RUBF88}.
This evidence was at that time interpreted as the result of stripping of
the dark matter halo surrounding galaxies during galaxy-galaxy and/or galaxy-IGM interaction,
or as the prevention of halo formation in the cluster environment
\citep{WHIF88}.\\
This result was however questioned by \cite{AMRS93} who analyzed more
accurate 2-D velocity fields of 21 cluster galaxies obtained with a Fabry-Perot spectrograph, 
finding flat (or smoothly declining) rotation curves. \\
More recently \cite{RUBW99}, in their analysis of long-slit rotation curves of 89  
spiral galaxies in the Virgo
cluster, reiterated that approximately one half of the Virgo galaxies show 
kinematic disturbances in their rotation curve (asymmetrical rotational velocities on the two sides
of the major axis, falling outer velocities, inner velocity peculiarities, dips in the 
rotation velocities at intermediate radii, velocities near zero for the near nuclear gas). 
These disturbances are consistent with recent ($\leq$ 1 Gyr, the time necessary 
to re-distribute the mass within the disks in a few rotation cycles) tidal encounters
or accretion events (see also \cite{BAR99}). Disturbances are not correlated with Hubble type, luminosity,
local galaxy density or HI deficiency. The authors claim that galaxies with perturbed
rotation curves have a Gaussian velocity distribution peaked at the average cluster velocity, 
but they are often located at the cluster periphery, possibly on radial orbits.\\
Using a sample of 510 cluster galaxies with long slit spectroscopy, 
\cite{DALG01} recently criticized this result.
They maintained that the residual of the Tully-Fisher relation does not depend on the 
location of galaxies within clusters. They only observed that the rotation
curve of galaxies in the periphery are more extended than in the galaxies close 
to the cluster core, consistent with their H$\alpha$ morphology.
By analyzing the H$\alpha$ rotation curves of late-type galaxies in nearby clusters \cite{VOG04a,CHEM05}
came to the similar conclusion that the distribution of HII regions along spiral disks is often 
truncated or asymmetric or perturbed. Asymmetric spirals are not observed beyond 1 $h^{-1}$Mpc, and are found
predominantly in the richest cluster cores. 
 We conclude that, if any, mild differences exist between the kinematic 
properties of cluster and ``field'' galaxies.

\subsection{ Large scale velocity distributions, evidences for infall?}
\label{infall}
 
\begin{figure*}[!htb]
\epsscale{0.7}
\plotone{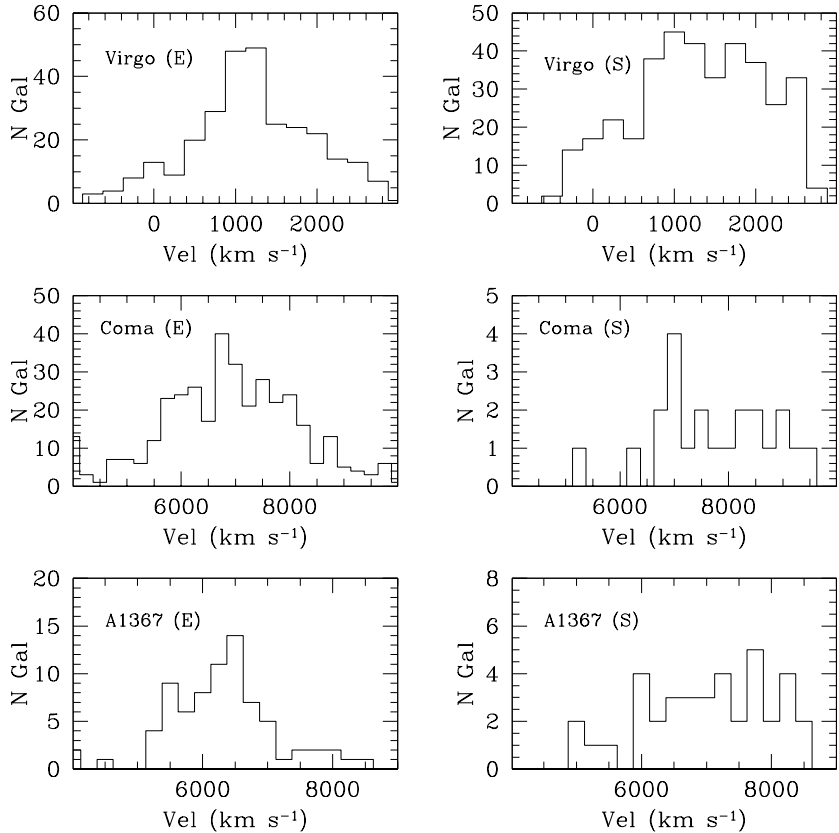}
\caption{The velocity distribution of galaxies in the A1367, Coma and Virgo clusters, divided
in early (left) and late-types (right).}
\label{vel_hist}
\end{figure*}

It is well known that the velocity distribution of cluster galaxies is type-dependent:
while that of the early-type component is virialized, the one of late-types hardly follows a Gaussian distribution,
being skewed toward highly deviating velocities \citep{BIVK97}, probably because of their 
 more elongated radial orbits, with higher elongations in late, star forming than in early-spirals \citep{BIVK04}.
By itself this provides an evidence for infall of late-type galaxies onto clusters 
(see e.g. \cite{COLD96,RING03}).\\
Using the updated redshifts of Virgo (taken from GOLDmine) and new redshift determinations 
\citep{CORG04} of galaxies in the central 1 deg regions of Coma and A1367
surveyed by \cite{IGLB02,IGLB03a} in the H$\alpha$ and $r'$ band down to faint ($r'<21$) magnitudes, 
we plot in Fig. \ref{vel_hist} the velocity distributions of galaxies in these three clusters.
For Virgo we divide early- from late-type galaxies according to the morphological classification 
given in the VCC. For Coma and A1367, where no such a reliable classification exists at this faint level,
we use the H$\alpha$ line to discriminate between star forming and non star forming galaxies.  
It is apparent from the figure that in the three clusters the velocity distribution of the early type 
objects is Gaussian:
$<V>_{A1367}$= 6347 km s$^{-1}$; $\delta V_{A1367}$ = 760 km s$^{-1}$, 
$<V>_{Coma}$ = 6994 km s$^{-1}$; $\delta V_{Coma}$ = 1149 km s$^{-1}$, 
$<V>_{Virgo}$= 1202 km s$^{-1}$; $\delta V_{Virgo}$ = 753 km s$^{-1}$,
while that of the late-type component is non-Gaussian, generally broader and with a significantly
larger mean redshift. For this component a realistic estimate of the velocity dispersion is the
one derived from the difference between the 25 and 75 percentiles of the redshift distribution: 
$<V>_{A1367}$= 7080 km s$^{-1}$; $\delta V_{A1367}$ =1415 km s$^{-1}$, 
$<V>_{Coma}$ = 7796 km s$^{-1}$; $\delta V_{Coma}$ = 1560 km s$^{-1}$, 
$<V>_{Virgo}$= 1340 km s$^{-1}$; $\delta V_{Virgo}$ =1150 km s$^{-1}$.
A consistent result for Virgo was obtained by \cite{BINT87, BINP93} and by 
\cite{CONG01} who also noticed
that the velocity distribution of dE galaxies resembles more that of late-type galaxies than of Es.
Similarly \cite{COLD96} (see also \cite{AND96}) showed that the velocity dispersion of the late-type galaxies in the Coma cluster 
is $\sim$ $\sqrt{2}$ that of the early type component, suggesting that spirals are free falling 
into the cluster core.\\
A direct proof of infall would require the knowledge of the galaxies trajectories
in 3-D. Unfortunately the velocity component perpendicular to the line of sight cannot be measured,
but, combining the line-of-sight component of the redshift with some independent distance estimate
(e.g. from the Tully-Fisher or fundamental plane relation), the 
peculiar motions can be reconstructed \citep{TULS84}). Using these methods 
peculiar galaxy motions (up to 2000 km s$^{-1}$), induced by the cluster potential well, 
have been detected in several nearby clusters such as 
Virgo \citep{TULS84, GAVS91b, GAVB99} and Coma \citep{GAVS91b}.\\
The Coma cluster has been considered for a long time as the prototype of old, relaxed clusters. 
X-ray observations and dynamical studies \citep{WHIB93, COLD96, BRIH01}, 
showed however that the structure of the cluster is  
formed by a massive main body centered on the 2 cD galaxies NGC 4874 and 4889, and a 
less massive substructure centered on NGC 4839 infalling onto the main cluster. 
It is still unclear whether 
the NGC 4839 subgroup already crossed the cluster core or it is falling into it for the first time \citep{COLD96}. 
The interaction between this substructure and the main cluster is probably the driver of the PSB galaxies 
observed by \cite{CALR93} and \cite{POGB04} (discussed in Section \ref{Surface brightness}).\\
 The dynamical study of A1367 \citep{CORG04} revealed the presence of a relatively compact 
group with low velocity dispersion ($\sim$ 170 km s$^{-1}$) and composed of highly 
perturbed, star forming objects, falling into the main cluster at $\sim$ 1700 km s$^{-1}$ (see Sect.
\ref{big}).
There is also evidence of individual galaxies falling onto nearby clusters, such as CGCG 97-073 
and CGCG 97-079 in A1367 (see sect. \ref{prototype}).\\
Because of its smaller distance (17 Mpc) and of its depth along the line of sight, the distance
determination of galaxies within the Virgo cluster using the Tully-Fisher and the fundamental plane relation
is sufficiently accurate to disentangle the 3-D structure 
of the cluster \citep{TULS84}; directly proving the presence of infalling galaxies. 
\cite{GAVB99} found that the distance of cluster A ($\mu_o=30.78\pm0.07$), associated with M87, is
consistent with the determination based on the Cepheid's method ($\mu_o$=31.0).
Cluster B, off-set to the south, is found at $\mu_o=31.76\pm0.09$. This
subcluster is falling onto A at about 500 km s$^{-1}$. 
Clouds W and M are at twice the distance of A. Their recessional velocity indicates
that these clouds are in Hubble flow, thus little perturbed by the presence of Virgo.
Galaxies on the North-West and South-East of the main cluster A belong to two clouds
composed almost exclusively of spiral galaxies with distances
consistent with A, but with significantly different velocity distributions, 
suggesting that they are falling onto cluster A at approximately 500 km s$^{-1}$ 
from the far and near-side respectively. 
Moreover a significant fraction of gas "healthy" and currently star forming galaxies
is unexpectedly found projected near the center of the Virgo cluster. Their average Tully-Fisher distance
is found approximately one magnitude further away ($\mu_o=$31.77) 
than that of their gas-deficient counterparts ($\mu_o=$30.85). 
It is suggested that these gas healthy objects belong to a cloud that in fact lies a few Mpc behind Virgo,
in the process of falling toward the Virgo cluster \citep{GAVB02b, SOLS02}.
The velocity dispersion of each individual substructure is $\sim$ 500 km s$^{-1}$ \citep{BINP93,GAVB99},
resembling that of a formed cluster rather than a group. \\


\subsection{The Spiral - S0 connection}
\label{Surface brightness}

\begin{figure}[!htb]
 \epsscale{0.7}
 \plotone{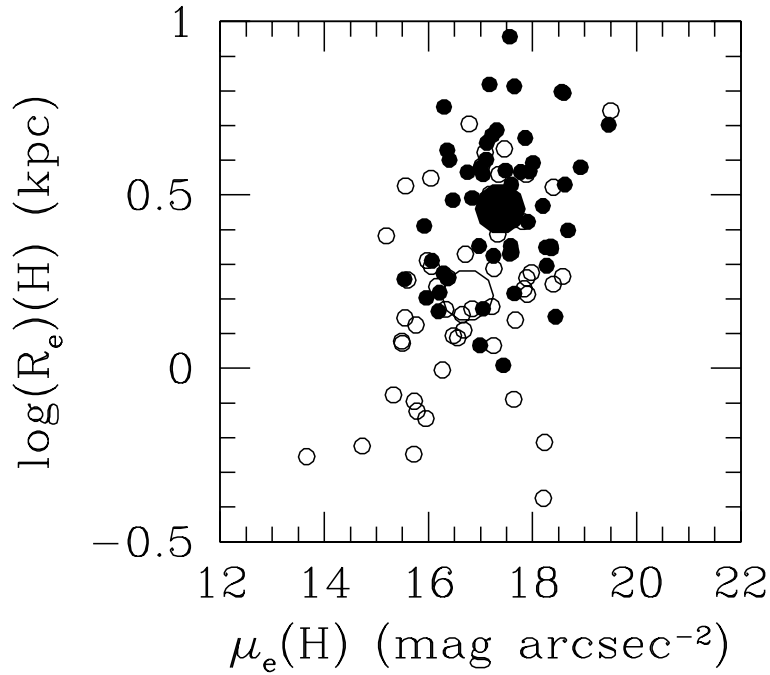}
 \caption{The relationship between the $H$ band effective radius and effective surface brightness in the Virgo cluster. 
 Open symbols are for S0 galaxies, filled symbols for Sa-Sb galaxies. Large symbols represents averages. }
 \label{anemic_korm}
 \end{figure}

When Van den Bergh discussed his morphology classification system, he pointed out 
the existence of "anemic" galaxies that might represent the missing link between 
normal spirals and lenticulars.
They are characterized by an arm-interarm contrast less pronounced than normal spirals,  
they are gas-poor, with low star formation activity, redder colors than normal spirals with 
similar bulge to disk ratios. Their fraction is significantly larger in
clusters such as Virgo than in the ``field'' \citep{BER76}. The "passive" spirals 
also inhabit several SDSS clusters at redshift 0.05$< z < $ 0.1, but they are 
preferentially found outside the cluster's cores \citep{GOTO03b}.
The first interpretation of their existence
was that, once the gas reservoir feeding the star formation is removed from
cluster spirals, they become lenticulars \citep{BER76}. 
In a detailed analysis of the star forming properties of anemic galaxies, \cite{ELME02}
concluded that their gas surface density is below the threshold for the star formation to take place. 
Because of the lack of supply of young stars with low velocity dispersion, the disk heats up,
dumping spiral waves on time scales of a few revolutions \citep{SELL84, FUCH98, ELME02}.
The anemic sequence would thus represent the intermediate
phase between spirals and S0 \citep{BER76}.
\cite{KOOK98} have quantitatively shown that the excess of early-type spiral galaxies 
in the Virgo cluster is partly due to misleading classifications of low-concentration (disk-dominated)
systems with reduced star formation activity rather than to a systematic
increase of the bulge to disk ratio with galaxy density. They interpreted this result as
an evidence that these low-concentration early-type spirals were blue, star forming late-type spirals
whose activity was quenched by gas removal induced by their interaction with the ICM.\\
There exist, however, several observational evidences indicating that
lenticular galaxies in nearby clusters are not gas stripped late-type galaxies.
\cite{DRE80} showed that the bulge size and the bulge-to-disk ratio of lenticulars
are systematically larger than those of spirals in all density regimes.
He thus concluded that ``since the tightly bound inner bulges should be unaffected by ablation, 
the dissimilarity in the bulge and bulge/disk distributions in all density regimes is inconsistent
with the idea that most S0 galaxies result from the removal of disk gas from a spiral by ram-pressure stripping
or evaporation'' \citep{DRE80, BUR79, GIS80}. \\
We checked this conclusion using our near-IR survey of the Virgo cluster. 
These $H$-band imaging data, beside a better photometric quality due to the linear response
of the NICMOS-3 detectors compared to the photographic plates of \cite{DRE80}, 
have the advantage of being free from dust obscuration and from relatively 
recent episodes of star formation which might alter the morphological properties of galaxies.
The $H$ band Kormendy relation between the effective surface brightness $\mu_e$ and the 
effective radii $R_e$  (radii including half of the total light) 
for lenticulars  and spirals of type Sa-Sb in the Virgo cluster 
is given in Fig. \ref{anemic_korm}.
Sa-Sbs have, on average $0.65$ $\rm mag~arcsec^{-2}$ lower surface brightness and 
almost a factor of 2 larger effective radii than lenticulars.
This confirms the results of \cite{DRE80} and makes 
the formation of lenticulars through gas sweeping in spirals very unlikely.
 This conclusion was recently confirmed by \cite{CHRI04} who found that the bulge to disk ratio
increases in cluster galaxies because of the increase of the bulge luminosity 
(as expected in a tidal interaction scenario) and not because of the fading of the disk luminosity 
(as expected in a ICM-IGM interaction). \\

Spectroscopic observations of 125 early-type galaxies in the Coma cluster revealed the presence
of a significant number of objects with  signs of recent star formation, such as
Balmer absorption and emission lines and negative CN/H8 indices \citep{CALR93}. 
The spectroscopic characteristics of these galaxies
are similar to those of post-starburst galaxies observed in intermediate-redshift clusters, the so called
E+A galaxies \citep{CALR93}. Interestingly most E+A galaxies 
were found in the south west periphery of the Coma cluster, probably associated with
the NGC 4839 group, but with a larger velocity dispersion than the remaining early-type
galaxies belonging to the same group \citep{CALR93}.\\
HST images of some starburst or post-starburst galaxies in the Coma periphery 
revealed that they have exponential disks, some with  
an embedded central spiral, whereas kinematic data showed rotation curves typical of
late-type galaxies \citep{CALR96, CALR99}.\\
The analysis of \cite{POGB01a, POGB01b} of $\sim$ 300
early-type galaxies in the Coma cluster confirmed most of the Caldwell et al. results. Using several 
spectral indices in the visible bands, these authors were able to discriminate between age and metallicity effects,  
and concluded that more than 40\% of the low luminosity ($M_B$ $>$ -19) S0s in 
Coma undergone star formation episodes in the 
last $\sim$ 5 Gyr, significantly later than elliptical galaxies.\\
\cite{POGB04} found that  k+a/a+k galaxies (post starburst/post-star forming galaxies
with no current star formation activity that were forming stars at a vigorous rate in the last 1.5 Gyrs,
as indicated by their high Balmer absorption lines (H$\delta$E.W. $\geq$ 5 \AA)) in Coma
are all low luminosity objects ($M_V$ $>$ -18.5), thus significantly different from 
the high luminosity ones ($M_V$ $\leq$ -20) observed in high redshift clusters. They have also shown that
these post starburst/post-star forming galaxies
spatially correlate with substructures in the hot IGM, as revealed by the XMM-Newton data,
are located in the central $\sim$ 1.4 Mpc of the cluster and have radial velocities significantly higher 
than the cluster mean. These evidences were interpreted as an indication that the activity
of this population of low-luminosity, post-star forming galaxies was quenched (and boosted) 
by the interaction with the dense ICM \citep{POGB04}.\\
How does the star formation history of early-type (Sa-Sb) spiral galaxies compares with that of normal (non-PSB) S0?
To try constraining their star formation histories 
by fitting the UV to near-IR SEDs with \cite{BRUC93} stellar population synthesis models
(as outlined in Section \ref{star formation}) \cite{GAVB02c} 
concluded that S0-S0a
have on average slightly smaller $\tau$s than early-type spirals
of similar luminosity, indicating that the bulk of the star formation activity in lenticulars is older 
than in spirals.\\
Kinematic studies of nearby cluster and ``field'' S0 argue for a different nature of spirals and lenticulars. 
The larger scatter and a small zero point offset in the Tully-Fisher relation observed in Virgo and Coma 
cluster S0 galaxies compared to spirals indicate that lenticulars can hardly be formed by simple gas
removal from healthy spirals. On the contrary it is argued 
that S0s are formed during minor mergers, slow encounters,
harassment, or some combination of these, making S0s an heterogeneous class of objects \citep{DRES83, NEIS99, HINZ03}.\\
 Other statistical considerations such as the extremely weak dependence of the morphology segregation effect 
on galaxy density (the fraction of spiral galaxies changes by a factor of $\sim$ 2 over a change of $\sim$ 1000 in 
galaxy space density) are hardly explained by a simple transformation of spirals into
S0 through ram pressure stripping, as noticed by \cite{DRE80} and \cite{DRES04b}, but seem to indicate
that the morphology of galaxies was settled shortly after their birth, before the dense
cluster environment was even formed.\\
In conclusion, while lenticular and quiescent early-type spirals have nearly indistinguishable
stellar populations, significant differences in their structural and kinematic properties 
as well as statistical considerations argue against
the origin of high mass S0s in nearby clusters from gas removal of spiral disks.
On the contrary, low-luminosity disk-dominated S0 might be the result of ram-pressure stripping
in late-type galaxies.

\section{The physical processes acting in clusters} 
\label{models}

\begin{table*}[-ht]
\caption{The adopted parameters of clusters Coma, A1367 and Virgo.}
\[
\scriptsize
\begin{array}{rccccccccccccc}
\hline
\noalign{\smallskip}
Cluster & <V>           & \delta V_{cluster}   & ref & \rho_{IGM}         & ref & T_{IGM} & ref & R_c      & ref & R_{vir} & ref &M_{cluster}    &ref\\
        &\rm{ km~s^{-1}}& \rm{ km~s^{-1}}&     &\rm{atom~cm^{-3}}  &     &\rm{keV} &     & \rm{kpc} &     &\rm{Mpc} & &\rm{M\odot}&   \\
\noalign{\smallskip}
\hline
\noalign{\smallskip}
Coma & 6960 & 880~(1560)^a  & 1 & 3.12~ 10^{-3} & 3 & 8.2 & 3 & 257 & 3 & 2.19 & 7 & 1.2~10^{15}& 2 \\
A1367& 6420 & 822~(1415)^a  & 1 & 1.25~ 10^{-3} & 3 & 3.5 & 3 & 240 & 3 & 2.13 & 7 & 6.9~10^{14}& 7 \\
Virgo& 951  & 886~(1150)^a  & 4 & 2.00~ 10^{-3} & 5 & 2.3 & 8 & 130 & 5 & 1.68 & 7 & 2.5 ~10^{14}& 6 \\ 
\noalign{\smallskip}
\hline
\end{array}
\]
\normalsize
References:
1: \cite{STRR91};
2: \cite{BRIH92};
3: \cite{MOHM99};
4: \cite{BINP93} (spirals in the core of cluster A);
5: B\"ohringer, private communication; 
6: \cite{SCHB99} (the total mass is here calculated as the sum of the M87, M49 and M86 total extended masses);
7: \cite{GIRG98};
8: \cite{BOHB94}\\
a: the velocity dispersion of the cluster as a whole and of the late-type component alone (in parenthesis), from this work.
\label{Tab1}
\end{table*}

The analysis presented so far has shown that late-type galaxies in nearby clusters 
differ systematically from their field counterparts mainly in two respects:
they are significantly HI-deficient and are characterized by a 
lower activity of star formation, with truncated HI and star forming disks. 
Other less clearcut differences are their molecular gas content,  metallicity, cold dust content, 
kinematical perturbations and radio-continuum synchrotron emission.
While in massive spirals the HI-deficiency and suppression of the star formation 
depend on the distance from the cluster center (provided that they are within 1-2 virial radii), in dwarf irregulars
they depend primarily on the galaxy luminosity and very little on the clustercentric position.\\
Which are the physical processes that produced such differences? 
Are the physical conditions in today's high density environments similar to 
the ones prevailing in the young universe when the morphology segregation was shaped? 
In other words, are the processes that transformed late-type galaxies into quiescent lenticulars in the past
still able to modify the late-type galaxies that are currently infalling nearby clusters? 
Trying to answer these questions we now briefly review the physical processes that are believed to describe
the interaction of galaxies in high density environments.
The comparison of model predictions with observations is particularly well suited in the nearby Universe
where the multifrequency observations have high spatial resolution. \\
Two broad classes of models are here considered: first the gravitational ones, including all
sorts of tidal interactions (galaxy-galaxy, galaxy-cluster, harassment), then
the hydrodynamical interactions taking place between the
galaxy's ISM and the hot intergalactic medium (ram-pressure, viscous stripping, thermal evaporation).
We also briefly review two hybrid processes: "starvation" (remotion of the gas feeding the star formation)
and "pre-processing" (taking place in groups of galaxies falling into clusters) that combine
gravitational with hydrodynamic mechanisms.

\subsection{Tidal interactions among galaxies}
\label{tidal}

Tidal interactions among galaxy pairs act  
on gas, dust and stars, as well as on dark matter, with an efficiency depending on
the gravitational bounding of the various components. 
This produces selective morphological transformations.
Since tidal forces act as $M/R^3$, if the typical galaxy radii are not too small 
compared to the average separation between 
galaxies, tidal interactions can be quite efficient at removing matter from galactic 
halos \citep{SPIB51, RIC76, FARS81, ICK85, MER83}.
As shown by the numerical simulations of \cite{VALJ90}, 
tidal interactions are more efficient in perturbing the loose peripheral 
or extra-planar HI gas, than the molecular gas, located in the inner potential well. 
For what concerns the star formation, both observations \citep{KEEK85, KENR87, HUMH90, CONC82} 
and simulations \citep{MIHR92, IONO04} of interacting pairs
show a major increase of the nuclear activity and a milder (if any) in the disk.\\ 
It is intuitive that tidal interactions among galaxies are boosted
in the dense cores of rich clusters of galaxies. 
However, due to the high relative velocities, tidal interactions  
among cluster galaxies, although more frequent, have significantly
shorter duration than in the ``field''  ($t_{enc}$ $\sim$ 10$^8$ yr, see 
Appendix), thus the effects of the perturbation are less severe. 
For instance, the formation of bars, that is expected to take place
in slow encounters, should be relatively rare in clusters \citep{MIHOS04}.\\
The simulations of \cite{BYRV90} (applied to spiral galaxies in clusters) show that 
tidal interactions produce enough  gas inflow from the disk to the circumnuclear regions,
provided that the perturbation parameter:
\begin{equation} 
{P_{gg}=(M_{comp}/M_{gal})/(d/r_{gal})^3}
\label{Pgg}
\end{equation}
\noindent
is $P_{gg}$ $\geq$ 0.006-0.1 (depending on the halo to disk mass ratio),
where $M_{comp}$ is the companion mass, $M_{gal}$ and $r_{gal}$ are the mass and the visible disk radius
of the spiral galaxy respectively and $d$ is the separation between the two galaxies.
A typical perturbation parameter $P_{gg}$ for $\sim$ 10 kpc radii galaxies in clusters 
can be roughly estimated assuming $M_{comp}$ $\simeq$ $M_{gal}$ and an 
average galaxy separation $\simeq$ 200 kpc inside a cluster 
of 2 Mpc radius including $\sim$ 1000 objects.
The resulting $P_{gg}$ $\simeq$ 10$^{-4}$ is significantly smaller than the critical $P_{gg}$ 
necessary for producing significant gas infall into the nucleus.
This simple estimate is confirmed by \cite{FUJ98}, who claims that the typical
perturbation induced by a single high-speed encounter among cluster galaxies 
is too small to significantly affect the star formation rate.
His model incorporates a more realistic modeling of the star formation process
based on clouds collisions, than the simple nuclear infall treated by \cite{BYRV90}.
Furthermore, as discussed by \cite{MER84} and by \cite{BYRV90}, 
the frequency of galaxy-galaxy encounters in rich clusters, 
as measured by the inverse of the relaxation time 
$t_{relax}$$\sim$ some 10$^{10}$ years (see Table \ref{tabtimescales} and section \ref{timescales}), 
comparable to the age of the Universe, is negligibly small (see however \cite{VALJ90}).\\
\cite{OKAN01} and \cite{DIAK01} with their 
hybrid $N$-body simulation and semi-analytical models, 
tried to reproduce the cluster morphology-density or morphology-radius relations 
observed by \cite{WHIG93}. They concluded that, 
while the distribution of elliptical galaxies in clusters can be obtained with major merging, 
this is not the case for the lenticular galaxies with intermediate bulge to disk ratios.
Bulges can be formed by the merging of two equal mass galaxies,
while the subsequent gas cooling can form disks \citep{DIAK01} by unequal mass mergers 
of disk galaxies, where disk destruction is not complete and some rotation is retained, or 
minor mergers between spirals and their companions (the disk is heated but not destroyed). 
The morphology segregation is qualitatively well reproduced by the semi-analytical 
simulation of \cite{SPRW01}. 
\cite{OKAN03}, however, remarked that this result is very sensitive to the assumed bulge 
to disk ratio of lenticular galaxies.\\
Typical examples of cluster galaxies which recently underwent a gravitational interaction 
are NGC 4438 and NGC 4435 in Virgo (see sect. \ref{prototype}). Tidally interacting objects 
in clusters are relatively difficult to identify since tidal tracers are very short-lived: while in
the field most of the ejected material in tidal tails remains bounded to the 
main galaxy, in clusters the tidal field strips the unbound material, maybe originating 
the diffuse intergalactic light \citep{MIHOS04}.

\subsection{Tidal interaction between galaxies and the cluster potential well}
\label{tidalcluster}

Given the large mass of clusters, exceeding 10$^{14}$ M$\odot$, tidal interactions between galaxies and 
the whole cluster potential well can effectively perturb cluster galaxies, inducing gas inflow, 
bar formation, nuclear and perhaps disk star formation \citep{MER84, MIL86, BYRV90}. 
Depending on the impact parameter, the disk structure can be perturbed, forming a spiral pattern 
(if the disk is parallel to the orbital plane), 
or developing a bulge (disk perpendicular or inclined with respect to the orbital plane, \citep{VAL93}).
In any case the thickness of the disk is expected to increase slightly \citep{VAL93}. 
Tidally perturbed disk galaxies in clusters, due to the increased non-circular velocities of stars 
can produce declining rotation curves at large radii \citep{VAL94}.\\
The models of \cite{FUJ98}, \cite{VAL93} and 
\cite{HENB96} show that tidal forces by the cluster potential well can
accelerate molecular clouds of disk galaxies falling toward the cluster center. 
The rise of the kinetic pressure in the interstellar medium 
induces star formation \citep{ELME97}.

\begin{figure}[!htb]
\epsscale{0.5}
\plotone{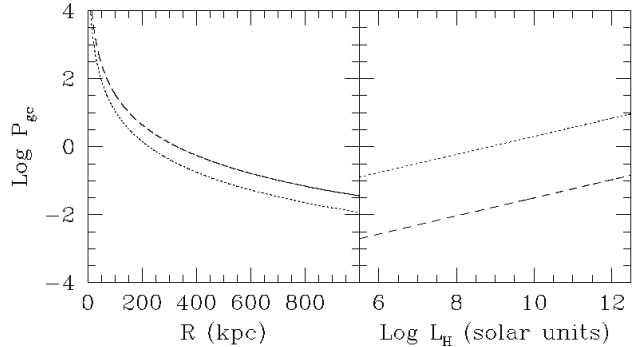}
\caption{The logarithm of the perturbation parameter, defined in eq. \ref{Pgc}, is plotted as a function of the 
distance from the cluster center (left panel) for Coma galaxies of linear radius 30 and 5 kpc respectively 
(top to bottom) 
and as a function of the $H$ band luminosity (right panel) for Coma galaxies at 200 
and 800 kpc (top to bottom) from the cluster center. 
}
\label{pert}
\end{figure}

The efficiency of the interaction can be quantified by the perturbation parameter 
$P_{gc}$ defined as \citep{BYRV90} \footnote{$P_{gc}$ as determined using eq. \ref{Pgc}
and $r_{tidal}$ given by eq. \ref{tidalr} must be considered as indicative 
since both equations are based on the hypothesis that the cluster potential is dominated by a 
unique dark matter halo and do not account for perturbations by the halos of individual galaxies.
A  more accurate estimate of both parameters requires N-body simulations to account for these
non-linear effects.}:
\begin{equation}
{P_{gc}=(M_{cluster}/M_{gal})\times(R/r_{gal})^{-3}}
\label{Pgc}
\end{equation}
where $R$ is the distance of the galaxy from the cluster center.
When the perturbation parameter becomes critical ($P_{gc}\sim$0.006-0.1), the gas in the 
disk is driven toward the center of the galaxy on time scales of 2-3 10$^8$ yr, triggering nuclear activity
\citep{BYRV90}.\\
For a perturbation parameter close to the critical value, however, the interaction is unable to remove the 
outer gas producing the observed HI-deficiency of cluster galaxies \citep{BYRV90}.
\cite{BYRV90} and \cite{HENB96} speculate that the poorly gravitationally 
bound HI would be ejected from the galactic plane
by supernova winds driven by the enhanced star formation activity.
The effects of the interaction on the inner gaseous component are less clear.
The molecular gas, shielded deeper inside the galaxy potential well would not be ejected, 
but consumed by star formation events on longer time scales.\\
The perturbation parameter for Coma, A1367 and Virgo galaxies
can be derived using the empirical mass vs. diameter relation:
\begin{displaymath}
{Log (M_{gal})~ (M\odot) = 8.46+2.37 \times Log (r_{gal})~~~~~ {\rm (kpc)}}
\end{displaymath}
\noindent
For the largest galaxies ($r_{gal}$=30 kpc) we find on average
$P_{gc}$=0.28 for Coma, 0.16 for A1367 and 0.06 for Virgo, 
at a typical distance of $R$=500 kpc from the cluster center.
Figure \ref{pert} shows that the perturbation parameter becomes critical ($P_{gc}$$\sim$ 0.1) first
for large 
galaxies passing within few hundred kpc of the cluster center. 
Notice also that the perturbation parameter,
at any given distance from the cluster center increases with the galaxy size (luminosity). 
Since the local tidal field varies as the inverse cube of the separation, 
and since this quantity is much larger than the typical size of spiral galaxies, 
disturbances are expected to be symmetric,  
contrary to galaxy-galaxy interactions \citep{BYRV90}. \\
Gas removal can take place only outside 
the tidal radius $r_{tidal}$ (the galactocentric distance up to which the perturbation is effective in removing material)  
which can be estimated for galaxies near the core radius $R_c$ \footnote{The lowest values for the tidal radius 
are expected close to the core radius \citep{MER84}} of the 
Coma, A1367 and Virgo clusters using the cluster parameters listed in Table \ref{Tab1} and \citep{MER84}: 
\begin{equation}
{r_{tidal}/R_c \sim 0.5 \times \Delta V_{gal}/\delta V_{cluster}}
\label{tidalr}
\end{equation}
\noindent
where $\Delta V_{gal}$ and $\delta V_{cluster}$ are the rotational velocity of the galaxy 
and the cluster velocity dispersion respectively.
Using an empirical relationship between the optical radius and the rotational velocity: 
\begin{displaymath}
{\Delta V_{gal}=21.55 \times r_{gal} ~~(\rm kpc) + 50.00~~~~~~~~~~~~~~{(\rm km~s^{-1})}} 
\end{displaymath}
\noindent 
we can estimate the tidal radius $r_{tidal}$ at the core radii of the three clusters as a function of
the galaxy optical radii:\\ 
$r_{tidal}(Coma)$=3.15 $r_{gal}$ + 7.30 kpc, 
$r_{tidal}(A1367)$=3.15 $r_{gal}$ + 7.30 kpc and 
$r_{tidal}(Virgo)$=1.58 $r_{gal}$ + 3.67 kpc.
The truncation radius of galaxies in Coma and A1367 is thus significantly larger than 
the optical and even of the HI radius (which is $\sim$ 1.8 times larger than the optical one, \cite{CAYK94}),
while it is comparable to the HI radius in Virgo. \\
In conclusion, tidal interactions with the cluster potential can induce an increase of the
nuclear activity of cluster galaxies and, eventually, a decrease of the total gas amount consumed through
star formation events. On the other hand, gas can be hardly removed directly by the interaction.

\subsection{Galaxy harassment}
\label{harassment}

\cite{MOOK96, MOOL98, MOOL99} proposed that the evolution of cluster galaxies is 
governed by the combined effect of multiple high speed galaxy-galaxy close ($\sim$ 50 kpc) 
encounters with the interaction 
with the potential of the cluster as a whole, a process that they named ``galaxy harassment''. 
Harassment depends on the collisional frequency, on the strength of the individual collisions, on 
the cluster's tidal field
and on the distribution of the potential within galaxies.\\
Moore et al. simulations show that, at a fixed mean orbital radius, galaxies on elongated orbits 
experience greater harassment than objects on circular orbits.
The multiple encounters heat the stellar component increasing the velocity dispersion and decreasing 
the angular momentum, 
meanwhile they make the gas to sink toward the galaxy center \citep{MOOK96}.\\
Because of their different potential distribution, massive and dwarf galaxies react differently to galaxy harassment. 
N-body (both pure gravitational and hydro-dynamical) simulations of low-mass
($L_*/5$ and $L_*/20$) spirals (represented by rotating exponential disks) in a Coma-like cluster
show that, at any given galaxy radius, dark matter is more easily stripped than stars because of the different 
orbital distribution of the two components. 
At early stages a large fraction (up to 50\%)
of the stars are removed; the subsequent increase of binding energy, caused by the increase of the central density, 
makes further star stripping less efficient. 
The obtained stellar profiles are exponentials, when only stars are included in the initial conditions, 
and nucleated exponentials when the gas is added \citep{MOOL98}. All these properties, including the observed 
increase of the velocity dispersion, mimic those of cluster spheroidals. 
The simulations also show that the smallest spheroidals are destroyed in the inner
(half of the virial radius) cluster. Only the densest and/or the nucleated objects would survive longer.
Another model prediction is that nucleated objects have 
higher velocity dispersion than normal spheroidals \footnote{The velocity dispersion of normal
dwarf ellipticals and spheroidals in Virgo cluster A is slightly ($\sim$  10 \%) smaller than that of their nucleated 
counterparts}.\\
The evolution of bright disk galaxies ($\sim$ $L_*$) in clusters differs from that of low-mass
systems, as it depends primarily on the depth of their potential wells and on the disk scale length.
High surface brightness galaxies, those with central steeply rising rotation curves, 
are found relatively stable to galaxy harassment. Beside minor star losses, the 
effect of the interaction is a small (0.5 mag arcsec$^{-2}$) increase of the central surface brightness, an increase
of the disk scale height (by a factor of 2-4) and of the central velocity dispersion, 
with the fading of spiral features 
\citep{MOOL99}. These structural and kinematic properties resemble those of bright lenticulars.\\
Low surface brightness galaxies, because of their low mass concentration (flat rotation curves and
large disk scale lengths), are strongly perturbed by the interaction. They are expected to loose most 
(up to 50-90 \%) of their stars
(the progenitors of the diffuse cluster light?), to
increase their central velocity dispersion and consequently their central surface brightness by $\sim$ 2 mag arcsec$^{-2}$
\citep{MOOL99}. Their resulting kinematic and structural properties resemble those of dE/dS0. \\
An increased star formation activity is expected because of the central accumulation of gas
and of the heating of 
molecular clouds, increasing the probability of cloud-cloud encounters, 
as shown by the models of \cite{FUJ98}.\\
The recent simulations of Virgo like clusters by \cite{GNE03} adopting various cosmologies 
($\Omega$$_0$=1, 0.4 with and without
$\Lambda$) show that tidal heating is more effective in low-$\Omega$$_0$ clusters. 
The maximum of the tidal forces do not always happen 
close to the cluster center, but during the encounters with massive galaxies or with
unvirialized remnants of infalling groups of galaxies. These simulations also show that the collision rate of galaxies 
increases by 10-50\% in the presence of substructures. 
In conclusion, galaxy harassment can effectively perturb low-luminosity galaxies because of their 
low-density cores and slowly rising rotation curves, thus contributing to the formation of cluster
dwarf ellipticals \citep{MOOL98}, to the fueling of low-luminosity AGNs \citep{LAKE98} and to
the destruction of low surface brightness galaxies in clusters \citep{MOOL99}. The effects on massive objects 
should be less pronounced, with a minor increase of the disk star formation activity \citep{MIHOS04} and 
an increase of the velocity dispersion in the bulge \citep{MOOK96}.

\subsection{Ram-pressure}
\label{Ram}

\cite{GUNG72} first proposed that the ISM could be removed from 
galaxies moving at $\sim$ 1000 km s$^{-1}$ through
the hot ($\sim$ 10$^7$-10$^8$ K) and dense ($\sim$ 10$^{-3}$-10$^{-4}$ atoms cm$^{-3}$) intergalactic medium by 
the ram-pressure mechanism. The interpretation of head-tail radio galaxies \citep{MILP72} in the ram-pressure scenario
brought to the prediction of a dense IGM in clusters of galaxies before X-ray observations actually detected it 
\citep{GURS72}.\\
Ram-pressure can effectively remove the
ISM if it overcomes the gravitational pressure anchoring the gas to the disk:
\begin{equation}
{\rho_{IGM}V_{gal}^2 \geq 2\pi G \Sigma_{star} \Sigma_{gas}}
\label{rampeq}
\end{equation}
where $\rho_{IGM}$ is the density of the IGM, $V_{gal}$ the galaxy velocity inside the cluster, 
$\Sigma_{star}$ is the star surface density and $\Sigma_{gas}$ the gas surface density.\\
A plethora of N-body and SPH simulations exists in the literature trying to investigate the role of ram-pressure 
on gas stripping of both cluster early- and late-type galaxies, 
and on the possible transformation of 
spiral into lenticular galaxies, or of dwarf irregulars into dSph. \\
The various models (SPH, N-body) differ in the way they account for  
the cluster gas distribution (density profile), 
the galaxy orbits within the cluster (radial, circular, galaxy inclination with respect to the orbit),
the potential distribution within the galaxy (disk vs. bulge, with or without dark matter), 
the star formation
(gas consumption and replenishment by recycled gas), 
the contribution of viscosity and/or thermal evaporation. In spite of these differences, 
assuming typical IGM densities and velocity dispersions observed in nearby clusters,
all variations of the model concur at establishing that
ram-pressure is sufficient to remove 
 part of the ISM from galaxies
on time scales comparable with their cluster crossing time (a few 10 $^9$ yr, see Sect. \ref{timescales}).\\
Radial orbits are more efficient because of the higher velocity, 
closer crossing to the cluster core \citep{ABAM99, QUIM00, VOLC01a}.
The efficiency of removal depends on the inclination of the galaxy disk with respect to the trajectory, 
with face-on interactions more efficient than edge-on or inclined encounters \citep{ABAM99, 
QUIM00, VOLC01a}.  
However spiral galaxies on radial orbits should end up stripped because
their interaction with the cluster IGM will sooner or later become 
face-on since the orientation of the galaxy rotation axis is conserved \citep{QUIM00}.
Because of their shallower potential well, gas removal is expected to act
more efficiently on dwarf irregular galaxies than on giant spirals \citep{MORB00, MARB03}.
 Models indicate that the ram-pressure efficiency is higher if the galaxy ISM has a multiphase 
structure, containing bubbles, shells and holes ranging in size from a few parsec to a kiloparsec, and if
it is taken into account that HI disks are centrally depleted even in isolated galaxies \citep{QUIM00}.\\
Before leading to a complete gas ablation, ram-pressure
produces significant compression ahead of the galaxy, with the possible formation of a bow-shock 
and of a low density gaseous tail behind, giving to
the galaxy a cometary shape \citep{BALL94, STEA99, MURB99, QUIM00, MORB00}. 
When the interaction is close to edge-on the central gas density can increase by a factor of $\sim$ 1.5
\citep{VOLC01a}, because the outer HI is captured by the inner disk. The distribution
of the HI gas can be very asymmetric, depending on the impact parameters \citep{VOLC01a}.
Only clouds with a surface density $<$ 4 10$^{-3}$ g cm$^{-2}$ can be swept out of the galaxy. Since molecular clouds have 
surface densities of the order of 10$^{-2}$ g cm$^{-2}$, they are mostly unaffected by the interaction 
\citep{QUIM00}, although the gas compression can facilitate the collapse of molecular clouds,
thereby increasing the star formation rate \citep{BEKC03}.\\
\cite{TOS94} shows that ram-pressure generates a single-armed spiral structure in the galactic gas, with the spiral
pattern having a retrograde rotation, while ring-like features can form in the inner disk 
\citep{SOFW93, SCHS01}. 
The models of \cite{SCHS01} also show
that the displacement of the gas with respect to the disk can trigger gravitational instabilities, with the formation
of flocculent spirals. These spiral arms transport angular momentum outward, compressing the inner disk and forming a 
prominent gas ring. 
Inclined galaxies experience less gas removal but a larger angular momentum loss then face-on events \citep{SCHS01}. \\
Gas can be accreted from the downstream side into the core \citep{BALL94, VOLC01a, SCHS01},
in particular in retrograde encounters, where the loss of angular momentum can create a galactic disk \citep{SOFW93}.
In the case of edge-on stripping more than 50\% of the stripped gas can be
re-accreted, whereas in face-on encounters the fraction of re-accreted gas is significantly smaller \citep{VOLC01a}.
This re-accreted gas might temporarily increase the disk surface density \citep{VOLC01a}.\\
The hydrodynamic interaction between the hot IGM and the cold ISM leads to an increase of the external pressure, the 
formation of bow-shocks, thermal instabilities and turbulent motions within the disk of the galaxy.
All these phenomena increase cloud-cloud collisions, cloud collapse, and can thus be responsible 
for an enhanced star formation activity of 
cluster galaxies \citep{EVR91, BEKC03}. 
By modeling pressure variations of the ISM due to ram-pressure, combined with HI to H$_2$ gas transformation (molecular
cloud formation and destruction) and HI sweeping, \cite{FUJ98} and \cite{FUJN99}
quantified the variations of the star formation activity following the formalism of \cite{ELME97}.
Their models show that on short time scales ($\sim$ 10$^8$ yr) in high density, rich clusters the star formation activity 
can increase up to a factor of 2 at most. On longer timescales, however, the removal of the HI gas reservoir
leads to a decrease of the fuel feeding the star formation, and galaxies become quiescent \citep{FUJ98, FUJN99,
OKAN01}.
This mild increase of the star formation activity is however not observed in models with low gas density and shallow 
potential clusters.\\
A decrease of the molecular gas content of cluster disk galaxies (up to $\sim$ 80\%) is
expected from the gas consumption due to the increase of star formation activity (from 0.1 to 0.5 M$\odot$ yr$^{-1}$ 8 Myr 
after the cloud collapse; \cite{BEKC03}) and not from gas removal during the interaction. 
It is interesting to note
that in clusters with typical gas temperature and velocity dispersion,
the increase of star formation is higher during the transit in the cluster outskirts, where the IGM density is
lower, than near the cluster center where gas removal by ram-pressure 
suppresses the star formation \citep{BEKC03}.\\
The old stellar component is unperturbed during ram-pressure interactions \citep{QUIM00}. For face-on
interactions, however, the gas escaping from 
the plane can induce tidal forces on the galaxy stellar disk shallowing the potential well on time scales 
comparable with the time scale for gas stripping
($\sim$ some 10$^8$ yr). These forces can increase the velocity dispersion perpendicular to the disk, thickening the stellar disk. 
The N-body simulations of \cite{FARS80} showed that the thickening of the disk is 
possible only in the outer disk since only there the gas is removed, leaving the inner disk
unaffected. It is thus unlikely that ram-pressure formed the thick and large bulges of S0s \citep{FARS80}.
\cite{SCHS01} speculated however that successive crossing of the cluster 
can produce cycles of annealing, which might form the prominent bulges typical of S0s.\\
\begin{figure}
\epsscale{0.5}
\plotone{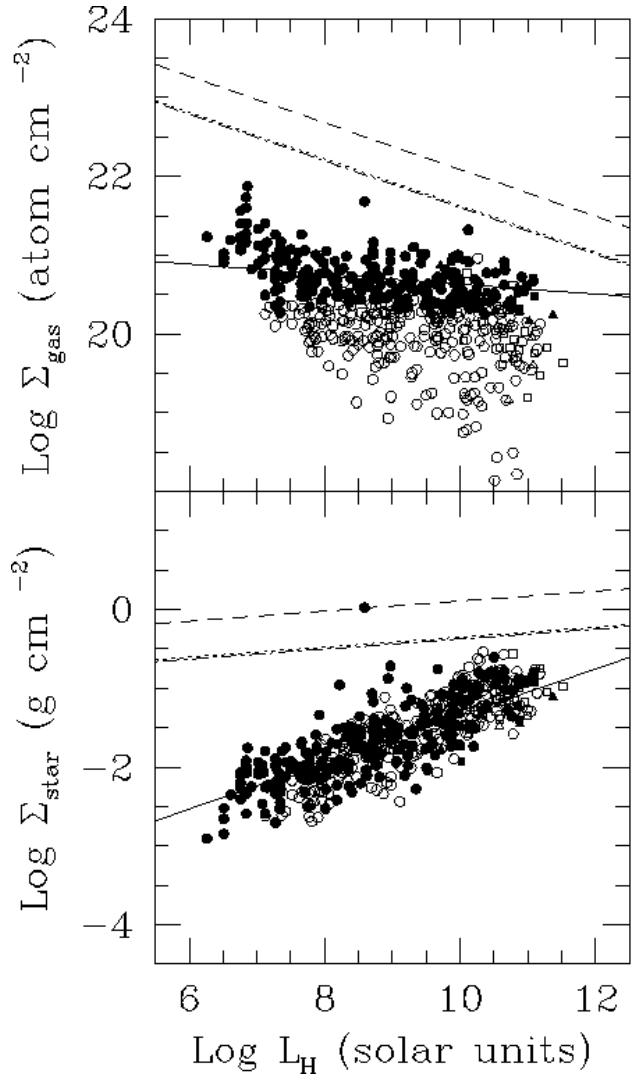}
\caption{The relationship between the logarithm of the gas surface density (upper panel), the stellar 
surface density (lower panel) and the
logarithm of the $H$ band luminosity. Filled symbols are for galaxies with a normal HI content 
(HI deficiency $\leq$ 0.3),
empty symbols for deficient objects (HI deficiency $>$ 0.3). Squares are for Coma, 
triangles for A1367 and circles for Virgo objects. 
The continuum line represents the best fit to the data (for the gas 
surface density only non-deficient galaxies are fitted). 
The dashed, dot-dashed and dotted lines indicate the upper limits
to the gas and stellar densities for an efficient ram-pressure stripping for Coma, A1367 and Virgo respectively.  
}
\label{surface}
\end{figure}
\begin{figure}[!htb]
\epsscale{0.5}
\plotone{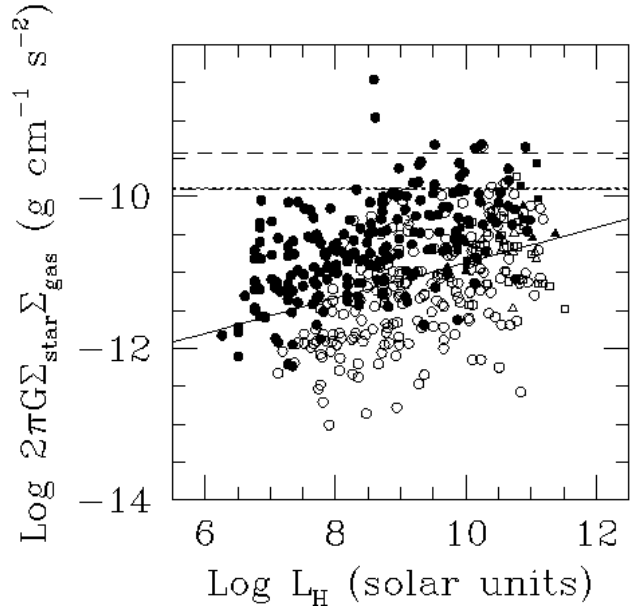}
\caption{The relationship between the gravitational forces per unit area  
and the $H$ band luminosity. Filled symbols are for galaxies with a normal HI content (HI deficiency $\leq$ 0.3),
empty symbols for deficient objects (HI deficiency $>$ 0.3), with squares for Coma, triangles for A1367 and circles for Virgo. 
The dashed, dotted-dashed and dotted lines indicate the upper limit to the gravitational pressure for an efficient ram-pressure stripping 
for Coma, A1367 and Virgo cluster galaxies respectively. The continuum line gives the fit to the data obtained by
combining the best fits in Fig. \ref{surface}.
}
\label{ram_vel}
\end{figure}
\begin{figure}[!htb]
\epsscale{0.5}
\plotone{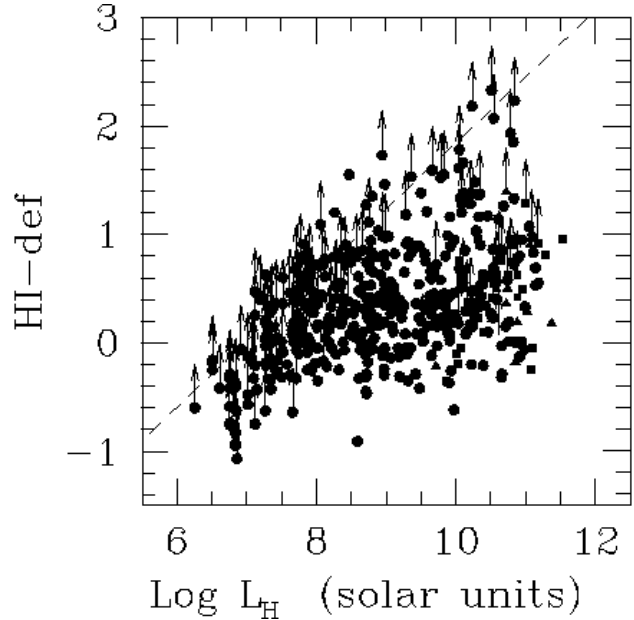}
\caption{The relationship between the HI-deficiency parameter and the $H$ band luminosity (in logarithmic scale) 
for Coma (squares), A1367 (triangles) and Virgo (circles) galaxies.
Filled dots are for HI-detected galaxies, arrows for HI-undetected objects. 
The dotted line indicates the region inaccessible to the
observations with a sensitivity of 1 mJy per channel for galaxies at the distance of Virgo. 
}
\label{ram_lum}
\end{figure}
If ram-pressure is responsible for the HI ablation in clusters
we would expect from Eq. \ref{rampeq} a correlation between the HI-deficiency
parameter and $\rho_{IGM}$$V_{gal}^2$/2$\pi G \Sigma_{star} \Sigma_{gas}$. 
Let's check the calculation for 3 nearby clusters, Coma, A1367 and Virgo (as usual using data from GOLDmine).
We assume $\delta V_{cluster}$ (for the late-type component) and $\rho_{IGM}$ from Table \ref{Tab1} and let
$V_{gal}$=$\sqrt{3 \delta V_{cluster}^2}$.\\
The stellar surface density within the optical diameter (in units of g cm$^{-2}$) can be 
estimated from the $H$ band luminosity, assuming a mass to light ratio of 4.6 
\citep{GAVPB96a}: 
\begin{equation}
{\Sigma_{star}  = \frac{4.6 L_H}{\pi (a/2)^2}}
\end{equation}
\noindent
where $L_H$ is the $H$ band luminosity and $a$ is the optical major diameter. The estimate of $\Sigma_{star}$ is simply
an average over the entire galaxy, without assuming a realistic stellar distribution.
As shown in Fig. \ref{surface}, the stellar surface density correlates with the total galaxy mass, as traced by
the $H$ band luminosity:
$Log \Sigma_{star}$=-4.310+0.296 $Log L_H$ (g cm$^{-2}$).\\
 The gas surface density $\Sigma_{gas}$ (in units of cm$^{-2}$) involved in Eq. \ref{rampeq} can be determined
adding the HI and H$_2$
mass estimates (when H$_2$ is available from \cite{BOSL02A}, or assuming that H$_2$ 
is on average 15\% of the total HI gas), and assuming that the HI gas diameter is $\sim$ 1.8 times larger than
the optical one \footnote{Since deficient galaxies have the HI gas mostly removed from the outskirts, 
the gas surface density is underestimated in galaxies with an HI deficiency parameter $\geq$  0.3.}:
\begin{equation}
{\Sigma_{gas}  = \frac{M(HI)+M(H_2)}{\pi (1.8 a/2)^2}}
\end{equation}
\noindent
The derived gas surface density $\Sigma_{gas}$ is a weakly decreasing function of the $H$ 
band luminosity when galaxies with a normal HI gas content (HI deficiency $\leq$0.3) are considered: 
$Log \Sigma_{gas}$=21.264-0.063 $Log L_H$ (atom cm$^{-2}$), 
as shown in Fig. \ref{surface} (filled dots).\\
The reader should not be surprised by the (mild) anti-correlation between $Log \Sigma_{gas}$ and $Log L_H$
since it is known that the gaseous mass fraction decreases with increasing mass \citep{BOSG01}. \\
At this stage we can estimate an empirical relationship
between the gravitational potential 2 $\pi$ $G$ $\Sigma_{star}$ $\Sigma_{gas}$ and the $H$-band luminosity and check if
this potential is sufficient to prevent the gas to be swept away from stars by ram-pressure, as a function of
H luminosity (see Fig. \ref{ram_vel}).
It turns out that in the three clusters ram-pressure can effectively remove the atomic HI from any object
with velocity similar or larger than the average cluster velocity dispersion\footnote {The efficiency 
of ram-pressure may be overestimated assuming the cluster central gas density $\rho_{IGM}$, as we did. However this should not 
represent a severe problem for most galaxies in radial orbits that will soon or later pass near the cluster center.}. 
This result is consistent with \cite{VOLC01a}, who were able to reproduce the 
observed radial truncation of the HI disks in Virgo cluster galaxies.
Given the dependence of the
gravitational potential from the total $H$ band luminosity, ram-pressure is expected to be more
efficient at removing gas from low-mass objects than from massive spirals. 
 This seems not confirmed observationally in Fig. \ref{ram_lum} where the expected inverse relationship
between the HI-deficiency parameter and the total mass of galaxies is not seen.
This, however, is due to an observational bias because,
at any given HI sensitivity, galaxies with larger HI deficiencies 
are more difficult to detect at low mass.
In order to detect high HI deficiencies in dwarf galaxies
it would be important to push the Arecibo observations of Virgo dwarf galaxies
to very low limits\footnote{This is in fact a high-priority project for the ALFA multi-beam system at Arecibo \citep{GIOV05}.}, 
beyond those reached by \cite{HOFS96} or \cite{GAVB04} and to better calibrate the
HI deficiency parameter for dwarf systems, as attempted by \cite{LEEM03}.  \\
 Examples of galaxies showing clear signs of undergoing ram-pressure stripping are CGCG 97-073 and 97-079
in A1367, NGC 4848 in Coma, NGC 4522 and 4654 in Virgo (see sect. \ref{prototype}).
 
\subsection{Viscous stripping}
\label{Viscous}

\cite{NUL82} proposed viscous stripping as a possible mechanism capable 
of dragging gas out of galaxies in clusters.
If a galaxy (of radius $r_{gal}$), carrying its cold and dense ISM, travels at speed $V_{gal}$ 
across the hot ($T_{IGM}$) and tenuous ($\rho_{IGM}$) intergalactic medium, the
outer layers of its ISM  experience a viscosity momentum transfer sufficient for dragging out gas at 
some rate depending on whether the flux is laminar or turbulent.\\ 
If the flow is laminar, i.e if the Reynolds number: 
\begin{displaymath}
Re=2.8\Big(\frac{r_{gal}}{\lambda_{IGM}}\Big)\times(\frac{V_{gal}}{c_{IGM}}) ~\leq 30
\end{displaymath}
and if the mean free path of the ions in the hot gas is sufficiently small for the classical viscosity to apply:
\begin{displaymath}
\lambda_{IGM}[kpc]\simeq11\Big(\frac{T_{IGM}[K]}{10^8}\Big)^2\times \Big(\frac{10^{-3}}{\rho_{IGM}[cm^{-3}]}\Big) \leq r_{gal} 
\end{displaymath}
(where $c_{IGM}$=$\sqrt{k_B T_{IGM}/m_H}$ is the sound speed in the IGM,
$k_B$ is the Boltzmann constant and $m_H$ is the proton mass), 
then the mass loss rate is given by:
\begin{equation}
{\dot{M}_{laminar} \simeq F_{laminar}/V_{gal}=\pi r_{gal}^2 \rho_{IGM} 
V_{gal} \times (12/Re)=4.3 \pi r_{gal} \rho_{IGM} \lambda_{IGM} c_{IGM}}
\end{equation}
where $F_{laminar}$ is the drag force.
If the flow is turbulent ($Re$ $\geq$30), then:
\begin{equation}
{\dot{M}_{turbulent} \simeq F_{turbulent}/V_{gal} = \pi r_{gal}^2 \rho_{IGM} V_{gal} }
\end{equation}
\noindent
In the turbulent case the drag force acting on the ISM is proportional to 
$\rho_{IGM}$$V_{gal}^2$, as in the case of ram-pressure.\\ 
In the case of the Coma, A1367 and Virgo clusters (whose properties
are listed in Table \ref{Tab1}), galaxies 
are subject to turbulent stripping whenever their linear sizes  
$r_{gal}$ $\geq$ 10, 6 and 15 kpc in the three clusters
respectively \footnote{These values have to be taken as upper limits since 
$\lambda_{IGM}$ might significantly decrease in the presence of weak tangled magnetic fields}. 
Smaller objects experience laminar viscosity  
\footnote {The condition $\lambda_{IGM}$$\leq$ $r_{gal}$ is satisfied for $r_{gal}$ $\geq$ 3.2, 1.4  
and 0.4 kpc in Coma, A1367 and Virgo respectively}.\\
As remarked by \cite{NUL82}, if the flow velocity $V_{gal}$ exceeds the 
adiabatic sound speed in the hot gas $\sqrt{5/3}$ $c_{IGM}$ 
(1450, 970, 730 km s$^{-1}$ for Coma, A1367 and Virgo respectively)
a shock will form ahead of the ISM. The stripping will occur 
preferentially on the front side of the galaxy, producing an asymmetric gas
distribution even on galaxies moving edge-on inside the IGM.\\
The time scales for complete gas stripping are relatively short: for both large (20 kpc; 5 10$^9$ M$\odot$) 
and small (5 kpc; 5 10$^8$ M$\odot$) galaxies in the Coma cluster, the time needed to completely
remove the atomic gas is $\sim$ 4 10$^7$ yr. Longer times are needed in A1367 and Virgo: $\sim$ 10$^8$ yr for
large galaxies, from 4 (A1367) to 10 (Virgo) times longer for smaller objects.\\
The signature of viscous stripping on the 
structural, kinematic properties,  molecular gas content and star formation activity of galaxies 
is expected to be similar to that of ram-pressure.\\
UGC 6697, proposed by \cite{NUL82} as the prototype of a galaxy ongoing turbulent viscous stripping,
is in fact a merging galaxy (see sect. \ref{prototype}). 

\subsection{Thermal evaporation}
\label{evaporation}

As shown by \cite{COWS77} thermal evaporation can efficiently remove gas in cluster galaxies. 
If the IGM temperature is high compared to the galaxy velocity dispersion,
at the interface between the hot IGM and the cold ISM the temperature of the ISM rises rapidly,
the gas evaporates and is not retained by the gravitational field.\\
The mass loss rate can be estimated using the relation:
\begin{equation}
{\dot{M}_{evaporation} = \pi r_{gal}^2 \rho_{IGM} c_{IGM} 4 \phi_s F(\sigma_o)}
\label{eq_evap}
\end{equation}
\noindent
with $\phi_s$ $\simeq$1, $\sigma_o$=1.84 $\lambda_{IGM}$/($r_{gal}$$\phi_s$) and 
$F(\sigma_o)$ =2$\sigma_o$ for $\sigma_o$ $\leq$1 \citep{NUL82}
($\lambda_{IGM}$ is defined in section \ref{Viscous}).
Given the properties of the three considered clusters, 
$\sigma_o$ $\leq$1 holds true for all galaxies, thus
we can assume $F(\sigma_o)$ =2$\sigma_o$, and eq. \ref{eq_evap} reduces to:
\begin{equation}
{\dot{M}_{evaporation} \simeq 14.7 \pi r_{gal} \rho_{IGM} c_{IGM} \lambda_{IGM}}
\end{equation}
where $c_{IGM}$ is defined in section \ref{Viscous}.\\
\begin{figure}
\epsscale{0.6}
\plotone{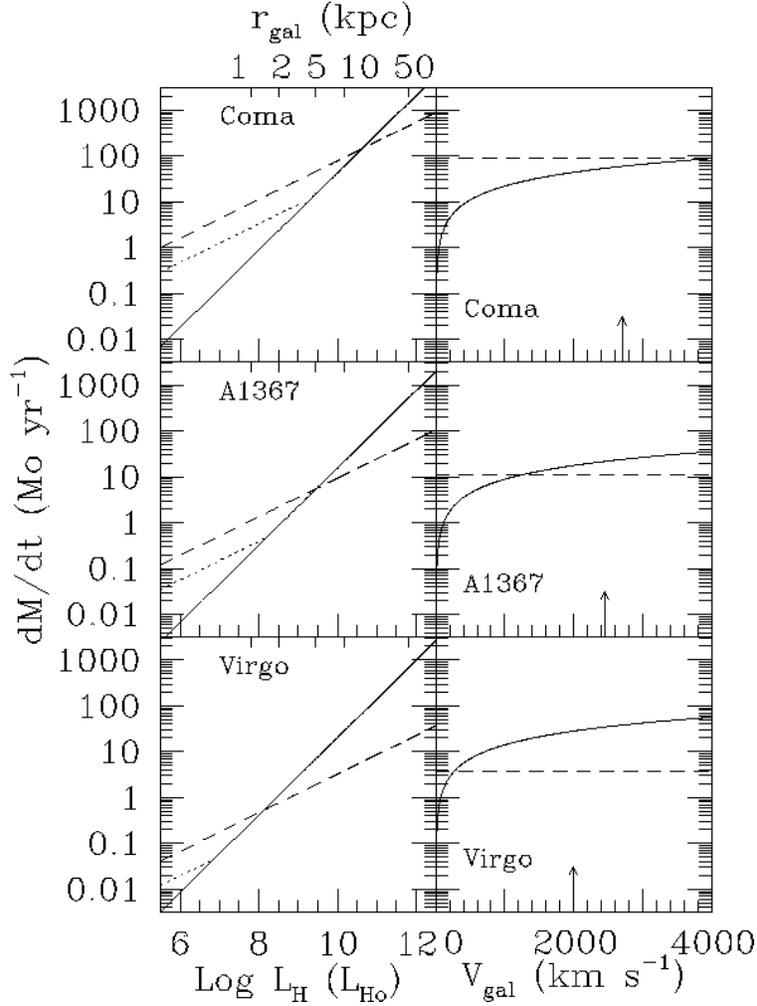}
\caption{The mass loss rate for thermal evaporation (without magnetic fields: dashed line) and ram-pressure
(or turbulent viscous stripping) for galaxies in Coma, 
A1367 and Virgo with radii $>$ 10, 4 and 1 kpc respectively  
(continuum line) and laminar viscous stripping (dotted line) are given
as a function of the $H$ band luminosity (linear size on the top axis) in the left panel.
The galaxy velocity within the cluster is assumed $V_{gal}$=$\sqrt{3\delta V_{cluster}^2}$.
Right panel: the comparison between the mass loss rates for thermal evaporation 
for galaxies of 10 kpc radii (dashed line) and ram-pressure 
(or turbulent viscous stripping)
(continuum line) as a function of the galaxy velocity within the cluster. 
The vertical arrow indicates $V_{gal}$=$\sqrt{3\delta V_{cluster}^2}$ for the late-type component.
}
\label{mstar}
\end{figure}
\begin{figure}
\epsscale{0.4}
\plotone{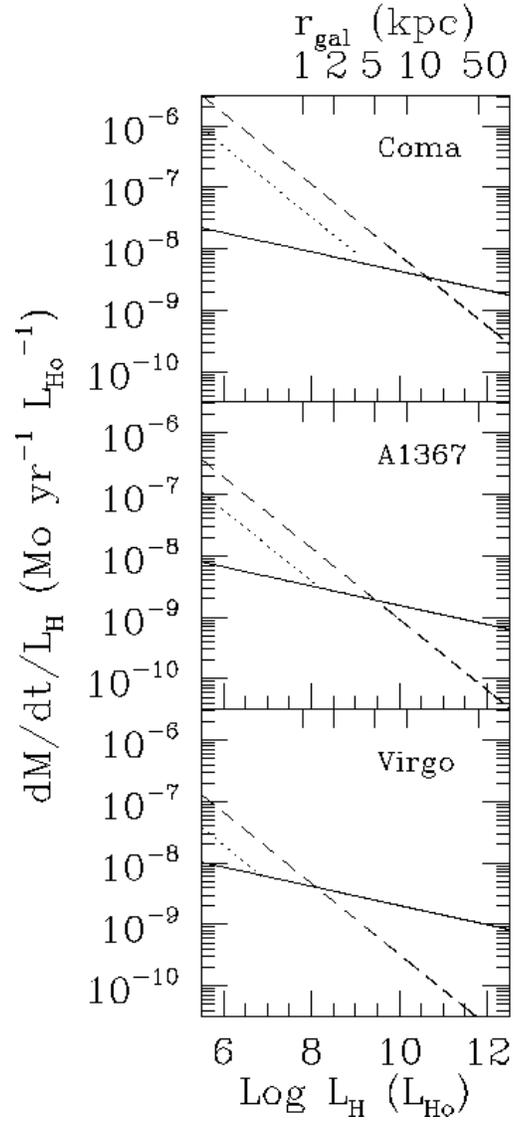}
\caption{The mass loss rate per unit mass as a function of the $H$ band luminosity. 
Same layout as in Fig. \ref{mstar} (left panel)}
\label{mstarn}
\end{figure}
Thermal evaporation is a sensitive 
function of the IGM temperature and of the magnetic field, and to a lesser extent of the density \citep{COWS77};
its efficiency can be reduced in presence of magnetic fields \citep{VIKF97}.  A typical galaxy of radius 15 kpc and 
5 10$^9$ M$\odot$ of atomic gas can be completely stripped (assuming no magnetic fields) in $\sim$ 4 10$^7$ yr in Coma, 3 10$^8$ yr 
in A1367 and $\sim$ 10$^9$ yr in Virgo.\\
As for the viscous stripping, the effects of the thermal evaporation on the 
kinematic and structural properties of cluster galaxies, as well as on the molecular gas content 
and star formation activity are difficult to quantify.\\

\subsection{Starvation}

Galaxy "starvation" or "strangulation", a process proposed more than 20 years ago by \cite{LARS80} to explain the 
transformation of spirals into lenticulars, has been recently invoked to explain 
the mild gradient in the morphology fraction found outside one virial radius 
in a cluster at $z$=0.4 \citep{TREU03}. Emphasis is put on the large scale on which such
a mechanism might be effective, contrary to
other mechanisms previously described
that, except galaxy harassment, are supposed to work on smaller scales \citep{TREU03,BALN00}. 
Since in normal galaxies the gas that feeds the star formation (on time scales as long as the Hubble time) 
comes from infall of an extended gas reservoir,
the effect of removing the outer galaxy halo would be that of preventing further infall of gas into the disk.
On time scales of a few Gyr the star formation would thus exhaust the available gas, 
quenching further star formation activity. \\
The seminal idea of \cite{LARS80} has been elaborated by \cite{BEKK02}. Their numerical simulations showed that 
even if a spiral orbits a cluster with a pericenter 
distance of $\sim$ 3 core radii, $\sim$  80\% of its halo is stripped within a few Gyr by the hydrodinamical interaction
with the ICM plus the global tidal field of the cluster, preventing gas accretion into the disk, 
and consequently
suppressing the star formation. In the end the spiral structure becomes less pronounced, and galaxies 
progressively becomes anemic, disk-dominated lenticulars. 
They might coincide with the 
small percentage of
passive, anemic galaxies found in the SDSS by \cite{GOTO03b} at large clustercentric distances.
Starvation, with its long time-scale, was invoked by \cite{BALN00} to explain the smooth transition 
in the star formation activity observed radially around density enhancements in the 2dF.

\subsection{Preprocessing}

According to the hierarchical scenario for the formation of large scale structures, 
groups of galaxies infalling onto clusters of galaxies 
represent the building blocks of today rich clusters of galaxies. 
Substructures seen in the X-rays and in kinematical studies of clusters of galaxies 
are the strongest evidence in favor of this cluster formation mechanism \citep{KODA05}. 
Galaxy groups may therefore represent natural sites for a 
\emph{preprocessing} stage in the evolution of cluster galaxies \citep{MIHOS04,FUJI04,DRES04b}
through tidal interactions, otherwise ineffective in high velocity dispersion environments.
Meanwhile ram-pressure, starvation and evaporation might be already effective 
in these groups at $z$ $\sim$ 0.5, as shown by the simulations of \cite{FUJI04}.
Preprocessing might represent the ideal circumstance for the formation of 
lenticulars through unequal-mass mergers or minor mergers in spirals \citep{KODAS01}. 
These relatively violent events, as opposed to starvation, are able to heat up the disks, thus producing high bulge 
to disk lenticulars, as those observed in clusters.
Since preprocessing occurs well outside the core of rich clusters, 
this mechanism has been invoked to explain 
why the atomic gas content and the star formation activity are found suppressed 
at large clustercentric distances ($\sim$ 1 virial radius) \citep{KODA01}.
Although preprocessing might be at the origin of the post-starburst galaxy population in the
SW extension of the Coma cluster \citep{CALR93}, this mechanism is rare in nearby clusters, because
in $\Lambda$ cosmologies the rate of accretion of small groups on clusters has decreased at the present epoch. 
As remarked by \cite{DRES04b}, the peak of groups in the building of cluster should happen at $z \sim 0.5$.
In the Virgo cluster, for instance, the substructures that are now collapsing into the main cluster
have already grown massive enough to prevent preprocessing.  
There exist however an example of ongoing preprocessing taking place in A1367 (see Sect. \ref{big}).
It should also be emphasized that local clusters contain significant examples of healthy 
spiral galaxies that made it to the inner cluster regions without being 
"wounded" by preprocessing and that are receiving their first and fatal kick directly from ram-pressure.  
Two are CGCG 97-073 and 079 in A1367 that are just as actively star forming as isolated objects and
many other HI and H$\alpha$ healthy galaxies that lie between the virial and the core cluster radii.

\subsection{Comparison between the various processes}

We conclude that the relative importance of the processes considered above in nearby clusters  
depends on parameters that are difficult to quantify for individual galaxies,
primarily depending on their 3-D position with respect the cluster center. 
Some general conclusions can however be summarized as follows: \\

1) Effects of the perturbation:\\
a) Gravitational interactions (galaxy-galaxy, galaxy-cluster, harassment) can induce nuclear gas infall that can
be at the origin of nuclear activity. They can also contribute to heating up stellar disks 
and thus increasing the bulge to disk ratio (while retaining some bulge rotation). These processes are more efficient in
low-luminosity and low-surface brightness galaxies, which might be easily transformed into dE. 
Gravitational interactions are thus able to transform spirals into lenticulars.\\ 
Tidal interactions with the cluster as a whole can 
hardly remove the outer HI disks since the truncation radius is larger 
than the HI disk radius. Thus they cannot be at the origin of the HI-deficient galaxy population in nearby clusters.\\
b) Interactions with the hot ICM can efficiently remove
the outer disk gas and quench the star formation (directly by gas removal or via starvation) but they can hardly
increase the bulge to disk ratio, as requested to explain the morphology segregation.\\

2) Time scales and probabilities:\\
Given the high velocity dispersion in clusters, the probability that spiral galaxies 
are perturbed by tidal interactions with nearby companions is extremely low. 
The time scale for tidal interactions (relaxation time) is some 10$^{10}$ yr.
Times scales for ram-pressure gas stripping ($\sim$ one crossing time, 
10$^9$ yr) are shorter than for harassment (several crossing times), where multiple encounters are necessary.
In relaxed, gas rich clusters such as Coma, the time scale 
for gas removal for thermal evaporation or viscous stripping are very short (less than 10$^8$ yr), while some 10$^8$ yr 
in unrelaxed, gas poor clusters such as A1367 or Virgo. Models indicate that the time  
scale for galaxy starvation are of the order of some Gyr.\\

3) Efficiency as a function of the clustercentric distance:\\
The galaxy - cluster IGM interactions are most efficient close to the cluster center where the density
and the temperature of the IGM (as well as the velocity of galaxies) reach their maxima. 
The perturbations induced by the cluster potential are also most efficient in the cluster center since 
the cluster tidal field is maximum at the core radius. 
Although the duration of the interaction might be shorter (because of the high relative velocity)
than in the cluster outskirts, the frequency of galaxy-galaxy interactions reaches the maximum in the densest central regions.
These processes seems thus confined to the inner part of the cluster, inside the virial radius.\\
Because of the combined action of galaxy-galaxy and galaxy-cluster gravitational interactions, galaxy
harassment might be effective also at the cluster periphery. For similar reasons starvation and 
pre-processing might be effective well outside the cluster core and even outside the virial radius.\\

4) Relative importance of hydrodynamical mechanisms:\\
The relative importance of the three interactions between galaxies and the cluster IGM: 
viscous stripping, thermal evaporation and ram-pressure can be quantified. 
The mass loss rates $\dot{M}$ $\simeq$ $F/v_{gal}$ are comparable for turbulent viscous and 
ram-pressure stripping, both being proportional to  $\pi r^2 \rho_{IGM} V_{gal}$. The laminar viscous 
stripping and the thermal evaporation have the same analytical dependence on the cluster and galaxy parameters, 
but the thermal evaporation is about three times more effective than the laminar viscous stripping.\\ 
At any given impact angle, position and transit velocity through the cluster 
(which are statistically known quantities), ram-pressure and viscous stripping dominate 
over thermal evaporation for large and fast galaxies. For small (dwarf) galaxies $\dot{M}_{evap} > \dot{M}_{lam} > \dot{M}_{ram}$ 
while for the giants it depends on the cluster: in Virgo 
$\dot{M}_{ram}$ dominates over the other processes, in Coma thermal evaporation dominates over ram-pressure.
Because of its higher IGM density and temperature,
mass losses are on average 10 times larger in the Coma than in the Virgo cluster.\\

\subsection{Prototypes of environmental disturbances}
\label{prototype}

We briefly review a few nearby, well studied objects that may be considered as representatives of specific
perturbation mechanisms.

\subsubsection{Tidal interactions}

Galaxies showing evident signs of an ongoing tidal interaction are relatively rare in nearby
clusters. In Virgo for example, where the morphological classification
is excellent owing to the available du Pont plates \citep{BINS85}, 
only NGC 4438 out of 1802
galaxies classified as cluster members has clear tidal tails or other signs of ongoing tidal interactions.\\
The peculiar Sb galaxy NGC 4438 has recently undergone a high velocity, off-center ($\sim$ 900 km s$^{-1}$) encounter
with the nearby (22 kpc projected separation) 
SB0 galaxy NGC 4435. NGC 4438 shows strong and extended tidal tails in both the optical \citep{ARP66},
near-IR \citep{BOST97A} and UV \citep{BOSG05}.
Both the atomic and molecular gas are displaced by $\sim$ 5 kpc,
toward the companion \citep{COMD88, CAYG90, KENR95}, coinciding in position with 
the peak of the X-ray and radio continuum emission.
The galaxy, with a mild nuclear activity, shows nuclear star formation  
and filaments of ionized gas originating from the plane of the disk \citep{KENR95, KENY02},
and has a perturbed H$\alpha$ rotation curve \citep{KENR95,CHEM05}.
Whether some cold dust frozen into the ISM was removed during the interaction is hard to tell: 
no 6.75 nor 15 $\mu$m emission is detected in mid-IR ISOCAM images at the level of a 
few $\mu$Jy arcsec$^{-2}$ neither associated with the HI/CO peaks nor along the tidal tails \citep{BOSS03b}.
By combining multifrequency spectro-photometric data with population synthesis and galaxy evolution
models \cite{BOSG05} showed that the galaxy recently ($\sim$ 10 Myr) underwent an instantaneous
burst of star formation, significantly younger than the tidal interaction with NGC 4435, dated 
by dynamical models at $\sim$ 100 Myr ago \citep{COMD88, VOLL04}.
It is interesting that the fraction of stellar mass produced by this starburst is $<$0.1 \% 
of the total galaxy stellar mass, extremely small even in such a violent interaction.\\
Another example of ongoing, although in a more advanced stage, dynamical interaction in clusters 
is the merging galaxy UGC 6697 in A1367.
This galaxy was originally taken as
prototype of ongoing turbulent viscous stripping by \cite{NUL82} or of ram-pressure by Gavazzi et al. (1984; 1995b)
because of its strong asymmetric HI \citep{GAV89, DICG91} and radio continuum 
morphology \citep{GAVJ87, GAVC95b}, and its strong star formation activity \citep{GAVC95b}.
Recent X-ray observations with Chandra \citep{SUNV04} reinforce the ram-pressure interpretation.   
An accurate kinematic study using both long slit and Fabry-Perot 2-D spectra suggests however the presence of
two merging galaxies (Gavazzi et al. 2001c).\\
 The presence of long filaments and plumes of ionized gas in the center and in the periphery of M87 (NGC 4486), 
 the dominant Virgo galaxy, 
 indicate a recent swallowing of a nearby gas rich object; the consequent matter infall 
 is the most likely feeder of nuclear activity in this radio galaxy \citep{SPAF93, GAVB00b}.\\
 M49 (NGC 4472), the other giant elliptical dominating the Virgo subcluster B, is in tidal interaction with the 
 dwarf irregular galaxy  UGC 7636 \citep{PATT92, HENS93, MCNS94, IRWS96}. 
 Most of the HI gas of the dwarf irregular galaxy is found at a location intermediate 
 between the dwarf irregular and the giant elliptical. Spatially coincident with the HI cloud
 \cite{LEER00} detected an HII region whose metallicity is similar to that of UGC 7636. 
  It is likely that tides from M49 first loosened the HI gas, then ram-pressure 
 completed the gas removal from UGC 7636.\\
NGC 4698 \citep{BERT99} and NGC 4772 \citep{HAYJ00}, two Sa galaxies in the Virgo cluster appear
regular in their morphology, but in fact they show signs of a past merging event, as
revealed by the geometric and kinematic decoupling of their disk and gas components.
It is worth stressing that the information on evolved merging systems lacks completeness, as it
requires ad-hoc high-resolution kinematic observations.

\subsubsection{Ram-pressure}
\label{ram_prot}

\begin{figure}
\epsscale{0.6}
\plotone{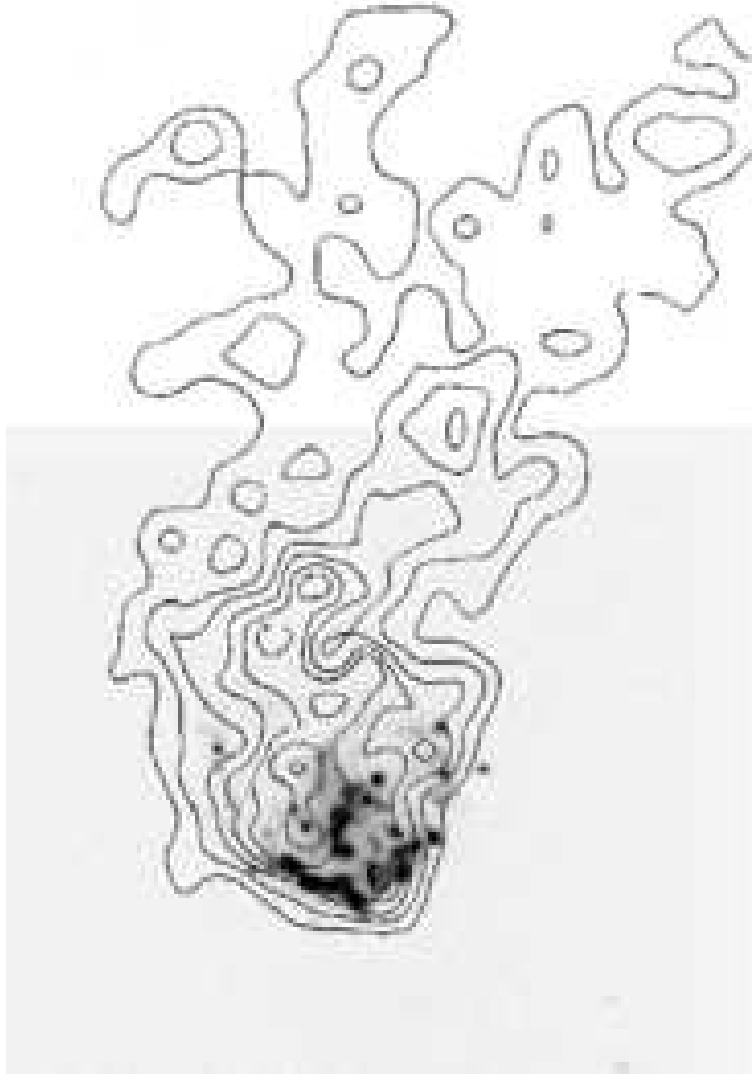}
\caption{Contours of the radio continuum emission at 1400 MHz of CGCG 97-073 superposed to the
 H$\alpha$ frame, showing the head-tail morphology of the extended component (adapted from \cite{GAVC95b}).}
\label{73radio}
\end{figure}

The prototype of strong ongoing ram-pressure stripping is the galaxy CGCG 97-073 in A1367 (Fig. \ref{73radio}). 
This galaxy, located at the N-W periphery of this spiral rich cluster, is characterized by an 
asymmetric radio morphology, with 75 kpc long radio continuum tails in the direction opposite to the cluster center,
similar to head-tail radio galaxies, and a steep gradient facing the cluster center 
\citep{GAVJ87, GAVC95b}. The galaxy has also an asymmetric morphology in H$\alpha$ and at short 
optical wavelengths (both tracing the young stellar population) with a bow-shock structure in the direction of the 
cluster center and a low surface brightness tail in the opposite direction (coincident with the 
direction of the radio continuum tail), 
while it is quite symmetric in the near-IR (old stellar population) \citep{GAVC95b}.
The HI gas, unresolved by the available VLA observations, is displaced toward the low surface 
brightness tail \citep{DICG91}, while the molecular gas is symmetrically distributed \citep{BOSG94}.
This galaxy has the radio to far-IR ratio higher than normal spiral galaxies, as well as an enhanced 
star formation activity per unit mass (as traced by its H$\alpha$ E.W.) compared 
to galaxies of similar type and luminosity
\citep{GAVC95b}. Deep H$\alpha$ observations revealed the presence of long (75 kpc) tails of ionized gas
associated to the radio continuum tail \citep{GAVB01b}.
All observational evidences indicate that this galaxy is undergoing ram-pressure
stripping while entering the cluster for the first time. The interaction removes the HI gas while leaving the molecular gas unaffected,
enhancing the star formation in the region facing the interaction with the cluster ICM 
and forming the radio continuum and H$\alpha$ trails \citep{GAVC95b, GAVB01b, BOSG94, CORG04}.
The ongoing interaction produced a weak starburst in the galaxy
a few 10$^7$ years ago, as shown by the excess  H$\alpha$ to UV ratio \citep{IGLBG03}, and triggered a shock 
compressing the magnetic field in the leading side of the galaxy.\\
The  edge-on galaxy CGCG 97-079 is physically and morphologically similar to its near companion CGCG 97-073 (long radio continuum 
\citep{GAVC95b} and ionized gas \citep{GAVB01b} tails, asymmetric HI profile \citep{DICG91},
unperturbed CO distribution \citep{BOSG94},
enhanced star formation activity and radio to far-IR ratio \citep{GAVC95b}).
It is worth mentioning that these two objects are at $\sim$ 0.4 virial radii from the cluster center, 
indicating that ram-pressure is effective not only in the highest density central regions.\\
Another candidate for currently ongoing ram-pressure stripping is 
NGC 4848 in the Coma cluster. Its HI and H$\alpha$ distributions 
are highly asymmetric \citep{GAV89, BRAC01, VOLB01b}. Some of the HI gas removed during the interaction might 
be falling back into the galaxy from the downstream side, as claimed by  \cite{VOLB01b}.\\
Well studied ram-pressure candidates in the Virgo cluster are NGC 4654, NGC 4522, NGC 4548, NGC 4569, NGC 4388
and NGC 4402.
The morphological asymmetries and the kinematical structure of NGC 4654, however, can be explained only
by a combined action of ram-pressure stripping (able to reproduce the extended gas) and
tidal interaction with the nearby NGC 4639 (responsible for the asymmetric stellar distribution) \citep{VOL03}. 
NGC 4522 has an H$\alpha$, HI, radio continuum and kinematical properties
indicating an ongoing ram-pressure event \citep{KENK99, VOLM00, VOLM04a, VOLM04b}. The comparison
of model predictions with multifrequency observations made \cite{VOLM00} to conclude that the galaxy passed through
the cluster center $\sim$ 6.5 10$^8$ yrs ago. Its perturbed H$\alpha$ morphology indicates that some
recent star formation has been induced by the extraplanar atomic gas, probably displaced by ram-pressure
stripping, falling back into the galaxy from the downstream side of its cometary trail \citep{VOLM00}.
NGC 4548 is similar to NGC 4522, except that
its face-on orientation makes the observation of the extraplanar gas more indirect \citep{VOLC99}.
The kinematic analysis of the anemic galaxy NGC 4569 in the center of Virgo has shown that 
its strong HI-deficiency is probably due to a ram-pressure stripping event occurred $\sim$ 300 Myr ago 
\citep{VOLM04c}. This result is in agreement with our recent analysis based on the comparison
of multifrequency 2-D images with chemo-spectrophotometric models of galaxy evolution \citep{BOSB05b}.
The complex kinematic of the inner kiloparsec brought however \cite{JOGE05} to conclude that
NGC 4569 is also in an early stage of bar-driven/tidally driven gas inflow able to perturb 
the central star formation activity.\\
In the Seyfert 2 Virgo galaxy NGC 4388, the ram-pressure stripped gas, seen in the form 
of extended HI filaments, is ionized by 
the UV radiation field of the active galactic nucleus, producing strong optical emission lines
\citep{YOS04, VEIB99}. An extended (110$\times$25 kpc) HI gas cloud of 3.4 10$^8$ M$\odot$ 
has been recently discovered at the north-east of NGC 4388 by \cite{OSVG05}. The authors claim that
this gas has been removed by the ram-pressure exerted by the halo of the nearby M86.
Another potential candidate for ram-pressure is NGC 4402, whose H$\alpha$ and HI morphology is similar 
to NGC 4522 \citep{CRKE05}.
Claims for ram-pressure stripping in early-type galaxies in Virgo, such as M86 (NGC 4406) are in
\cite{RANW95, WHIF91, NULC87} (see however \cite{ELME00}).\\
Another issue that we wish to stress is that several galaxies that can be considered prototypes
of the ram-pressure mechanism, either because their morphology indicates it
or because they were successfully modeled by \cite{VOLM00, VOLM04c, VOLM04a, VOLM04b} as being 
subject to ram-pressure stripping
are found between 0.3 and 0.75 projected virial radii, i.e. at rather large clustercentric distances.
Even though the mechanism itself operates at least up to 0.4 virial radii, its effects 
can last sufficiently to affect galaxies that have moved up to $\sim$ 1 virial radius. 

\subsubsection{Pre-processing}
\label{big}

\begin{figure}[!h]
\epsscale{1.5}
\plottwo{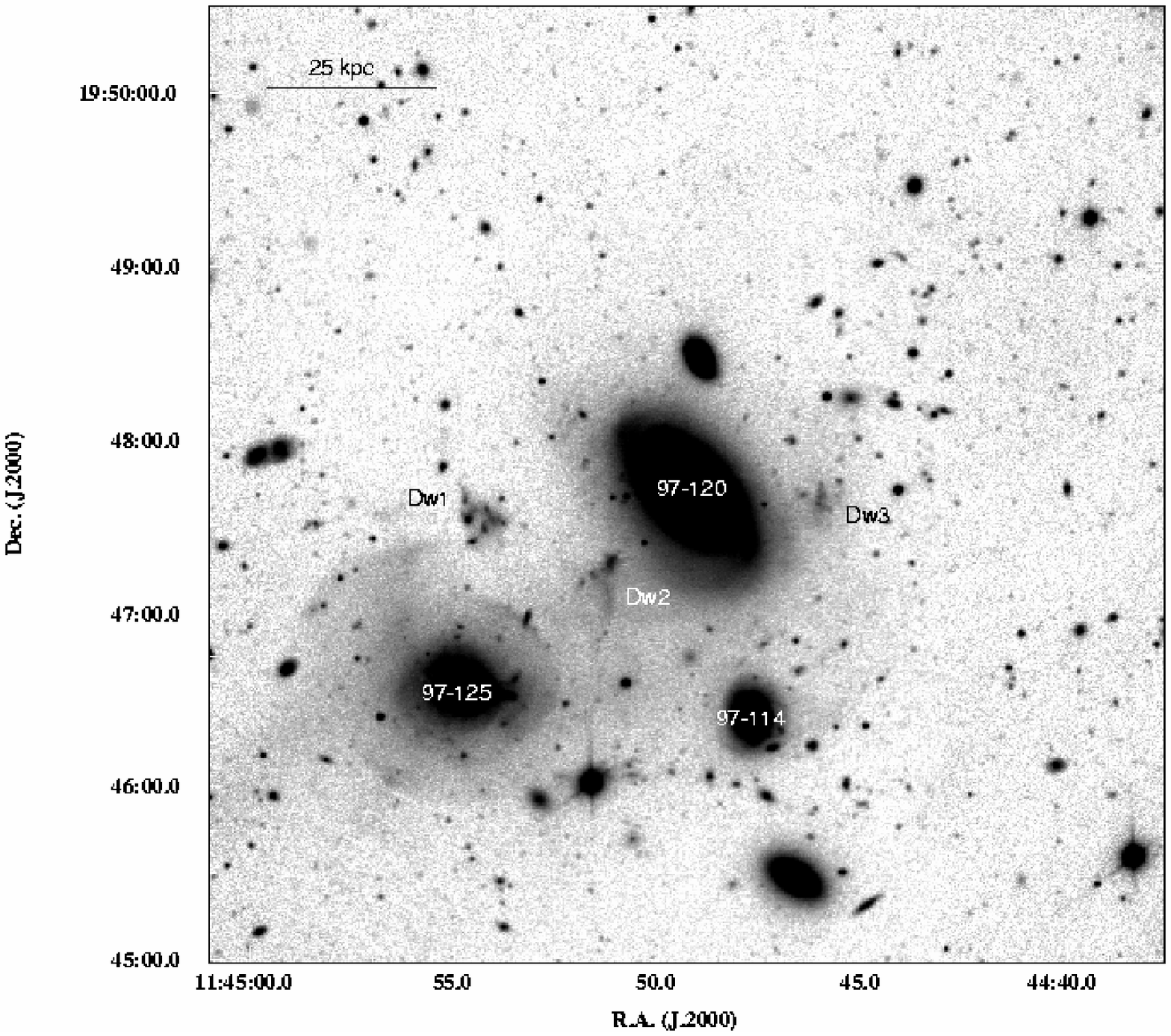}{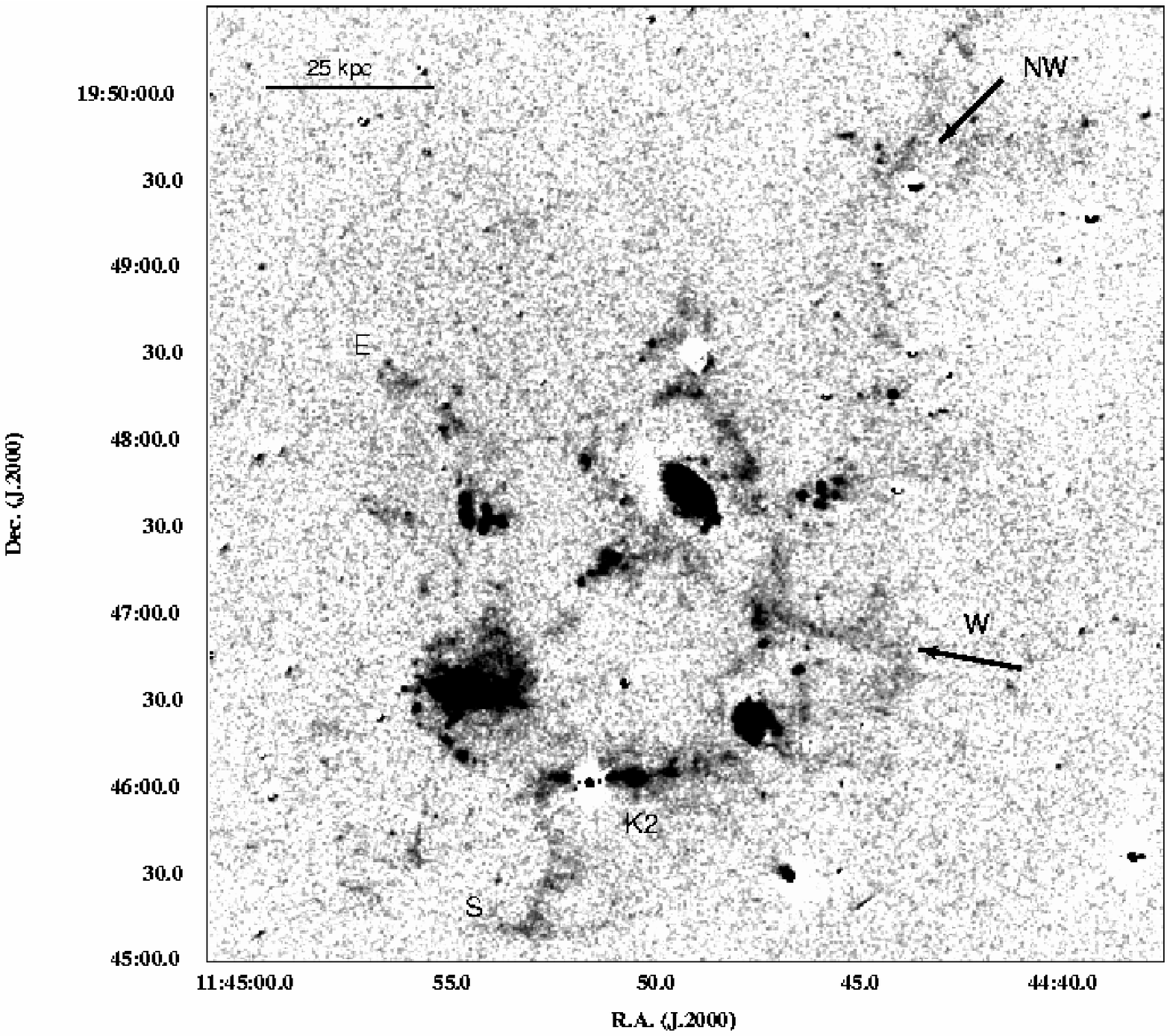}
\caption{$r$ band continuum (top) and H$\alpha$ net (bottom) images of the Blue Infalling Group in A1367.
Notice the large-scale filamentary
trails detected in the H$\alpha$ line (adapted from \cite{COGA06}).}
\label{bigfig}
\end{figure}
 The Blue Infalling Group in A1367 independently discovered by \cite{SAKK02} and by \cite{GAVC03} and 
recently studied in depth by \cite{COGA06}
is a typical (and to our knowledge unique) example of a pre-processing event in the nearby Universe (see Fig. \ref{bigfig}).
This group is composed of two $\sim$ $M^*$ galaxies (CGCG 97-114 and 97-125) and several
star forming dwarfs and extragalactic HII regions, all forming stars at a prodigious rate. 
The group has a small velocity dispersion ($\sim$ 170 
kms$^{-1}$) and dynamical considerations indicate that it is falling at high 
velocity ($\sim$ 1700 kms$^{-1}$) into the cluster A1367 \citep{GAVC03,CORG04}.
All members show perturbed morphologies and enhanced star formation activity that seem produced by ongoing tidal interactions.
High metallicity is determined in the group dwarfs and extragalactic HII regions, suggesting that
they were produced by the violent tidal interaction between the two massive galaxies \citep{SAKK02},
contrary to what it would be expected if they were independently evolved dwarfs. More evidence for 
gravitational interactions among the group members is offered by the early-type galaxy 
CGCG 97-125 (see Fig. \ref{bigfig}) whose evident stellar shells witness a recent merging event. 
\cite{COGA06} discovered long ($\sim$ 150 kpc) H$\alpha$ trails associated with this group that are
interpreted as due to ram-pressure stripping exerted by the cluster ICM on the 
ISM of the high velocity infalling galaxies. The thermal energy in the IGM is invoked to keep the gaseous trails
ionized for a time much longer ($10^8$ yr) than the recombination time ($10^7$ yr).

\section{Discussion and conclusion}
\label{Discussion}

Conclusive evidence built up in the last 25 years that, beside the morphological segregation effect, 
late-type galaxies belonging to rich nearby clusters differ significantly from their ``field'' counterparts. 
Primarily they are found deficient in atomic gas with respect to isolated galaxies (see sect. \ref{The atomic gas}).
Large HI deficiency parameters provide the clearest signature of their cluster membership.   
Because of the lack of gas, their star formation is significantly quenched (see sect. \ref{star formation}), 
despite the fact that the molecular gas component seems little perturbed (see sect. \ref{The molecular gas}).
Additional, but less clear-cut differences are: 
an excess of radio continuum activity, a higher metallicity and possibly a lower
dust mass content of galaxies in high density environments. 
Kinematic properties instead (i.e. rotation curves), witnessing the dark matter distribution, 
seem little affected in clusters (see sect. \ref{constituents} and \ref{obs_diff}).

 Which physical processes, among the ones reviewed in sect. \ref{models}, are affecting
the evolution of late-type galaxies falling on today's clusters?
The gravitational perturbations seem unlikely.
Tidal interactions among galaxies are uneffective simply because 
in today's high velocity dispersion clusters they are too short-lived 
to produce severe disturbances (sect. \ref{models} and \ref{timescales}).
The gravitational interaction with the cluster potential might induce some nuclear gas infall, 
but it can hardly remove the outer disk HI and thus originate the cluster HI-deficient galaxy 
population (sect. \ref{tidalcluster}). Gas stripping in HI-deficient galaxies seems too a recent 
event (some 10$^8$ yr) for being produced by galaxy harassment, whose time scale 
is $\sim$ ten times longer (sect. \ref{harassment}).\\
In today's universe the interactions with the IGM are likely to be dominant.
Among interactions with the cluster IGM, evaporation is effective in hot Coma-like 
clusters, especially for small galaxies,
unless they are shielded by relevant magnetic fields (sect. \ref{evaporation}).
Ram pressure seems however the most relevant process whenever the gas 
is not too hot (A1367, Virgo) 
and provided that the cluster crossing velocity is high (see sect. \ref{Ram}).
This condition seems satisfied since the orbits of spirals in nearby clusters are predominantly radial, 
as indicated by their non-Gaussian velocity distribution (see sect. \ref{infall}). 
Dynamical simulations confirm this 
picture also for the young, unrelaxed Virgo cluster, whose IGM density is not particularly high.
Gas stripping might be locally more effective if the IGM has a non-uniform, 
clumpy distribution. Considering that the effects of ram-pressure, thermal evaporation 
and viscous stripping are probably additive,
on time scales comparable to a crossing time,
any late-type galaxy infalling into a rich cluster is expected to loose 
most of its atomic gas reservoir, in particular the weakly gravitationally bound one located 
in the outer disk.\\
Observations corroborate the idea that galaxy - 
IGM interactions are the dominant phenomena affecting the evolution of late-type cluster galaxies in nearby clusters.
The location of most severely HI-deficient late-type galaxies, with the smallest HI disks 
near the  peaks of the 
IGM gas density argues in favor of the hydrodynamic scenario (sect. \ref{The atomic gas}). 
\begin{figure}[!ht]
\epsscale{0.8}
\plotone{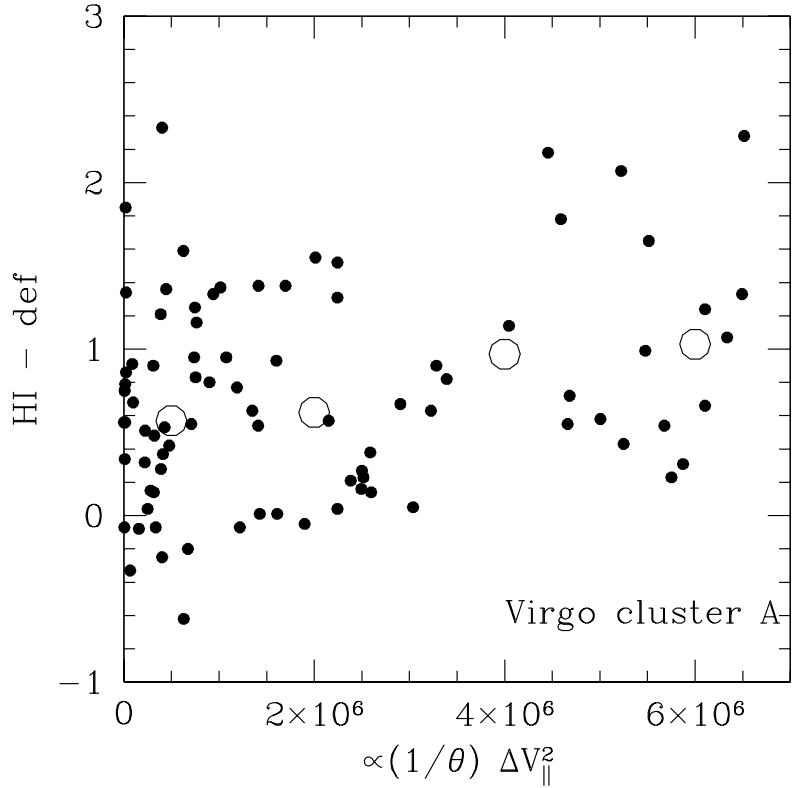}
\caption{The correlation between the HI deficiency parameter and the
square of the velocity deviation along the line of sight, divided by the angular distance
from the center of cluster A in Virgo. Empty dots give averages in regularly spaced bins.}
\label{defagg30}
\end{figure}
There is also evidence that spirals in the Virgo main cluster 
A show a weak correlation (see Fig. \ref{defagg30}) between
the HI deficiency parameter and the quantity $(1/\theta) \times \Delta V_{||}^2$, proportional 
to the square of the velocity along the line of sight divided by the angular distance
from the cluster center (a quantity that is expected to increase with the the ram-pressure $\rho~V^2$),
arguing in favor of the ram-pressure hypothesis (notice that $V^2$ is very poorly determined from the 
observable component of the velocity parallel to the line of sight). 
Certainly this evidence does not support the tidal interaction scenario where $HI-def$ 
should rather decrease with $\Delta V_{||}$.\\ 
The truncation of the 
star forming disk in cluster HI-deficient spirals (see sect. \ref{star formation}) 
can also be explained in the ram-pressure scenario because 
the star formation is progressively suppressed once the gas is removed from the outer disk, 
as in NGC 4569. 
The interaction with the cluster IGM might sometime increase, on short time scales
and in specific configurations, the total star formation activity of galaxies, 
as observed (i.e. CGCG 97-073 in A1367) and predicted by models (sect. \ref{Ram}), but 
on long timescales the galaxy global activity should however decrease.\\ 
On the contrary gravitational interactions are expected to trigger star formation, 
in particular in the nuclei (sect. \ref{tidal}), while  
the observations
have unambiguously shown that on average the star formation activity 
of giant late-type galaxies decreases towards the cluster center.
There are also evidences that cluster galaxies did not undergo 
strong bursts of star formation, as their
constant H$\alpha$ to UV flux ratio implies, and that not all 
cluster galaxies with a truncated gas or young stellar disk
have enhanced nuclear activity (gravitational interactions are expected 
first to trigger nuclear gas infall and only in the 
strongest cases to produce disk truncation) (see sect. \ref{star formation}). \\
Several prototypical galaxies provide convincing cases
that the interactions with the cluster IGM are predominant in the nearby universe.
NGC 4569, the typical HI-deficient, anemic galaxy in the Virgo cluster
has kinematic and spectro-photometric disk properties that mimic a recent (300-400 Myr) 
ram-pressure stripping event (see sect. \ref{prototype}). 
We cannot exclude that some nuclear gas infall has been  
induced by the gravitational interaction with the potential of the cluster. 
Broadly speaking, the fact that most of the HI-deficient galaxies have truncated 
 HI and H$\alpha$  disks,  but not at longer wavelengths 
confirms that in most cases the interaction 
responsible for the gas depletion is relatively recent ($\leq$ 5 10$^8$ yr), 
as in NGC 4569.
The existence of several marginally HI-deficient cluster spirals with asymmetric 
HI-profiles provides further evidence that the 
galaxy-IGM interaction is a recent event. The long, ionized trails 
behind CGCG 97-073 and CGCG 97-079 
impose even stronger time constraints, indicating that in a few cases the 
interaction is still undergoing (sect. \ref{prototype} and sect.\ref{timescales}).

The hydrodynamical galaxy - IGM interaction scenario poses some problems that might 
however only be apparent. 
It has been claimed that observing
the threshold in the star formation activity  
at $\sim$ one virial radius, i.e. at clustercentric 
distances where the IGM gas density is low \citep{TREU03}, 
argues against the ram-pressure stripping scenario.
This argument has been used in favor of other processes dominating the evolution of cluster 
gas-rich galaxies, such as pre-processing and starvation \citep{BALN00, LEWB02, BOWB04, DRES04b, NICH04}.
First of all this result is based on wide field surveys, 
such as SDSS and 2dF, where HI data are unavailable and the optical 
galaxy classification is based on light profiles and/or spectra,
without direct plate inspection.
The threshold in the mean star formation activity of late-type cluster
galaxies might be contaminated by the morphology segregation effect.
 
Furthermore we have stressed that the evidence offered by 
CGCG 97-079 and 97-073 in A1367 that ram-pressure is active at least up to
$\sim$ 0.8 Mpc (0.4 virial radius) combined with other  examples showing that
the effects of stripping might last until $\sim$ 1 virial radius,
reduces the necessity of invoking other processes such as pre-processing 
or starvation acting in the outskirt of clusters at $z$=0.
Compact groups accreting on clusters, such as the Blue Infalling Group in A1367 (sect. \ref{prototype}),
where effective pre-processing might take place, are extremely rare at $z$=0. 
We have in fact evidence that galaxies infall as individuals
(CGCG 97-073 and CGCG 97-079), or belonging
to extended sub structures with velocity dispersion comparable to already 
formed clusters ($\geq$ 500 km s$^{-1}$, sect. \ref{infall}).

A unified picture that combines near and far, present and past can be drawn coherently
if one takes into account that the environmental conditions and the physical properties
of galaxies have evolved significantly during the cosmic time, making
the ranking of the mechanism that prevailed in shaping the Hubble sequence to change with time. 
The number of low velocity dispersion infalling groups, 
where pre-processing is effective, increased with $z$; 
the density of the IGM and its temperature decreased with $z$. 
The gas fraction and the activity of star formation of galaxies, independently 
of their environmental conditions, increased significantly from $z$=0 at least up to $z$=1. 

Given that the interaction with the cluster IGM cannot easily 
thicken the disk of a spiral galaxy and
transform it into an S0 (sect. \ref{Surface brightness})\footnote{In the ram-pressure scenario
bulges can form only under particular conditions,
such as through the triggering of cycles of annealing after several crossings of 
the cluster \citep{SCHS01} or
edge-on stripping, when up to 50\% of the gas can be re-accreted by the disk 
\citep{VOLC01a} (see sect. \ref{Ram})},  
the formation of bulges is obtained straightforwardly
from disk heating (perturbing or even stopping the rotation)
during gravitational interactions with nearby 
companions and/or with the cluster potential
(galaxy harassment)
(sect. \ref{tidal}, \ref{tidalcluster}, \ref{harassment}).
 The observed kinematical, structural and spectrophotometrical properties of bright lenticulars 
in nearby clusters are consistent with this interpretation. 
For low-luminosity lenticulars, however, as stated by 
\cite{POGB04}, there are some evidences that the suppression of the star 
formation was due to their interaction with the IGM (sect. \ref{Surface brightness}). 
These considerations have induced groups of researchers
working on clusters at high $z$ to identify in gravitational pre-processing 
the most likely mechanism at the origin of the morphological segregation \citep{DRES04b},
although others invoked smoother processes such as starvation \citep{BALN00}. 
We refer the reader to several papers that appeared in the 
Carnegie Observatories Astrophysics Series, Vol. 3: 
Clusters of Galaxies: Probes of Cosmological Structure and Galaxy Evolution (2004)
elaborating on this idea.

Several problems concerning the evolution of galaxies in clusters remain unsolved in this unified scenario. 
One deals for example with the origin of dwarf galaxies.
Although the strong HI-deficiency of 
cluster dwarf irregulars expected in the hydrodynamical interaction scenario cannot be 
proved because of the lack of sensitive HI observations
(sect. \ref{The atomic gas}), there is 
evidence that their star formation activity does not decrease 
with clustercentric distances as for massive objects (sect. \ref{star formation}),
maybe due to continuous replenishment of gas-rich dwarfs from outside, or
maybe because dwarf irregulars, after gas removal, quickly fade into dEs,
thus they drop out of late-type surveys.   
A close link between dIs and dEs, first proposed by 
\cite{LINF83}, is strengthened by the existence of rotationally 
supported dEs \citep{VANZ04a,VANZ04b}, 
by the residual/past star formation activity of dEs observed in UV \citep{BOSC05c}
or in optical spectra (i.e. VCC1499, \cite{GAVZ01, CONS03b}), 
by the existence of dEs not totally  devoid of gas
\citep{CONS03a}, by the shape of the faint 
end of the cluster luminosity functions \citep{CONS02} and by the
velocity distribution as broad as that of dwarf irregulars \citep{CONG01}.

It seems fair to conclude that 
gravitational perturbations in infalling groups are 
at the origin of local lenticular galaxies. 
However the process dominating the evolution of gas-rich, 
late-type galaxies falling-in at the present epoch is the interaction with the hot 
and dense IGM. 
The fate of late-type galaxies will be slightly different from what it would
have been a Gyr ago: they will become HI-deficient,
anemic, rotationally-supported spirals. On long timescales 
(several crossing times), because of the lack of gas supply, they will lose
angular momentum, somewhat heating their stellar disks. 
They will probably become disk dominated, quiescent systems (\ref{Surface brightness}), but 
they will never make it to become real S0s.\\
If this evolutionary picture is correct, we still have to understand at 
which epoch and under what environmental conditions the
hydrodynamical and the gravitational processes 
interchanged their prevailing role and, among other things,
compare it with the inset of the Buchler-Oemler effect. 
These observational goals are within the reach of last generation ground based 
optical telescopes and of the HST. However we must wait another decade before 
neutral hydrogen measurements of galaxies at significant $z$ with SKA will shade the
final light on the evolutionary processes that took place in clusters of galaxies.

\newpage 
\section{Appendix: Characteristic time scales}
\label{timescales}

The age and the duration of certain interactions, their probability and their frequency 
can be relatively well quantified  from the observations. 
In this appendix we summarize some time scales relevant to the processes acting on nearby clusters of galaxies:

\subsection{Galaxy revolution time scale: $t_{rev}$}

The time required by a galaxy to make a complete revolution around the rotation axis, 
$t_{rev}$, is given by the relation:

\begin{equation}
{t_{rev}=2 \pi r_{gal}/\Delta V_{gal}}
\end{equation}
\noindent
For normal late-type galaxies, with typical luminosities in the range $8<log L_H<12$ (L$_{H\odot}$)
$t_{rev}$ ranges between 5 $\times 10^7$ and 3 $\times10^8$ yr.\\
Some cluster galaxies have
asymmetric HI distributions, with most of the gas located on one side of the galaxy, as observed using both 
interferometers and single dish radiotelescopes\footnote{The HI profiles of asymmetric galaxies with
an inclination larger than $\geq$ 30 degrees are ``one-horned'' instead of ``two-horned'' as
observed in isolated unperturbed galaxies \citep{GAV89}}.
Whenever the atomic gas is partly removed from the disk of the galaxy, the resulting
asymmetric HI distribution lasts for a few times $t_{rev}$. 
On longer time scales differential rotation re-distributes uniformly the gas over the disk.
Thus observing an HI asymmetry implies
that the gas removing mechanism has been acting since a very short time,
or that gas sweeping is an ongoing process: whatever the interaction is, it ended
no more than $t_{rev}$ years ago.   
Typical galaxies with an asymmetric HI distribution are NGC 4654 \citep{PHOM95}
and NGC 4438 \citep{CAYG90} in Virgo, NGC 4848 (CGCG 160-055),
NGC 4921 (CGCG 160-096) in Coma \citep{GAV89,BRAC01}, UGC 6697 (CGCG 97-087) and CGCG 97-079 in A1367
\citep{GAV89, DICG91}. These are marginally HI-deficient galaxies: the ongoing interaction
did not have enough time yet to remove the majority of the atomic gas.\\

\subsection{Cluster crossing time: $t_{cross}$}

It is more difficult to date the stripping epoch of HI-deficient galaxies (HI deficiency $\ge$ 0.3)
since several processes can be responsible for gas removal. 
Ram-pressure, for instance,has a time scale comparable with the
cluster crossing time $t_{cross}$.\\


The cluster crossing time $t_{cross}$ at the virialization radius $R_{vir}$ is given by the relation:
\begin{equation}
{t_{cross} = 2 \times R_{vir}/V_{gal} \simeq 2 \times R_{vir}/\sqrt{3}\delta V_{cluster}}
\end{equation}
Typical crossing times for galaxies in the clusters Coma, A1367 and Virgo range between
2 and 3 $\times 10^9$ yr (see Table \ref{tabtimescales}).\\


\subsection{Duration of tidal encounters: $t_{enc}$}

Tidal encounters  have shorter
timescales, of the order of $t_{enc}$ $\sim$ 10$^8$ years, 
the time during which two galaxies moving at high speed in the
potential well of the cluster remain close enough to get tidally perturbed.
The time scale defining the effective duration of high velocity encounters $t_{enc}$ such as
galaxy-galaxy tidal interactions can be estimated using the relation 
\citep{BINTRE87}:
\begin{equation}
{t_{enc} \simeq max(r(1)_{gal},r(2)_{gal},b)/\Delta V}
\end{equation}
\noindent
where $r(1)_{gal}$ and $r(2)_{gal}$ are the radii of the two galaxies, 
$b$ their separation at the closest approach and $\Delta V$ the relative speed.
For galaxies belonging to the three clusters listed in Table \ref{Tab1}, for $\Delta V$ $\simeq$ $\delta V_{cluster}$, 
$t_{enc}$ $\sim$ 10$^8$ yr. Another very short time-scale.\\

\subsection{Relaxation time: $t_{relax}$}

The frequency of encounters, however, is significantly smaller  ($\sim$ $1/t_{relax}$).
The relaxation time $t_{relax}$, the time necessary for a galaxy to have its orbit perturbed by a tidal interaction, 
is given by the relation:
\begin{equation} 
{t_{relax} = 0.1 \times (D_{cluster}/\delta V_{cluster}) \times (N_{gal}/ln(N_{gal}))=0.2 \times (R_{vir}/\delta
V_{cluster}) \times (N_{gal}/ln(N_{gal})) }
\end{equation}
where $D_{cluster}$ is the size and $\delta V_{cluster}$ the velocity dispersion of the cluster, 
and $N_{gal}$ the number of members \citep{BINTRE87, BYRV90}. 
The typical relaxation time for the three clusters analyzed in this work can be estimated using 
the data listed in Table \ref{Tab1} and using a rough estimate of $N_{gal}\sim$ 1000.
They are $\sim 5 \times 10^{10}$ yr, comparable to the age of the Universe (see Table \ref{tabtimescales}).\\

\subsection{Merging time: $t_{mer}$}

\cite{MAK97} studied the merging rate of galaxies of similar mass in clusters.
Merging processes are relatively rare in rich clusters since the probability of a merger
is significantly higher in low speed encounters. If $N_{gal}$ is the 
number of galaxies randomly distributed in a cluster of radius $D_{cluster}$, $r_h$ is the half-mass radius
of the galaxy, $\Delta V_{gal}$ and $\delta V_{cluster}$ the galaxy and cluster velocity dispersions,
the merging time $t_{mer}$ is given by the relation \citep{MAK97}:

\begin{equation}
{t_{mer} \sim \frac{2000}{N_{gal}^2}(\frac{D_{cluster}}{1 Mpc})^3(\frac{0.1 Mpc}{r_h})^2(\frac{100 km s^{-1}}{\Delta V_{gal}})^4(\frac{\delta
V_{cluster}}{300 km s^{-1}})^3 Gyr}
\end{equation}
\noindent
As shown in Table \ref{tabtimescales} the time scales for merging of 
$L^*$ galaxies inside the cluster virial radius are relatively high in Coma, A1367 and Virgo.
Given the short length of time of $t_{enc}$ and the low frequency of encounters ($\sim$ $1/t_{relax}$),
it is not surprising that in the whole Virgo cluster only NGC 4438 shows 
clear evidence for ongoing tidal interactions.
$t_{merg}$ must have been much shorter in the past because of the strong dependence $\sim (1+z)^3$ 
of the density on $z$. 

\begin{table}
\caption{The typical time scales for galaxies in the Coma, A1367 and Virgo clusters.} 
\label{Tab4}
\[
\begin{array}{rccc}
\hline
\noalign{\smallskip}
Cluster & t_{cross} & t_{relax} & t_{merging}\\
        & yr        &  yr     &  yr   \\
\noalign{\smallskip}
\hline
\noalign{\smallskip}
Coma &  1.6~10^9 &2.4~10^{11}  &  6.9~10^{11} \\
A1367&  1.7~10^9 &7.2~10^{10}  &  1.7~10^{12} \\
Virgo&  1.7~10^9 &3.6~10^{10}  &  3.4~10^{11} \\
\noalign{\smallskip}
\hline
\end{array}
\]
\label{tabtimescales}
\end{table}

\subsection{Recombination time: $t_{rec}$}

The presence of asymmetric radio continuum tails, such as those observed in 
CGCG 97-073 and CGCG 97-079 in A1367
\citep{GAVC95b}, indicate that the interaction is still ongoing. 
These galaxies also present long ($\sim$ 75 kpc) ionized gas tails \cite{GAVB01b}. The recombination time $t_{rec}$
is given by the relation \citep{OST89}:
\begin{equation}
{t_{rec} = 1/(n_e*\alpha_A)}
\end{equation}
\noindent
where $n_e$ is the electron density and $\alpha_A$=4.2 10$^{-13}$ cm$^3$ s$^{-1}$. If the gas density is known, 
and in the hypothesis that the gas is not ionized in situ, this relation can be used, combined with the
tail length and the galaxy speed within the cluster, to estimate a lower limit to the first interaction 
responsible for the formation of the ionized tails. \cite{GAVB01b} estimate that this time scale is a
few times $10^7$ yrs.\\
\begin{figure}[!htb]
\epsscale{0.5}
\plotone{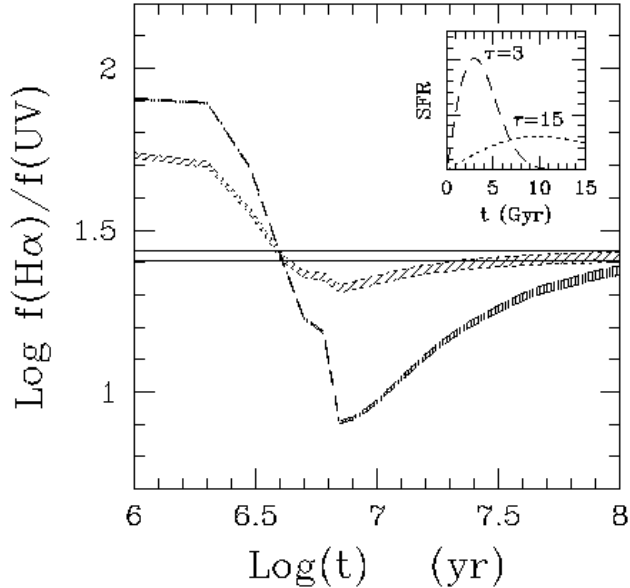}
\caption{The time behavior of the H$\alpha$ to UV flux ratio after an instantaneous burst of star formation that
occurred at $t=0$ of 
intensity 100 (thick shaded region) and 10 (thin shaded region) times higher than the continuum formation activity,
adapted from \cite{IGLBG03}. 
}
\label{HAUV}
\end{figure}
\subsection{Dating recent starbursts: $ t_{starburst}$}

Beside fitting the SED's continuum from the UV to the near-IR  with population synthesis models 
\citep{GAVB02c}, several spectral features, combined with population synthesis models, are used to estimate the 
age of the underlying stellar population (see \cite{WOR94, POGB96, BRUC03}). The debated issue here
is which combination of indexes should be used to break the age-metallicity degeneracy.\\
Many applications of these methods are found in the literature aimed at
estimating the age of galaxies in distant clusters. A local application to Coma PSB galaxies 
is the one by \cite{POGB04}. 
Spectral indices can be used to date the age of starbursts. The most generally used are 
the Balmer emission and underlying absorption lines. The draw-back of these methods is that
the results are strongly dependent on the assumed star formation history: significant
age difference arise from the same index 
if one or more instantaneous or continuum starbursts are assumed.\\
A  promising clock for dating recent bursts of star formation comes from the 
H$\alpha$ to UV flux ratio, assuming that the IMF is universal.
The emission at these wavelengths is associated with young stars. 
However, the stars producing the ionizing photons responsible of the H$\alpha$ emission 
are O-B stars, younger ($t$ $<$ 4 10$^6$ yr) and more massive ($m$ $>$ 10 M$\odot$) than the A stars 
($m$ $\sim$ 2-5 M$\odot$; $t$ $\sim$ 3 10$^8$ yr) emitting in the UV \citep{BOSG01}. 
After an instantaneous starburst 
the H$\alpha$ to UV ratio would thus increase on a time scale comparable to the life time of the O-B stars 
(0$<$ $t_{starburst}$ $\simeq$ 4 10$^6$ yr) and decrease on timescales 
comparable to the age of the UV emitting stars (10$^7$ $<$ $t_{starburst}$ $\simeq$ 3 10$^8$ yr), 
as shown in Fig. \ref{HAUV} (adapted from \cite{IGLBG03}). 
However, the variation of the H$\alpha$ to UV flux ratio is significant only for
intense starbursts, where the star formation activity increases by at least a factor of 10 with respect 
to the standard activity. This promising technique needs
high quality UV and H$\alpha$ photometric data, as well as Balmer decrement measurements and far-IR data 
necessary for an accurate dust extinction correction \citep{IGLBG03}.

\begin{acknowledgements}

We wish to thank J.M. Deharveng for encouraging us to write this review.
C. Adami, H. B\"ohringer, S. Boissier, L. Cortese, J. Iglesias, C. Mendes de Oliveira, H. Wozniak, A. Zaccardo
for their contributions. M. Colpi, L. Cortese, J.M. Deharveng, M. Haynes, R. Kennicutt, J. Lequeux, 
B. Poggianti and H. Wozniak for discussions and comments. We want to thank an anonimous 
referee whose comments and suggestions were precious for improving the present manuscript.

\end{acknowledgements}

\end{document}